\documentclass{aa}  
\usepackage{natbib,epsfig,graphicx,color,psfrag,amsmath,pdflscape}
\usepackage[varg]{txfonts}
\bibpunct{(}{)}{,}{a}{}{,}
\newcommand{\msun}{\ensuremath{M_{\odot}}}

\newcommand{\rsun}{\ensuremath{R_{\odot}}}

\newcommand{\Teff}{\ensuremath{T_{\rm eff}}}
\newcommand{\vinf}{\ensuremath{v_{\infty}}}
\newcommand{\mdot}{\ensuremath{\dot{M}}}

\newcommand{\msunyr}{\ensuremath{M_{\odot} {\rm yr}^{-1}}}

\newcommand{\beq}{\begin{equation}}
\newcommand{\eeq}{\end{equation}}
\newcommand{\beqa}{\begin{eqnarray}}
\newcommand{\eeqa}{\end{eqnarray}}

\newcommand{\nbeq}{\begin{equation*}}
\newcommand{\neeq}{\end{equation*}}

\newcommand{\kms}{\ensuremath{{\rm km}\,{\rm s}^{-1}}}
\newcommand{\seconds}{\ensuremath{{\rm s}}}

\newcommand{\dd}{{\rm d}}

\newcommand\OI{O\,{\sc i}}

\newcommand\SiIV{Si\,{\sc iv}}

\newcommand{\Ha} {H$_{\rm \alpha}$}

\newcommand{\Rstar}{\ensuremath{R_{\ast}}}
\newcommand{\Rmax}{\ensuremath{R_{\rm max}}}

\newcommand{\logg}{\ensuremath{\log g}}
\newcommand{\YHe}{\ensuremath{Y_{\rm He}}}

\newcommand{\Trad}{\ensuremath{T_{\rm rad}}}

\newcommand{\vesc}{\ensuremath{v_{\rm esc}}}
\newcommand{\vth}{\ensuremath{v_{\rm th}}}

\newcommand{\vrot}{\ensuremath{v_{\rm rot}}}

\newcommand{\vsini}{\ensuremath{v{\thinspace}\sin{\thinspace}i}}
\newcommand{\vturb}{\ensuremath{v_{\rm turb}}}

\newcommand{\chibar}{\ensuremath{\bar{\chi}}}

\newcommand{\Jbar}{\ensuremath{\bar J}}

\newcommand\ie{\hbox{i.e.}}
\newcommand\eg{\hbox{e.g.}}

\newcommand{\vecown}[1]{\ensuremath{\vec{#1}}}
\newcommand{\matown}[1]{\ensuremath{\vec{#1}}}

\newcommand{\iincv}{\ensuremath{\vecown{I}_{\rm inc}}}
\newcommand{\scontv}{\ensuremath{\vecown{S}^{\rm (c)}}}
\newcommand{\slinev}{\ensuremath{\vecown{S}^{\rm (L)}}}

\newcommand{\alo}{\ensuremath{\Lambda^\ast}}
\newcommand{\alom}{\ensuremath{\matown{\Lambda}^\ast}}
\newcommand{\unitym}{\ensuremath{\matown{1}}}

\newcommand{\epsc}{\ensuremath{\epsilon_{\rm c}}}
\newcommand{\epsl}{\ensuremath{\epsilon_{\rm L}}}
\newcommand{\scont}{\ensuremath{S_{\rm c}}}
\newcommand{\sline}{\ensuremath{S_{\rm L}}}
\newcommand{\scontijk}{\ensuremath{S_{\rm ijk}^{\rm (c)}}}
\newcommand{\slineijk}{\ensuremath{S_{\rm ijk}^{\rm (L)}}}
\newcommand{\kcont}{\ensuremath{k_{\rm c}}}
\newcommand{\kline}{\ensuremath{k_{\rm L}}}
\newcommand{\kapline}{\ensuremath{\kappa_{\rm 0}}}
\newcommand{\chicijk}{\ensuremath{\chi_{\rm ijk}^{(\rm c)}}}
\newcommand{\chibarijk}{\ensuremath{\bar{\chi}_{\rm ijk}^{\rm (L)}}}
\newcommand{\chic}{\ensuremath{\chi_{\rm c}}}
\newcommand{\chil}{\ensuremath{\chi_{\rm L}}}
\newcommand{\chith}{\ensuremath{\chi_{\rm Th}}}
\newcommand{\chiinf}{\ensuremath{\chi_{\rm \infty}}}

\newcommand{\profile}{\ensuremath{\Phi_x}}
\newcommand{\profilenu}{\ensuremath{\Phi_\nu}}

\newcommand{\nx}{\ensuremath{N_{\rm x}}}
\newcommand{\ny}{\ensuremath{N_{\rm y}}}
\newcommand{\nz}{\ensuremath{N_{\rm z}}}
\newcommand{\nnz}{\ensuremath{N_{\rm NZ}}}
\newcommand{\nnu}{\ensuremath{N_{\rm F}}}
\newcommand{\ntheta}{\ensuremath{N_{\rm \theta}}}
\newcommand{\nphi}{\ensuremath{N_{\rm \phi}}}

\newcommand{\vmin}{\ensuremath{v_{\rm min}}}
\newcommand{\hei}{\ensuremath{I_{\rm He}}}

\newcommand{\Jnu}{\ensuremath{J_{\rm \nu}}}
\newcommand{\Bnu}{\ensuremath{B_{\rm \nu}}}

\newcommand{\eqrt}{\hbox{EQRT}}

\newcommand{\ralf}{\ensuremath{R_{\rm A}}}
\newcommand{\rapex}{\ensuremath{r_{\rm m}}}
\newcommand{\rkep}{\ensuremath{R_{\rm K}}}
\newcommand{\Bpole}{\ensuremath{B_{\rm p}}}
\newcommand{\etastar}{\ensuremath{\eta_{\rm \ast}}}
\newcommand{\mdotz}{\ensuremath{\dot{M}_{\rm B=0}}}

\newcommand{\vthfid}{\ensuremath{v_{\rm th}^{\rm \ast}}}
\newcommand{\ddopfid}{\ensuremath{\Delta \nu_{\rm D}^{\rm \ast}}}
\newcommand{\ddop}{\ensuremath{\Delta \nu_{\rm D}}}

\newcommand{\wrt}{w.r.t.}
\newcommand{\sigmae}{\ensuremath{\sigma_{\rm e}}}
\newcommand{\nel}{\ensuremath{n_{\rm e}}}

\newcommand{\mbh}{\ensuremath{M_{\rm BH}}}

\newcommand{\hdname}{\textit{}}
\newcommand{\omp}{\textsc{OpenMP}}

\newcommand{\Cul}{\ensuremath{C_{\rm ul}}}
\newcommand{\Aul}{\ensuremath{A_{\rm ul}}}

\newenvironment{rcases}{\left.\begin{aligned}}{\end{aligned}\right\rbrace}

\begin{document}
\title{3D radiative transfer: Continuum and line scattering in non-spherical
  winds from OB stars}

\author{L. Hennicker\inst{1}, J. Puls\inst{1}, N. D. Kee\inst{2}, and J. O. Sundqvist\inst{3}
}

\institute{LMU M\"unchen, Universit\"atssternwarte, Scheinerstr. 1, 81679
M\"unchen,
           Germany, \email{levin@usm.uni-muenchen.de}
           \and
           Institut f\"ur Astronomie und Astrophysik, Universit\"at T\"ubingen,
           Auf der Morgenstelle 10, 72076 T\"ubingen, Germany
           \and
           Instituut voor Sterrenkunde, KU Leuven, Celestijnenlaan 200D,
           3001 Leuven, Belgium
}

\date{Received 30 August 2017; Accepted 29 April 2018}

\abstract 
{State of the art quantitative spectroscopy utilizes synthetic spectra to
  extract information from observations. For hot, massive stars, these
  synthetic spectra are calculated by means of 1D, spherically symmetric, NLTE
  atmosphere and spectrum-synthesis codes. Certain stellar atmospheres,
  however, show strong deviations from spherical symmetry, and need to be
  treated in 3D.}
{We present and test a newly developed 3D radiative transfer code, tailored to
  the solution of the radiation field in rapidly expanding stellar
  atmospheres. We apply our code to the continuum transfer in wind-ablation
  models, and to the UV resonance line formation in magnetic winds.}
{We have used a 3D finite-volume method for the solution of the
  time-independent equation of radiative transfer, to study continuum- and
  line-scattering problems, currently approximated by a
  two-level-atom. Convergence has been accelerated by coupling the formal
  solver to a non-local approximate $\Lambda$-iteration scheme.  Particular
  emphasis has been put on careful tests, by comparing with alternative
  solutions for 1D, spherically symmetric model atmospheres. These tests
  allowed us to understand certain shortcomings of the methods, and to
  estimate limiting cases that can actually be calculated.}
{The typical errors of the converged source functions, when compared to 1D
  solutions, are of the order of $10-20\%$, and rapidly increase for optically
  thick ($\tau \gtrsim 10$) continua, mostly due to the order of accuracy of
  our solution scheme.  In circumstellar discs, the radiation temperatures in
  the (optically thin) transition region from wind to disc are quite similar
  to corresponding values in the wind.  For MHD simulations of dynamical
  magnetospheres, the line profiles, calculated with our new 3D code, agree
  well with previous solutions using a 3D-SEI method. When compared with
  profiles resulting from the so-called analytic dynamical magnetosphere (ADM)
  model, however, significant differences become apparent.}
{Due to similar radiation temperatures in the wind and the transition region
  to the disc, the same line-strength distribution can be applied within
  radiation hydrodynamic calculations for optically thick circumstellar discs
  in `accreting high-mass stars'.  To properly describe the UV line
  formation in dynamical magnetospheres, the ADM model needs to be further
  developed, at least in a large part of the outer wind.}

\titlerunning{3D radiative transfer in non-spherical winds from OB stars}

\authorrunning{L. Hennicker et al.}

\maketitle
%
%
\section{Introduction} 
\label{sec:intro}
Hot, massive stars are key tools to interpret the Universe. For example, they
both enrich the interstellar medium with highly processed material due to
their strong stellar winds, as well as end their lives as supernovae,
enriching the surroundings with metals even further. Additionally, the latter
produce strong shocks within the interstellar medium, thus triggering star
formation.

Massive stars are thought to be the progenitors of massive black holes ($\mbh
\gtrsim 25\, \msun$) when their final fate is a direct collapse into a black
hole instead of a supernova explosion. The collapse into a black hole,
however, is only possible when the progenitor stars exhibit a weak wind, or,
more generally, do not lose too much mass during their
lifetime. \cite{Belczynski16} proposed a low-metallicity environment to
explain the massive black hole merger GW150914 ($M_{\rm 1}\approx 36\, \msun,
M_{\rm 2}\approx 29\, \msun$, \citealt{Abbott16}) observed at the Laser
Interferometric Gravitational-Wave Observatory (LIGO), whereas \cite{Petit17}
showed that magnetic surface fields could also lower the mass loss to form
such progenitor stars.

The physical properties of massive stars are generally derived by means of
quantitative spectroscopy, that is by modelling their atmospheres numerically
and comparing the calculated synthetic spectra with observations. When using
one-dimensional (1D), spherically symmetric codes (\eg~{\sc CMFGEN}
\citep{hilliermiller98}, {\sc PHOENIX}\footnote{There is also a 3D version of
  this code, see below.} \citep{Haus92}, {\sc PoWR} \citep{Graf02}, WM-{\sc
  basic} \citep{pauldrach01}, {\sc FASTWIND} \citep{Puls05, rivero12}, as a
non-exhaustive list for massive star atmospheric models), this has meanwhile
become a routine job and is widely applied. Within the last decade, however,
it became more and more evident, from both the theoretical and observational
side, that many stars differ from spherical symmetry, rendering the results
from 1D codes questionable for such objects. The deviations can be of
different origin and shape:

(i) Magnetic winds: \cite{Wade12} showed within the Magnetism in
  Massive Stars (MiMeS) survey, that about 7\% of all (Galactic) OB-stars
have detectable magnetic fields. Magneto-hydrodynamic (MHD) calculations from
\cite{udDoula02} and \cite{udDoula08} revealed, that large scale magnetic
fields, possibly of fossil origin \citep{Alecian13}, have a direct impact on
the stellar wind by channeling the wind outflow along magnetic-field lines,
often producing disc-like structures around the magnetic equator.  Depending
on the competition between wind energy and magnetic-field energy
(characterized by the so called Alfv\'en-radius, \ralf), and the rotational
velocities (characterized by the Kepler-radius, \rkep), a dynamical
($\ralf < \rkep$) or centrifugal ($\ralf > \rkep$) magnetosphere may form (see
also \citealt{Petit13}). In the former case, confined material in the
equatorial disc falls back onto the stellar surface, whereas in the latter
case, material is supported by centrifugal forces, forming a quite stable and
strongly confined disc structure.  \cite{Townsend05a} were able to explain the
observed Balmer-line variability in $\sigma$ Ori E, a magnetic Bp
star, by applying the oblique-rotator model. Recently, \cite{Owocki16}
developed a simplified model, the `analytic dynamical magnetosphere' (ADM), in
order to provide a framework for the analysis of magnetic winds, and were able
to reproduce the observed \Ha-line variations of \hdname{HD191612}. However, a
test of the ADM model for UV resonance lines is still missing, and will be
addressed in this paper.

(ii) Accretion and decretion discs: Both pre-main-sequence and a significant
fraction of massive main-sequence stars are observed to have circumstellar
discs, from which material is either accreted onto the star, or decreted by
different mechanisms, respectively. For classical Be stars, \cite{Kee16}
proposed that the destruction of the circumstellar disc by wind-ablation, that
is by radiative forces along the disc surface, is the major mechanism, which
possibly plays also a significant role for stars that already have ignited
hydrogen burning, however are still accreting material through a disc. These
objects will be named `accreting high-mass stars' within this
paper. Although \cite{Kee15} applied reasonable approximations in order to
derive the incident intensity at the disc's surface layers, the local
ionization stages are not treated explicitly. Since the line-force depends on
the line-strength distribution function (\eg~\citealt{CAK},
\citealt{Puls00}), which crucially depends on the local ionization stages of
the ions and therefore also on the mean intensity, a consistent treatment of
the (pseudo) continuum radiative transfer (RT) is needed.

(iii) Fast rotation: Stellar rotation is a natural consequence of the
classical star-formation scenario, that means of the collapse of an
initially rotating molecular cloud. Within the VLT-FLAMES Tarantula survey
(VFTS), \cite{RamirezAgudelo13} derived the distribution of projected
rotational velocities of (LMC) O-type stars, and found that it peaks at
relatively low values ($\vsini \approx 50-100$~\kms), and has an extended tail
towards higher ones, thought to be due to binary interaction. For example, the
very rapidly rotating O-star, VFTS102, rotates at a nearly critical rotation
rate \citep{Dufton11}.  Nearly critically rotating stars are highly distorted,
with a ratio of equatorial to polar radius approaching a value of
three half. Additionally, the emergent flux at the stellar surface becomes
latitude-dependent due to gravity darkening (see \citealt{Zeipel24} for
uniform rotation, and \citealt{MaederIV}, \citealt{MaederVI} for shellular
rotation).

(iv) Binary interaction: By a detailed investigation of binarity within the
VFTS, \cite{Sana13} found that about 50\% of the observed O-star population in
the Tarantula nebula have binary companions, that will interact with the
primary star during its lifetime via Roche-lobe-overflow, or even
merging. Roughly 30\% - 40\% of these stars have periods less than six days (see
also \citealt{Sana11}), which influences already the main-sequence
evolution. Due to the mass transfer, the donor star will be stripped and spins
down, whereas the companion will spin up due to the transport of angular
momentum (\eg~\citealt{deMink13}). In case of two massive stars being part of
the binary system, their winds will collide and lead to strong shocks and
X-ray emission (\eg~\citealt{Prilutskii1976}, \citealt{Cherepa1976},
\citealt{Stevens92}, \citealt{pittard09}), in addition to phase-locked
variations of recombination lines (see \citealt{Sana01}).

Since all these effects break the symmetry of many observed objects, a 3D
treatment of the RT has to be applied to derive the (possibly dynamical)
spectral energy distributions. This, however, is computationally expensive due
to the dependence of the radiation field on frequency and direction, in
addition to the three spatial dimensions.  Nevertheless, solution schemes have
been developed already from the early 70s on, starting with the
long-characteristics (LC) method (see \citealt{Jones73a} and
\citealt{Jones73b}), over the short-characteristics (SC) method (see
\citealt{Kunasz88}), to finite-volume methods (FVM, see \citealt{Adam90}).
Within the LC method, the equation of radiative transfer (\eqrt) is solved for
a given direction along a ray intersecting a considered grid point. Thus, for
a 3D grid with $N^3$ grid points, and assuming on average $N/2$ grid points
along the ray, $N^4/2$ operations have to be performed to obtain the
intensities at all grid points for a single direction. Each operation consists
of interpolating the physical properties from the 3D grid onto the ray, and
solving the formal integral along all points on the ray.  The SC method solves
the \eqrt ~only across each grid cell, with incident intensities on the cell
surfaces obtained by interpolation from the already known grid points. The
number of operations is thus reduced by a factor of $N/2$. For both methods,
the accuracy of the applied interpolation scheme is crucial for the final
performance of the obtained solution.

Instead of considering discrete rays, as in the above methods, the FVM solves
the EQRT for a single direction, by considering the complete set of rays with
the given direction entering the volume of a grid cell, and assuming that the
physical properties at the corresponding grid point are representative
averages for the complete cell. Though this still gives $N^3$ operations for
the solution for a single direction, all interpolations are avoided. Thus,
when compared to the LC or SC method, the amount of work needed for each
operation is reduced considerably. Interestingly, the development of methods
underwent an evolution from computationally most expensive methods (LC) to
cheaper methods (SC and FVM).

The above mentioned tools are called formal solvers, since they calculate the
radiation field for given source- and sink-terms. However, the most important
problem arises in scattering dominated atmospheres, where the sources (and
sinks) depend on the radiation field, and vice versa. Such atmospheres need to
be calculated by applying, for instance, the so-called $\Lambda$-iteration and
associated acceleration techniques (\eg~\citealt{Cannon73}).

Meanwhile, various 2D and 3D codes exist, with widely differing assumptions on
the underlying geometry and designated applications. In this respect, one
important point is the assumption of local thermodynamic equilibrium (LTE),
which cannot be applied in expanding atmospheres, when the radiation field
dominates the level populations. For the latter problem, the coupling between
the non-LTE (NLTE) rate equations and radiation field is mostly performed
using the aforementioned acceleration techniques (\eg~the accelerated
$\Lambda$-iteration, ALI). We briefly mention specific codes designated to the
multi-D RT in stellar atmospheres (for other applications, there are many more
such codes; \eg~for multi-D codes related to the ionization balance in the
interstellar medium, see \citealt{Weber2013} and references therein):

\textsc{Wind3D} \citep{Lobel08} is a 3D FVM code, which has been developed to
calculate the line transport in scattering dominated environments using
Cartesian coordinates, and with level populations approximated by a
two-level-atom (TLA). It adopts the classical $\Lambda$-iteration scheme,
which has poor convergence properties for optically thick,
scattering-dominated lines. \textsc{Wind3D} is thus restricted to the
treatment of weak lines. For those, however, \cite{Lobel08} were able to model
the time variations of discrete absorption components (DACs), as observed for
the \SiIV ~1400\,\AA\, doublet of the B0.5Ib supergiant
\hdname{HD\,64760}. Together with the hydrodynamic code \textsc{Zeus3D}, they
were able to reproduce these DACs, supporting the suggestion of
\cite{Mullan84}, that they arise from co-rotating interaction regions (see
also \citealt{CO96}).

\textsc{ASTAROTH} (\citealt{Georgiev06} and \citealt{Zsargo2006}) is a 2D SC
code, which is capable of solving the RT in parallel with the NLTE rate
equations for axisymmetric problems with non-monotonic velocity fields. It
uses spherical coordinates and includes a (local) ALI scheme. Many tests have
been performed by a comparison to 1D spherically symmetric models from {\sc
  CMFGEN}, giving errors of a few percent only. \cite{Zsargo08a} applied the
code to investigate the H and He ionization stages in the envelopes of B[e]
supergiants, and modelled, among others, the \Ha ~line.

\textsc{IRIS} \citep{Ibgui13} is a 3D SC code, which, to our knowledge, has
been only applied for studying laboratory-generated radiative shocks
\citep{Ibgui13a} thus far.  \cite{Ibgui13} performed test cases (searchlight
beam test and 1D plane-parallel models), which show an astonishing accuracy
for the solution of intensity, mean intensity, radiative flux, and radiation
pressure tensor, on non-uniform Cartesian coordinates and including velocity
field gradients. By now, the code assumes LTE. However, the inclusion of
scattering terms is planned for the future.

\textsc{MULTI3D} \citep{Leenaarts09} is a 3D SC code developed to accurately
solve the NLTE rate equations in cool FGK-type stars. \cite{Amarsi16} used
this code to predict, among other lines, the \OI ~777nm line for a grid of 3D
hydrodynamical models, by setting up a 23-level model atom. This code,
however, has been developed (and optimized) for the application to cool stars
(with subsonic velocity fields, at most), and cannot be used in our context.

\textsc{Phoenix/3D} (\citealt{Haus06} and other papers in this series) is a 3D
LC solver, which is capable of solving the RT together with the multi-level
NLTE rate equations. They use spherical, cylindrical or Cartesian coordinates,
and implemented a non-local ALI scheme. With the extension from
\cite{Seelmann10}, arbitrary velocity fields can be included as well. To our
knowledge, \textsc{Phoenix/3D} has not been applied, thus far, to real,
multi-dimensional, problems.

In this paper, we develop a 3D solution scheme to calculate continuum and line
source functions in expanding atmospheres. As a first step, we implement a FVM
to determine the formal solution, since this method has already been applied
in the context of 3D structures (\eg~\citealt{Adam90}, \citealt{Stenholm91},
\citealt{Lobel08}), and is supposed to have the lowest computational cost,
when compared to the SC and LC methods. To calculate optically thick continua
and strong lines, an ALI technique is required, and developed for the first
time within the 3D FVM framework by means of a non-local approximate
$\Lambda$-operator (ALO, see Sect.~\ref{sec:numerical_methods}). In
Sect.~\ref{sec:err_analysis}, we present, also for the first time, an error
analysis of the 3D FVM for the case of 1D spherical atmospheres, both for
optically thin and optically thick environments. This error analysis allows us
to understand certain shortcomings of the methods, and to estimate limiting
cases that can actually be calculated. As first applications for our 3D
solver, we study specific aspects related to the circumstellar discs of
accreting high-mass stars (Sect.~\ref{sec:wind_ablation}), and to the UV line
formation in magnetic winds (Sect.~\ref{sec:dm}).
\section{Basic assumptions}
\label{sec:eqrt}
The problems considered within this paper assume an (almost) stationary
atmospheric structure, meaning that the densities, velocities and boundary
conditions are assumed to be constant in time, or are at least changing much
slower than the radiation field. Thus, we use the time-independent EQRT
(\eg~\citealt{mihalasbook78}):
\beq
\label{eq:eqrt}
   \vecown{n} \vecown{\nabla} I(\vecown{r}, \vecown{n}, \nu) =
   \chi(\vecown{r}, \vecown{n}, \nu) \bigl(S(\vecown{r}, \vecown{n}, \nu) -
   I(\vecown{r}, \vecown{n}, \nu) \bigr) \,.
\eeq
In the following, we skip the explicit notation for the spatial (\vecown{r})
and directional (\vecown{n}) dependencies, and write the frequency dependence
as subscript, or, when appropriate, skip also this notation. The source
function, $S_\nu$, and the opacity, $\chi_\nu$, consist of the sum of
continuum and line processes.  By splitting the continuum emissivity and
opacity into true\footnote{By true processes, we mean all those processes,
  which interact with the thermal pool of the gas, for example
  photo-ionization and recombination.} and scattering processes, the
corresponding source function can be parameterized by a single parameter,
\epsc, the ratio of true absorption to total opacity, which is called the
thermalization parameter:
\beq \label{eq:scont}
\scont = (1-\epsc) \Jnu + \epsc \Bnu \,,
\eeq
where \Jnu ~is the mean intensity, and \Bnu ~the Planck function.  As a first
approach, we treat the line transfer similarly by considering a TLA only. This
is well suited to describe resonance lines, but for future applications the
complete rate equations need to be taken into account, of course. The line
source function in the TLA approach reads
\beqa \label{eq:sline}
\sline &=& (1-\epsl) \Jbar + \epsl B \\
\epsl &=& \dfrac{\epsilon'}{1+\epsilon'} , \quad
\epsilon'=\dfrac{\Cul}{\Aul}\Biggl[\exp\Bigl(-\frac{h \nu}{k_{\rm B} T} \Bigr)
  \Biggr] \,,
\eeqa
with \Cul ~and \Aul ~being the collisional rate (from the upper to the lower
level) and the Einstein-coefficient for spontaneous emission,
respectively. $\Jbar=1/4\pi \int I_\nu \profile \dd x \dd \Omega$ ~is the
profile-weighted and frequency-integrated, angle-averaged intensity, the
so-called scattering integral. The profile function, \profile, has been
approximated by a Doppler profile. Further, we did not calculate the profile in
frequency space, but rather in the variable $x_{\rm CMF}$, describing the
comoving frame (CMF) frequency shift from line centre, in units of a fiducial
Doppler width, \ddopfid:
\beq
x_{\rm CMF} = \dfrac{\nu_{\rm CMF} - \nu_{\rm 0}}{\ddopfid};
\quad
\ddopfid=\dfrac{\nu_{\rm 0} \vthfid}{c} \,,
\eeq
where $\nu_{\rm 0}$ and \vthfid ~are the line-centre transition frequency, and
the fiducial thermal velocity, respectively. The fiducial width is required to
enable a depth-independent frequency grid. $x_{\rm CMF}$ is related to the
corresponding observer's frame quantity via $x_{\rm CMF} = x_{\rm OBS} -
\vecown{n}\cdot \vecown{V}$, with $\vecown{V}=\vecown{v}/\vthfid$ ~the local
velocity vector in units of \vthfid. In most cases, we do not label $x$
explicitly to distinguish between comoving and observer's frame (nor from the
spatial $x$-coordinate), since the meaning should be clear wherever it occurs.

The profile function (in frequency-space) is
\beq
\profilenu = \dfrac{1}{\sqrt{\pi}\ddop} \exp \Biggl[-\Bigl(\dfrac{\nu_{\rm
      CMF}-\nu_{\rm 0}}{\ddop} \Bigr)^2 \Biggr] = \dfrac{1}{\sqrt{\pi}\ddop}
\exp \Biggl[-\Bigl( \dfrac{x_{\rm CMF}}{\delta}\Bigr)^2 \Biggr] \,,
\eeq
where 
\beq
\label{eq:vmicro}
\delta = \sqrt{\dfrac{2 k_{\rm B} T}{m_{\rm A}} + \vturb^2} \Bigg/ \vthfid
\eeq
is the ratio between the local thermal velocity (including micro-turbulence,
see Sect.~\ref{subsec:fvm}) and the fiducial velocity, $T$ is the local
temperature, and $m_{\rm A}$ is the mass of the considered ion.  With the
profile function, normalized in x-space,
\beq
\label{eq:profile}
\profile = \dfrac{1}{\sqrt{\pi}\delta} \exp \Biggl[-\Bigl(\dfrac{x_{\rm
      CMF}}{\delta} \Bigr)^2\Biggr] =
\dfrac{1}{\sqrt{\pi}\delta} \exp \Biggl[- \Bigl(\dfrac{x_{\rm OBS} - \vecown{n}
    \cdot \vecown{V}}{\delta} \Bigr)^2\Biggr] \,,
\eeq
the line opacity, which needs to be described in frequency-space (because the
RT and Planck function are formulated \wrt ~frequency), is given by
\beqa
\chil(\nu) &=& \bar{\chi}_{\rm 0} \profilenu = \bar{\chi}_{\rm L} \profile \\
\bar{\chi}_{\rm 0} &=& \dfrac{\pi {\rm e}^2}{m_{\rm e} c} (gf)
\Biggl[\dfrac{n_{\rm l}}{g_{\rm l}}-\dfrac{n_{\rm u}}{g_{\rm u}}\Biggr]
\approx \dfrac{\pi {\rm e}^2}{m_{\rm e} c} \cdot f \cdot n_{\rm l} \,,
\eeqa
with $(gf)$ the gf-value of the considered transition, and $n_{\rm l}$,
$n_{\rm u}$, $g_{\rm l}$, $g_{\rm u}$ the occupation numbers and statistical
weights of the lower and upper level, respectively. Since $n_{\rm l} \gg
n_{\rm u}$~for resonance lines, we have neglected stimulated emission on the
right-hand side.  $\bar{\chi}_{\rm 0}$ is the frequency integrated
opacity\footnote{ We stress that the frequency integrated opacity is often
  written as $\bar{\chi}_{\rm L}$, and must not be confused with the quantity
  defined by the same symbol as defined here (different normalization!).}, and
is related to $\bar{\chi}_{\rm L}$ by
\beq
\bar{\chi}_{\rm L} = \dfrac{\delta}{\ddop}\bar{\chi}_{\rm 0} =
\dfrac{\bar{\chi}_{\rm 0}}{\ddopfid} \,.
\eeq
We finally parameterize the continuum and line opacities in terms of the
Thomson-opacity, $\chith=\nel \sigmae$,
\beqa
\label{eq:kline}
\chil &=& \kline \cdot \chith \cdot \profile \quad \Longleftrightarrow \quad \kline :=
\dfrac{\bar{\chi}_{\rm L}}{\chith} = \dfrac{\bar{\chi}_{\rm 0}}{\ddopfid \nel
  \sigmae} \\
\chic &=& \kcont \cdot \chith \,,
\eeqa
where we use \kline ~as a depth-independent parameter, since the ratio $n_{\rm
  l}/\nel$ remains almost constant in the atmosphere for resonance lines
(almost frozen-in ionization).

The radiation field depends on the source functions via the \eqrt, and the
source functions depend on the radiation field via equations \eqref{eq:scont}
and \eqref{eq:sline}. Although this coupled problem could be solved directly,
at least in principle, we note already here that this would require the
inversion of a very large matrix, which is computationally prohibitive in 3D
calculations (see Sect.~\ref{sec:ali}).  Thus, we apply a $\Lambda$-iteration,
for which a formal solution (FS) of the \eqrt\, is obtained, given the
continuum and line source functions. These are subsequently updated according
to Eq. \eqref{eq:scont} and \eqref{eq:sline}, given the formal solution of the
previous iterate.  For large optical depths and low \epsc ~(\epsl), however,
the strong non-local coupling between the source functions and the radiation
field results in the well known convergence-problem of the classical
$\Lambda$-iteration. This problem is based on the fact that, due to scattering
processes, photons can travel over many mean-free-paths before being destroyed
or escaping from the atmosphere. On the other hand, information is propagated
in each iteration step only over roughly one mean-free-path. Therefore, a
large number of iteration steps would be required, until a consistent solution
was obtained (if at all). Thus, acceleration techniques are urgently needed,
and developed in this work in terms of an ALI scheme using operator-splitting
techniques \citep{Cannon73}.
\section{Numerical methods} 
\label{sec:numerical_methods} 
The ALI scheme can be split into two parts, namely the calculation of the FS,
and the construction of a new iterate, by means of an approximate
$\Lambda$-operator. There are various solution schemes to obtain the FS,
differing in required computing power, and using different assumptions on the
geometry and physical properties of the considered problems, as outlined in
the introduction. Among those, the FVM \citep{Adam90} is expected to be the
fastest and most simple one, and is used here as a first step to tackle 3D
radiative transfer problems.

Since we aim at modelling non-monotonic velocity fields, for which a
comoving-frame formulation, if at all, is very complicated to implement, we
solve the \eqrt ~in the observer's frame. We also use a Cartesian coordinate
system, which has the advantage of constant angular directions \wrt ~the
spatial grid. Anyhow, a description of the atmospheric structure in spherical
coordinates loses its advantages for the problems considered in this (and
future) work.

In the first part of this section, we describe the FVM, and the angular
and frequency integration methods. Main focus is put on the development
of the ALI scheme tailored to the FVM (Sect.~\ref{sec:ali}).
\subsection{Finite-volume method}
\label{subsec:fvm}
The main ideas of the FVM originate from heat transfer \citep{patankarbook80},
and have been already applied to radiative transfer problems in accretion
discs by \cite{Adam90}, as well as to the formation of discrete absorption
components in hot star winds \citep{Lobel08}.  Although the derivation of the
discretized \eqrt ~can be found, for instance, in \cite{Adam90}, we outline the
basic ideas in more detail in Appendix~\ref{app:fvm}. The final solution
scheme reads:
\beq \label{eq:eqrt_disc}
I_{\rm ijk} = a_{\rm ijk} \scontijk + b_{\rm ijk} \slineijk + c_{\rm
  ijk}I_{\rm i-\alpha jk} +
         d_{\rm ijk}I_{\rm ij-\beta k} + e_{\rm ijk}I_{\rm ijk-\gamma}
\eeq
\begin{eqnarray*}
f_{\rm ijk} &=& \chicijk + \chibarijk \profile^{\rm (ijk)} + \\
           & & + \dfrac{2 n_x}{x_{\rm i+\alpha} - x_{\rm i-\alpha}} + \dfrac{2
                 n_y}{y_{\rm j+\beta} - y_{\rm j-\beta}} + \dfrac{2
                 n_z}{z_{\rm k+\gamma} - z_{\rm k-\gamma}} \\ 
a_{\rm ijk} &=& \dfrac{\chicijk}{f_{\rm ijk}} \\
b_{\rm ijk} &=& \dfrac{\chibarijk \profile^{\rm (ijk)}}{f_{\rm ijk}} \\
c_{\rm ijk} &=& \dfrac{2n_x}{ (x_{\rm i+\alpha} - x_{\rm i-\alpha})\cdot f_{\rm ijk}} \\
d_{\rm ijk} &=& \dfrac{2n_y}{ (y_{\rm j+\beta} - y_{\rm j-\beta})\cdot f_{\rm ijk}} \\
e_{\rm ijk} &=& \dfrac{2n_z}{ (z_{\rm k+\gamma} - z_{\rm k-\gamma})\cdot
  f_{\rm ijk}} \,,
\end{eqnarray*}
with $\alpha, \beta, \gamma$ set to $\pm 1$ for direction-vector components
$n_x,n_y,n_z \gtrless 0$. All quantities except \chic, $\chibar_{\rm L}$, and
the source terms, depend on the considered direction $\vec{n}$ as well as on
frequency. Eq.~\eqref{eq:eqrt_disc} represents a pure upwind scheme, with
projected $\Delta \tau$-steps\footnote{The projected $\Delta \tau$-steps
  represent the optical depth of the cell for a given direction, and are
  easily obtained from Eq.~\eqref{eq:eqrt_disc} and the definition of the
  coefficients $c_{\rm ijk}\ldots e_{\rm ijk}$, \eg~ $c_{\rm ijk} = :
  1/(1+\Delta \tau_x)$.} calculated from a central-differencing approach
(following \citealt{patankarbook80}, see also Appendix~\ref{app:fvm}). Due to
the upwind scheme, and because the coefficients $a_{\rm ijk}\ldots e_{\rm ijk}
\in [0,1]$, the solution method is unconditionally stable (\citealt{Adam90}).

Contrasted to our central-differencing approach, \cite{Adam90} and
\cite{Lobel08} used backward differences for the calculation of the $\Delta
\tau$-steps. We have tested both methods, and found superior results when
using central differences.

Instead of using an equidistant grid, a grid-construction procedure is applied
in order to resolve the important regions of the atmosphere. In the case of
nearly spherical winds, this basically consists of setting up a 3D spherical
grid in the first octant, with uniformly distributed latitude- and
azimuth-angles, and with radial coordinates such that $\Delta \tau_{\rm c}$
and $\Delta v_{r}$ are (nearly) equidistant.  $\Delta \tau_{\rm c}$ and
$\Delta v_{r}$ are the step-sizes in continuum optical-depth and radial
velocity, respectively.  The final (Cartesian) $x$ ($y,z$) -coordinates are
chosen such that they correspond to the distribution of the $x_s$ ($y_s,z_s$)
-coordinates from the spherical grid. If not indicated explicitly, we use
$\nx=\ny=\nz=133$ grid points for the continuum, and $\nx=\ny=\nz=93$ grid
points for the line formation, distributed over the complete range, $[-\Rmax,
  \Rmax]$. We allow for a higher grid resolution within the continuum
calculations, since only one frequency point needs to be considered, and we
thus have lower computation times anyhow. As a comparison, \cite{Lobel08} used
only $\nx=\ny=\nz=71$ grid points for their (optically thin) models. A typical
grid in the x-z-plane is shown in the top panel of
Fig.~\ref{fig:searchlight}. The applied grid resolution is required to
properly resolve the resonance zones (where the CMF line opacity and
corresponding profile function is non-negligible), and to obtain reasonable
optical-depth steps within our (first-order\footnote{Actually, our FVM is a
  first-order scheme only for large optical-depth steps. For small
  optical-depth steps, the method becomes second-order accurate.}) method.  We
emphasize that a large number of grid points would be required to properly
describe the radiation field in the optically thick limit, since a first-order
scheme is generally unable to reproduce the (second-order) diffusion
equation. This problem, however, arises only at large optical depths (\eg~when
considering the winds of Wolf-Rayet stars), and is only of minor importance
for the winds of OB stars, where the total continuum optical depth is of the
order unity. To consider only the regions where we actually know the structure
of the atmosphere, we calculate the RT within a predefined `calculation
volume', defined by the constraint $r(x_i,y_j,z_k) \in [\Rstar, 1.1\Rmax]$.
\paragraph{Angular integration.}
\begin{table}
\begin{center}
\caption{Mean relative errors of the mean intensity for a zero-opacity model,
  using either Gauss-Legendre quadrature, or the trapezoidal rule with
  integration nodes from \citet{Lobel08}. The theoretical solution has been
  calculated from the dilution factor and the intensity emitted from the
  stellar core (see text). $\nx=\ny=\nz=133$ spatial grid points have been used.}
\label{tab:err_angles}

\begin{tabular}{c|ccc|ccc}

\multicolumn{1}{l|}{} & 
\multicolumn{3}{c|}{Legendre-Integration} &
\multicolumn{3}{c}{Trapezoidal Rule} 
\\
\hline\hline
\noalign{\vskip 0.5mm}
\multicolumn{1}{l|}{$\ntheta \cdot \nphi$} &
\multicolumn{2}{c}{>968} &
\multicolumn{1}{c|}{512} &
\multicolumn{1}{c}{2105} &
\multicolumn{1}{c}{1037} &
\multicolumn{1}{c}{544}
\\
\hline
\noalign{\vskip 0.5mm}
\multicolumn{1}{l|}{$\bar{\Delta J} [\%]$} &
\multicolumn{2}{c}{9.4} &
\multicolumn{1}{c|}{9.9} &
\multicolumn{1}{c}{9.6} &
\multicolumn{1}{c}{10.4} &
\multicolumn{1}{c}{12.4}

\end{tabular}
\end{center}
\end{table}
The mean intensity at each grid point is obtained from the solutions of the
\eqrt ~for many directions, where the distribution of these directions over
the unit sphere depends on the used quadrature formula.  The resulting
intensities are numerically integrated via
\beq
\label{eq:mint}
J_{\rm ijk} = \dfrac{1}{4\pi} \int I_{\rm ijk} \dd \Omega \approx \sum_{\rm
  l}w_{\rm l} \sum_{\rm m} w_{\rm m} I_{\rm ijk}(\theta_{\rm m}, \phi_{\rm l})
\,,
\eeq
with $\theta_{\rm m}, \phi_{\rm l}$ being the co-latitude (measured from the
z-axis) and the azimuth (measured from the x-axis), respectively, and $w_{\rm
  m}, w_{\rm l}$ the corresponding integration weights. The projection factor
$\sin(\theta_{\rm m})$, and the normalization have been included into $w_{\rm
  m}$, $w_{\rm l}$. Several different techniques and directional distributions
have been tested, including the standard trapezoidal rule with equidistant
$\theta,\phi$-grids, as well as using the approach of \cite{Lobel08}, who
basically use an ansatz $\dd \Omega =$~const, and obtain different
$\phi$-grids on each $\theta$-level.  To obtain a fair angular resolution of
the unit sphere, that means without preferring certain directions, we split
the integration into a sum over all octants, with the same nodes and weights
within each octant. The mean relative errors\footnote{The mean relative error
  of any quantity q is defined throughout this paper by $\bar{\Delta q} :=
  \dfrac{1}{N} \sum_{i=1}^N\dfrac{\vert q_i - q_i^{\rm (theo)} \vert}{q_i^{\rm
      (theo)}}$, with $N$ the number of grid points within the calculation
  volume.} of the mean intensity for a zero-opacity model, using
$\nx=\ny=\nz=133$ grid points and different angular grids, are summarized in
Table \ref{tab:err_angles}.  The corresponding theoretical solution has been
calculated from $J^{\rm (theo)}=WI_{\rm c}$, with
$W=1/2\bigl[1-\sqrt{1-(\Rstar / r ) ^2} \bigr]$ the dilution factor, and
$I_{\rm c}$ the intensity emitted from the stellar core.

The best results have been obtained using a Gauss-Legendre quadrature, for
which the integration nodes are fixed, and cannot be chosen arbitrarily.
Using this quadrature, already $\ntheta \cdot \nphi = 512$ grid points would
be sufficient for an accurate integration, at least in principle.  We note,
however, that the solution in the outer part of the atmosphere oscillates
around the theoretical value, because the star is not resolved by the angular
grid in these regions. This effect becomes even worse for the trapezoidal
rule. The corresponding errors, however, contribute only weakly to the mean
error, because most grid points are located near to the star. Additionally,
the dependence of the profile function on the local projected velocity (see
Eq. \ref{eq:profile}) requires a much finer angular resolution in the line
case. Thus, to safely avoid artefacts from the angular integration scheme, we
use $\ntheta \cdot \nphi=2048$ throughout this paper. This is a factor of
three lower than the number of angular integration points used by
\cite{Lobel08}, $(\ntheta \cdot \nphi)^{\rm Wind3D} =6400$.  We finally note
that the calculation of integration nodes for the Gauss-Legendre quadrature
requires the inversion of a large matrix, which directly shows the advantage
of a Cartesian grid, for which the angular nodes and weights are the same at
all grid points, and can be calculated prior to the complete solution
scheme. Using a spherical grid, Gaussian quadrature would be computationally
very time-consuming, because the involved angles \wrt ~the coordinate system
depend on position in the atmosphere.
\paragraph{Frequency integration.}
The integration over frequency is only performed when solving for \Jbar ~in
the line case. Due to its simplicity, we apply the trapezoidal rule with
equidistant steps,
\beqa
\label{eq:jbar}
\Jbar_{\rm ijk} &=& \dfrac{1}{4\pi} \int \dd \Omega \int_{-\infty}^{\infty}
                   \dd x I_{\rm ijk} \profile^{\rm (ijk)} \approx \nonumber \\ 
         &\approx& \dfrac{1}{4\pi} \int \dd \Omega \sum w_{\rm x} I_{\rm ijk}
                   \profile^{\rm (ijk)} \,.
\eeqa
The integration is performed over the complete frequency range, $x_{\rm OBS} =
\Bigl[\dfrac{-\vinf-3\vth(\Rmax)}{\vthfid}, \dfrac{\vinf+3\vth
    (\Rmax)}{\vthfid}\Bigr]$, with maximum velocity, \vinf, corresponding
thermal velocity, $\vth(\Rmax)$, and a rapidly vanishing Doppler profile for
$\vert x_{\rm CMF}/ \delta \vert > 3$ \footnote{$\profile(\vert x_{\rm CMF} /
  \delta \vert =3) \approx 10^{-4}\profile(x_{\rm CMF} / \delta=0)$}. To
resolve the profile function everywhere in the atmosphere, we require 
$\Delta x_{\rm CMF} / \delta \lesssim 1/3$ at each position, which is achieved
by defining $\vthfid := \vth^{\rm (min)}$, and choosing the number of
frequency points, \nnu, such that $\Delta x_{\rm OBS} = (x_{\rm OBS}^{\rm
  (max)}-x_{\rm OBS}^{\rm (min)})/(\nnu-1) \lesssim 1/3$. $\vth^{\rm (min)}$
is the minimum thermal velocity (including micro-turbulence) of the
atmosphere. However, since the comoving frame frequency depends on the
projected velocity, the corresponding integration nodes may not be centred
around the profile maximum anymore. This issue is only of minor importance and
has been checked by a comparison to model calculations that use $\Delta x_{\rm
  OBS} \lesssim 1/6$.  In atmospheres with very low micro-turbulent
velocities, a number of $\nnu\approx 1200$ frequency points would be required,
for typical values of $\vth^{\rm (min)}=10$~\kms ~(if no micro-turbulent
velocities were present), and typical wind terminal speeds, \vinf\, = 2000
\kms.  When using a rather high micro-turbulent velocity, \vturb\, = 100 \kms,
$\nnu$ is reduced considerably, by a factor of ten. Fortunately, such high
values are not un-typical in the winds of hot stars (see next paragraph).
Again, when compared to \cite{Lobel08}, who used $\nnu=100$ frequency points
for a thermal velocity of $\vth = 30$~\kms, we have a much finer resolved
frequency grid, with $\approx 15$ frequency points distributed over the
complete line profile, whereas they resolved the profile function with only
three frequency points.
\paragraph{Microturbulence.}
To correctly treat the radiative transport in the line, we need to resolve the
resonance zones, and demand that
$\Delta(\vecown{n}\cdot\vecown{v}/\vth)\approx 1/3$. Again, for low
micro-turbulent velocities, we would require a resolution of at least 1200
grid points per spatial dimension (for a spherical wind accounting for both
hemispheres). On the other hand, including a large micro-turbulent velocity
(\vturb\, = 100~\kms), as applied here, results in a much lower required
resolution, both in space and frequency. Due to the linear (cubic) scaling of
computation time with the number of frequency (spatial) grid points, this
results in a reduction of computation time by a factor of $10^{4}$, when
compared to models without large micro-turbulent velocities. Putting it the
other way round, and already mentioning here that typical model-calculations
take about 50 minutes of wallclock time per iteration (and using 16
processors, see next paragraph), a large micro-turbulent velocity is needed
thus far to keep the computation time on a reasonable scale.

\citet{Hamann81a} showed that such micro-turbulent velocities can indeed be
used to correctly model the black absorption troughs observed in the P~Cygni
profiles of hot star winds. From a theoretical point of view, a large velocity
dispersion mimicks the effects of multiply non-monotonic velocity fields, as
originating from the line-driven instability (see, \eg~\citealt{Lucy83},
\citealt{POF93}).
\paragraph{Parallelization and timing.}
In the line case, a number of $\ntheta \times \nphi \times \nnu$ formal
solutions of the \eqrt ~are required to calculate the scattering integrals. In
order to obtain accurate angular and frequency integrals, the minimum number
of integration nodes is basically fixed, and the computation time can only be
reduced further by parallelization. In our case, the most simple and
straightforward procedure is a parallelization in frequency, which was done
here using \omp.  The computation time of a typical model with
$\nx=\ny=\nz=93$, $\ntheta\cdot\nphi=2048$, $\nnu=139$ spatial, angular and
frequency grid points, respectively, is about 50 minutes of wallclock time per
iteration on a 16 CPU \textsc{Intel Xeon X5650 (2.67 GHz)} machine. As a
reference, \textsc{Wind3D} \citep{Lobel08} requires about 30 minutes per
iteration for their grid parameters ($N_{\rm x}^{\rm Wind3D}=N_{\rm y}^{\rm
  Wind3D} = N_{\rm z}^{\rm Wind3D}=71$, $N_{\rm \theta}^{\rm Wind3D} \cdot
N_{\rm \phi}^{\rm Wind3D} = 6400$, $N_{\rm \nu}^{\rm Wind3D}=100$, $N_{\rm
  CPU}^{\rm Wind3D}=16$). To obtain a more meaningful comparison, we scale the
computation time by the number of applied angular and frequency grid points,
and perform a test calculation, which uses the same spatial grid as
\cite{Lobel08}.  Although our CPUs are more modern and faster, our code
performs only slightly better than \textsc{Wind3D}, with computation times
$t_{\rm FS}^{\rm Wind3D} \approx 0.045 \seconds$ vs. $t_{\rm FS}^{\rm our}
\approx 0.037 \seconds$, per iteration, per angular and frequency point, and
per CPU. We note, however, that our algorithm requires at least a factor of
two more operations per formal solution, since we additionally compute the
(non-local) approximate $\Lambda$-operator in parallel (see next section).
\subsection{$\Lambda$-iteration}
\label{sec:ali}
By now, we are able to construct a formal solution for a given source
function. In this section, we discuss the iteration procedure. The following
discussion considers the line case alone, with a frequency-independent
background continuum (\ie~constant continuum opacity and source function),
assumed to be known (either in form of an optically thin continuum, or from
previous calculations in the absence of the line). For convenience, we
summarize the basic ideas and corresponding acceleration techniques via the
ALI from first principles.
\paragraph{Matrix equation.}
To show that the $\Lambda$ formalism can also be applied to our 3D
formulation, we derive a matrix equation for the scattering integral, and show
that this equation is consistent with an affine representation of the
$\Lambda$ operator. Since, however, the final ALO will be calculated
differently, a detailed description of the involved matrices is skipped
in the following.

All 3D quantities are expressed as vectors of length $\nx\times\ny\times\nz$,
by introducing a unique ordering of the $(i,j,k)$-triple:
\beq
\label{eq:mindx}
   m:=i+\nx\cdot(j-1) + \nx \ny \cdot (k-1) \,,
\eeq
where $i,j,k\in [1,N_{\rm x,y,z}]$, and $m\in[1,\nx \ny \nz]$.  Replacing the
$(i,j,k)$-indices in Eq. \eqref{eq:eqrt_disc} with this unique index, $m$, and
after collecting all intensity terms on the left-hand side, we obtain:
\beq
\label{eq:eqrt_matrix0}
\dfrac{1}{b_{\rm m}} I_{\rm m} - \dfrac{c_{\rm m}}{b_{\rm m}} I_{\rm m-\alpha}
- \dfrac{d_{\rm m}}{b_{\rm m}} I_{\rm m-\beta\nx} - \dfrac{e_{\rm m}}{b_{\rm
    m}} I_{\rm m-\gamma \nx\ny} = \dfrac{a_{\rm m}}{b_{\rm m}} S_{\rm m}^{\rm
  (c)} + S_{\rm m}^{\rm (L)} \,.
\eeq
This equation is written in matrix form, 
\beq
\label{eq:eqrt_matrix1}
\matown{T}\cdot\vecown{I} = \matown{Q}\cdot \scontv + \slinev + \iincv \,,
\eeq
where \iincv ~refers to the boundary condition.  Since we use a TLA with given
opacity (\ie~the opacity does not change during the iteration), and consider
time-independent boundary conditions, we combine the constant vectors from
Eq. \eqref{eq:eqrt_matrix1}, $\tilde{\vecown{I}}_{\rm inc} := \matown{Q} \cdot
\scontv + \iincv$. Inverting the matrix $\matown{T}$ and integrating over all
angles and frequencies gives:
\beqa
\label{eq:lambdamat}
4\pi\vecown{\Jbar} &=& \int \matown{\profile} \cdot \vecown{I} \dd \Omega \dd x =
\nonumber \\
&=& \int \matown{\profile} \cdot \matown{T}^{-1} \cdot \slinev \dd \Omega \dd x + \int
\matown{\profile} \cdot \matown{T}^{-1} \cdot \tilde{\vecown{I}}_{\rm inc} \dd
\Omega \dd x = \nonumber \\ 
&=& \int \matown{\profile} \cdot \matown{T}^{-1} \dd \Omega \dd x
\cdot \slinev + 4\pi \vecown{\Phi}_{\rm B} = \nonumber \\
&=:& 4 \pi \Bigl[\matown{\Lambda} \cdot \slinev + \vecown{\Phi}_{\rm B}\Bigr] \,.
\eeqa
The diagonal matrix $\matown{\profile}$ and the vector $\vecown{\Phi}_{\rm B}$
describe the local profile function, and the contribution of the boundary
conditions and the background continuum to the scattering integral,
respectively. As shown below, an explicit calculation of these quantities is
not required to obtain the finally used ALO. We note that also for the
continuum case, which is calculated close (\wrt~frequency) to the line,
Eq.~(\ref{eq:lambdamat}) is applicable, with a different
$\matown{\Lambda}$-matrix and boundary contribution though. A comparison of
Eq. \eqref{eq:lambdamat} with the common $\Lambda$-operator formalism
(\ie~formally writing $\Jbar = \Lambda [\sline]$) directly shows that the
$\Lambda$-operator is an affine operator, that means a linear operator given
by the $\matown{\Lambda}$-matrix plus a constant displacement vector
$\vecown{\Phi}_{\rm B}$ (see also \citealt{Puls91}), also for our 3D method.

From equations \eqref{eq:sline} and \eqref{eq:lambdamat}, we could formulate
an explicit solution of the radiation field already now. This, however, would
require the calculation, storage and inversion of the complete
$\matown{\Lambda}$-matrix, which is computationally prohibitive: 

Firstly, the $\matown{\Lambda}$-matrix is a full matrix with $\nx \ny \nz
\times \nx \ny \nz$ elements, which would require, for typical grid sizes of
$N:=\nx = \ny = \nz = 93$, $N^6 \approx 6.5 \cdot 10^{11}$ numbers, equivalent
to 5.2 TB data to be stored in memory, when double-precision numbers are
used. Secondly, the $\matown{\Lambda}$-matrix elements can be obtained, at least in
principle, by inversion (see also \citealt{Puls91}),
\beq
\label{eq:lelements}
\Lambda_{m,n} = \Jbar_{m} (\slinev = \vecown{e}_n, \vecown{\Phi}_{\rm
B} = 0 ) \,,
\eeq
with $\vecown{e}_n$ being the n-th unit vector. Thus, $\nx \times \ny \times
\nz$ formal solutions would be needed to calculate the complete
$\matown{\Lambda}$-matrix, which again is computationally prohibitive on
reasonably well-resolved grids.

An iterative solution is therefore the only possibility to solve problems of
this kind. Due to the well known convergence problems of the classical
$\Lambda$-iteration, we directly focus on the ALI. For completeness, let us
mention here a similar approach, the `non-linear multi-grid method'
(see \citealt{FB97} and references therein), which has even better convergence
properties, and, in contrast to the ALI method, does not depend on the spatial
resolution of the grid. For simplicity, however, we only implement an ALI
scheme. The ALI is an operator-splitting technique, which splits the original
$\Lambda$-operator into the combination
\beq
\label{eq:split}
\Lambda = \alo + (\Lambda - \alo) \,,
\eeq
with an appropriately chosen ALO, $\alo$ (see
\citealt{Cannon73}). Appropriately chosen means that \alo\, should be easily
calculated (preferentially in parallel with the formal solution), and easily
inverted. Moreover, the ALO should reflect the basic physical properties of
the original $\matown{\Lambda}$-matrix, in order to significantly boost the
convergence.
\paragraph{ALI.}
Using Eq.~(\ref{eq:split}), where now the first term acts on the current
iterate of the source function, and the second one on the previous iterate, we
obtain in combination with Eq. \eqref{eq:sline}
\beq
\vecown{S}^{\rm (n)} = \matown{\zeta} \cdot \vecown{J}^{\rm (n)} +
\vecown{\Psi} \approx \matown{\zeta} \cdot \alo[\vecown{S}^{(n)}] + \matown{\zeta}
\cdot (\Lambda-\alo) [\vecown{S}^{\rm (n-1)}] + \vecown{\Psi} \,, 
\eeq
where the approximate relation becomes an exact one for the converged
solution. Here we have used the notation and definitions of the diagonal
matrix, $\matown{\zeta} := \unitym - \matown{\epsilon}_{\rm L}$, and the
thermal contribution vector, $\vecown{\Psi}:=\matown{\epsilon}_{\rm L} \cdot
\vecown{B}_{\rm \nu} (\vecown{T})$, from \cite{PH88}. Again, all quantities
are ordered according to Eq. \eqref{eq:mindx}. After some algebra, we obtain
\begin{multline}
\label{eq:itrule}
(\unitym - \matown{\zeta} \cdot \alom) \vecown{S}^{\rm (n)} \approx
\matown{\zeta} \cdot (\Lambda - \alo)[\vecown{S}^{\rm (n-1)}] + \matown{\zeta} \cdot
\vecown{\Phi}_{\rm B} + \vecown{\Psi} = \\
\matown{\zeta} \cdot \bigl(\vecown{\Jbar}^{\rm (n-1)} - \alom\cdot\vecown{S}^{\rm
  (n-1)} \bigr) + \vecown{\Psi}.
\end{multline}
Since also the ALO is an affine operator (analogous to the original
$\Lambda$-operator), the displacement vector $\vecown{\Phi}_{\rm B}$ only
cancels if it remains constant over subsequent iteration steps, that is, when
the background and the boundary conditions remain constant over the
iteration. (In realistic, multi-level NLTE calculations, this means in
practice that the ALI cycle shows a fast convergence only when the continuum
is close to convergence).
 
Given a previous iterate of the source function, $\vecown{S}^{\rm (n-1)}$, and
the corresponding formal solution, Eq. \eqref{eq:itrule} is used to calculate
the next iterate, $\vecown{S}^{\rm (n)}$. The detailed choice of the ALO is
the crucial point, and finally determines the convergence behaviour.  We
could, for instance, choose $\alo = \Lambda$, which would result in the direct
solution via inversion, and is computationally not feasible, as discussed
above. On the other hand, choosing $\alo=0$ would result in the classical
$\Lambda$-iteration, with the known convergence problems. \cite{OAB86} showed
that an ALO containing only the diagonal of the exact
$\matown{\Lambda}$-matrix is very efficient, because the matrix $(\unitym -
\matown{\zeta} \cdot \alom)$ becomes diagonal, and Eq. \eqref{eq:itrule} could
be solved by a simple scalar division. Furthermore, the diagonal, that means
the local part, already contributes most to the radiative transfer (at least
in the critical optically thick case), and thus, is quite a good approximation
for the original $\Lambda$-operator. Such an ALO corresponds to the well known
Jacobi-iteration (see also \citealt{Bueno1995} for a thoughtful discussion,
also about a Gauss-Seidel method with successive overrelaxation in the context
of the ALI).

In 3D calculations, however, a diagonal ALO will not converge fast enough (see
Sect.~\ref{subsec:convergence}). To achieve faster convergence rates, a
multi-band ALO is favourable, as already shown by \cite{Olson87} for 1D cases,
and extended to a 3D, long-characteristics solver by \cite{Haus06}. For such
ALOs, the matrix $(\unitym - \matown{\zeta} \cdot \alom)$ is sparse, whereas
its inverse is a full matrix, and cannot be stored due to the $N^6$ scaling of
required memory. Therefore, we have already formulated Eq. \eqref{eq:itrule}
as a fix-point iteration, $\matown{A}\cdot \vecown{S}^{\rm(n)} = \vecown{b}$,
which can be solved for the new iterate by applying Jacobi or Gauss-Seidel
methods. We found that a Jacobi-iteration, coupled with the storage of the
iteration-matrix in coordinate-format (COO)\footnote{In COO, all non-zero
  entries are stored, together with the row and column indices of the non-zero
  elements.}, is particularly fast and easy, because its computationally most
expensive term is a matrix-vector multiplication, which reduces to $\nnz$
operations only, where $\nnz$ is the number of non-zero elements
(\eg~\citealt{tessem13}).
\paragraph{Constructing the ALO.}
To construct a multi-band ALO as aimed at above, we need to to calculate the
corresponding elements of the exact $\matown{\Lambda}$-matrix. This could be
done, in principle, by using Eq. \eqref{eq:lambdamat}, which would require the
inversion of \matown{T}. Due to the upwind scheme, however, we can simply use
Eq. \eqref{eq:lelements}, in combination with Eq. \eqref{eq:eqrt_disc} and
\eqref{eq:jbar}. Since the $(m,n)$-th element of the $\matown{\Lambda}$-matrix
describes the impact of a non-vanishing source term at point $n
\leftrightarrow (i'j'k')$ onto a point $m \leftrightarrow (ijk)$, the local
contribution is given by $n=m$, whereas the coupling with directly
neighbouring points is found from:
\begin{eqnarray*}
   \makebox[11ex][l]{$n(i-1,j,k)$} \makebox[16ex][l]{$= m-1$}
   \makebox[4ex][l]{for}  \makebox[6ex][l]{$n_x > 0$} \makebox[1ex][l]{} \\
   \makebox[11ex][l]{$n(i,j-1,k)$} \makebox[16ex][l]{$= m-\nx$}
   \makebox[4ex][l]{for} \makebox[6ex][l]{$n_y > 0$} \makebox[1ex][l]{} \\
   \makebox[11ex][l]{$n(i,j,k-1)$} \makebox[16ex][l]{$= m-\nx\ny$}
   \makebox[4ex][l]{for} \makebox[6ex][l]{$n_z > 0$} \makebox[1ex][l]{} \\
   \makebox[11ex][l]{$n(i+1,j,k)$} \makebox[16ex][l]{$= m+1$}
   \makebox[4ex][l]{for}  \makebox[6ex][l]{$n_x < 0$} \makebox[1ex][l]{} \\
   \makebox[11ex][l]{$n(i,j+1,k)$} \makebox[16ex][l]{$= m+\nx$}
   \makebox[4ex][l]{for} \makebox[6ex][l]{$n_y < 0$}  \makebox[1ex][l]{} \\
   \makebox[11ex][l]{$n(i,j,k+1)$} \makebox[16ex][l]{$= m+\nx\ny$}
   \makebox[4ex][l]{for} \makebox[6ex][l]{$n_z < 0$} \makebox[1ex][l]{.}
\end{eqnarray*}
One big advantage of our method is that the exact elements of local and
neighbouring terms can be easily calculated from Eq. \eqref{eq:eqrt_disc},
\beqa
\label{eq:lambda_diag}
\Lambda_{\rm m,m} &=& \dfrac{1}{4\pi} \int \int b_{\rm ijk} \profile^{\rm (ijk)} \dd \Omega \dd x \\
\Lambda_{\rm m,m-1} &=& \dfrac{1}{4\pi}\int \int_{n_{\rm x} > 0} b_{\rm i-1 jk} c_{\rm ijk}
                       \profile^{\rm (ijk)} \dd \Omega \dd x \\
\Lambda_{\rm m,m-\nx} &=& \dfrac{1}{4\pi}\int \int_{n_{\rm y} > 0} b_{\rm ij-1 k} d_{\rm ijk}
                       \profile^{\rm (ijk)} \dd \Omega \dd x \\
\Lambda_{\rm m,m-\nx\ny} &=& \dfrac{1}{4\pi}\int \int_{n_{\rm z} > 0} b_{\rm ijk-1} e_{\rm
                       ijk} \profile^{\rm (ijk)} \dd \Omega \dd x \\
\Lambda_{\rm m,m+1} &=& \dfrac{1}{4\pi}\int \int_{n_{\rm x} < 0} b_{\rm i+1 jk} c_{\rm ijk}
                       \profile^{\rm (ijk)} \dd \Omega \dd x \\
\Lambda_{\rm m,m+\nx} &=& \dfrac{1}{4\pi}\int \int_{n_{\rm y} < 0} b_{\rm ij+1 k} d_{\rm ijk}
                       \profile^{\rm (ijk)} \dd \Omega \dd x \\
\Lambda_{\rm m,m+\nx\ny} &=& \dfrac{1}{4\pi}\int \int_{n_{\rm z} < 0} b_{\rm ijk+1} e_{\rm
                       ijk} \profile^{\rm (ijk)} \dd \Omega \dd x \,.
\eeqa
We call this ALO `direct neighbour' (DN)-ALO, to discriminate from the
`nearest neighbour' ALO from \cite{Haus06}, who use all 26 surrounding grid
points and the local term, whereas we are using the local term and the
contribution from the six direct neighbours only\footnote{For tests of the
  convergence properties in Sect.~\ref{subsec:convergence}, we also calculated
  a purely diagonal ALO by means of Eq. \eqref{eq:lambda_diag}
  alone.}. Although it would be possible to include also the other
neighbouring terms in our calculations, we note that the calculation of the
ALO elements in parallel to the formal solution requires already 50\% of the
calculation-time in our case, which would increase rapidly when including even
more terms for the ALO. On the other hand, the inversion of the ALO, that means the
calculation of the new iterate via Jacobi-iterations, requires only about
0.5\% of the calculation time needed for the complete FS. 
We emphasize that the ALO actually needs to be calculated only once, because
the opacity of the (simplified) TLA remains constant over subsequent iteration
cycles. When considering multi-level atoms (as planned in the future), the
situation changes, and the opacity depends on the occupation numbers, and
thus, also on the radiation field. We therefore implemented the calculation of
the ALO in parallel to the formal solution at each iteration step already at
the current stage of our code.

To accelerate the iteration scheme further, we implemented the extrapolation
technique from \citet[see also \citealt{OAB86}]{ng74}. In order to use
independent extrapolations, the Ng-acceleration is applied in every fifth
iteration step.  We finally note that the convergence behaviour depends on the
grid resolution, which determines the optical-depth steps, $\Delta
\tau_{x,y,z}$, and affects the coefficients $c_{\rm ijk}, d_{\rm ijk}, e_{\rm
  ijk}$. The finer the grid, the poorer the convergence behaviour
\citep{Kunasz88b}.

To summarize, we have implemented a 3D FVM in Cartesian coordinates, to
determine the FS for a given source function, and to calculate, in parallel, a
DN-ALO applied within an ALI scheme.
\section{Spherically symmetric models}
\label{sec:err_analysis}
Before applying our method to first non-spherical test problems, we have
checked its reliability by investigating the convergence properties, and by
comparing our 3D solution for a spherical wind with an accurate 1D
solution. The 1D solution for the line case has been found by a ray-by-ray
solution scheme in p-z-geometry, which has been formulated in the comoving
frame and accelerated by an appropriate ALO \citep{Puls91}. The 1D solution
for the continuum transport has been found by applying the Rybicki-algorithm
(\eg~\citealt{mihalasbook78}).

Three major error sources of our 3D code will be discussed in the following:
Firstly, errors occurring due to false convergence
(Sect.~\ref{subsec:convergence}), secondly, errors originating from the
incorporation of the boundary conditions and from numerical diffusion
(Sect.~\ref{subsec:numdiff}), and finally, errors occurring for optically thick
media, due to the order of accuracy of our method
(Sect.~\ref{subsec:varodepth}).

The spherically symmetric models to be compared with have been calculated from
a prescribed $\beta$-velocity law, the equation of continuity, and a
temperature structure from the 1D code (which plays almost no role in our test
cases).
\beqa
v(r) &=& \vinf \Bigl(1 - b \dfrac{\Rstar}{r} \Bigr)^{\beta} \\
b &=& 1-\Bigl(\dfrac{\vmin}{\vinf}\Bigr)^{1/\beta} \nonumber \\
\rho(r) &=& \dfrac{\mdot}{4 \pi r^2 v(r)} \,.
\eeqa
For opacities and source functions, see Sect.~\ref{sec:eqrt}, and the required
electron density, \nel, has been derived from a completely ionized H and He plasma, with
$N_{\rm He}/N_{\rm H} = 0.1$. For all following model calculations, we used a
fixed set of prototypical input parameters, summarized in Table
\ref{tab:model_ss}. These parameters roughly correspond to the wind from
$\zeta$ Pup, when assuming an unclumped wind. Within the line transfer, we
considered a generic UV resonance transition, with different line-strengths
following Eq. \eqref{eq:kline}. To calculate the thermal width, we used $m_{\rm
  A}=12 \, m_{\rm p}$, $m_{\rm p}$ being the proton mass. The total width of
the line profile, however, is mainly controlled by the (large) turbulent
velocities. Different optical depths,
\beq
\tau_{\rm r} = \int_{\Rstar}^{\Rmax}
  \chith \, \kcont \, \dd z \approx 0.17 \cdot \kcont \,, 
\eeq
(for the model considered in Table \ref{tab:model_ss}), and scattering
properties of the model atmosphere were simulated by varying the scaling
factors, \kcont, \kline, and the thermalization parameters, \epsc, \epsl, in
the continuum and line case, respectively.
\begin{table}
\begin{center}
\caption{Input parameters for the spherically symmetric models used for our
  test calculations.}
\label{tab:model_ss}
\begin{tabular}{c|c|c|c}
\multicolumn{1}{c}{\Teff [kK]} & 
\multicolumn{1}{c}{\Rstar [\rsun]} &
\multicolumn{1}{c}{\vmin [\kms]} &
\multicolumn{1}{c}{\mdot [\msunyr]}
\\
\multicolumn{1}{c}{40} & 
\multicolumn{1}{c}{19} &
\multicolumn{1}{c}{10} & 
\multicolumn{1}{c}{$5 \cdot 10^{-6}$}
\\
\hline\hline
\multicolumn{1}{c}{$\beta$} & 
\multicolumn{1}{c}{\Rmax [\Rstar]} &
\multicolumn{1}{c}{\vinf [\kms]} &
\multicolumn{1}{c}{\vturb [\kms]} 
\\
\multicolumn{1}{c}{1} &
\multicolumn{1}{c}{12} &
\multicolumn{1}{c}{2000} &
\multicolumn{1}{c}{100}
\end{tabular}
\end{center}
\end{table}
\subsection{Convergence behaviour}
\label{subsec:convergence}
\begin{figure*}[t]
\resizebox{\hsize}{!}{
   \begin{minipage}{0.5\hsize}
      \resizebox{\hsize}{!}{\includegraphics{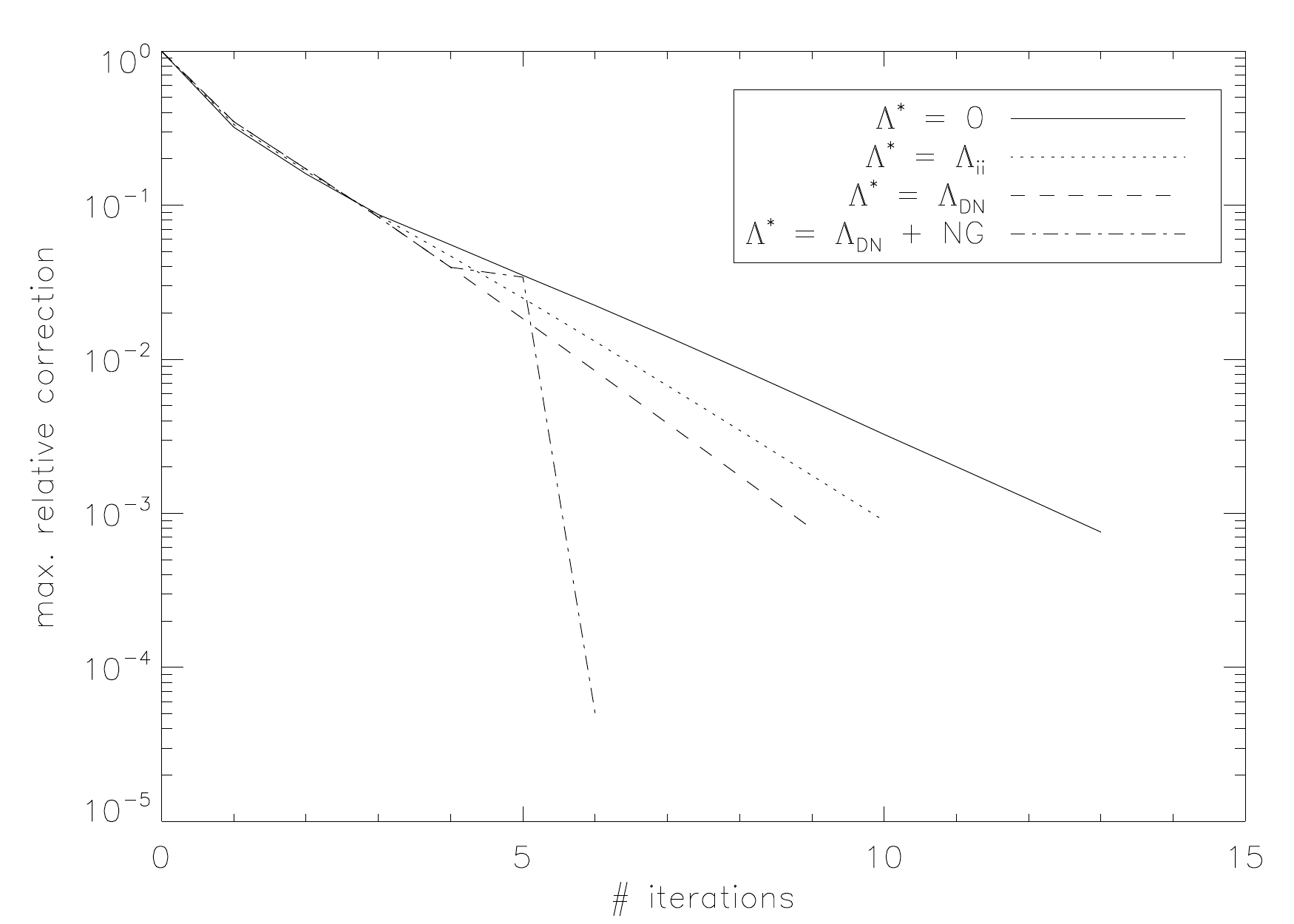}}
   \end{minipage}
   \begin{minipage}{0.5\hsize}
      \resizebox{\hsize}{!}{\includegraphics{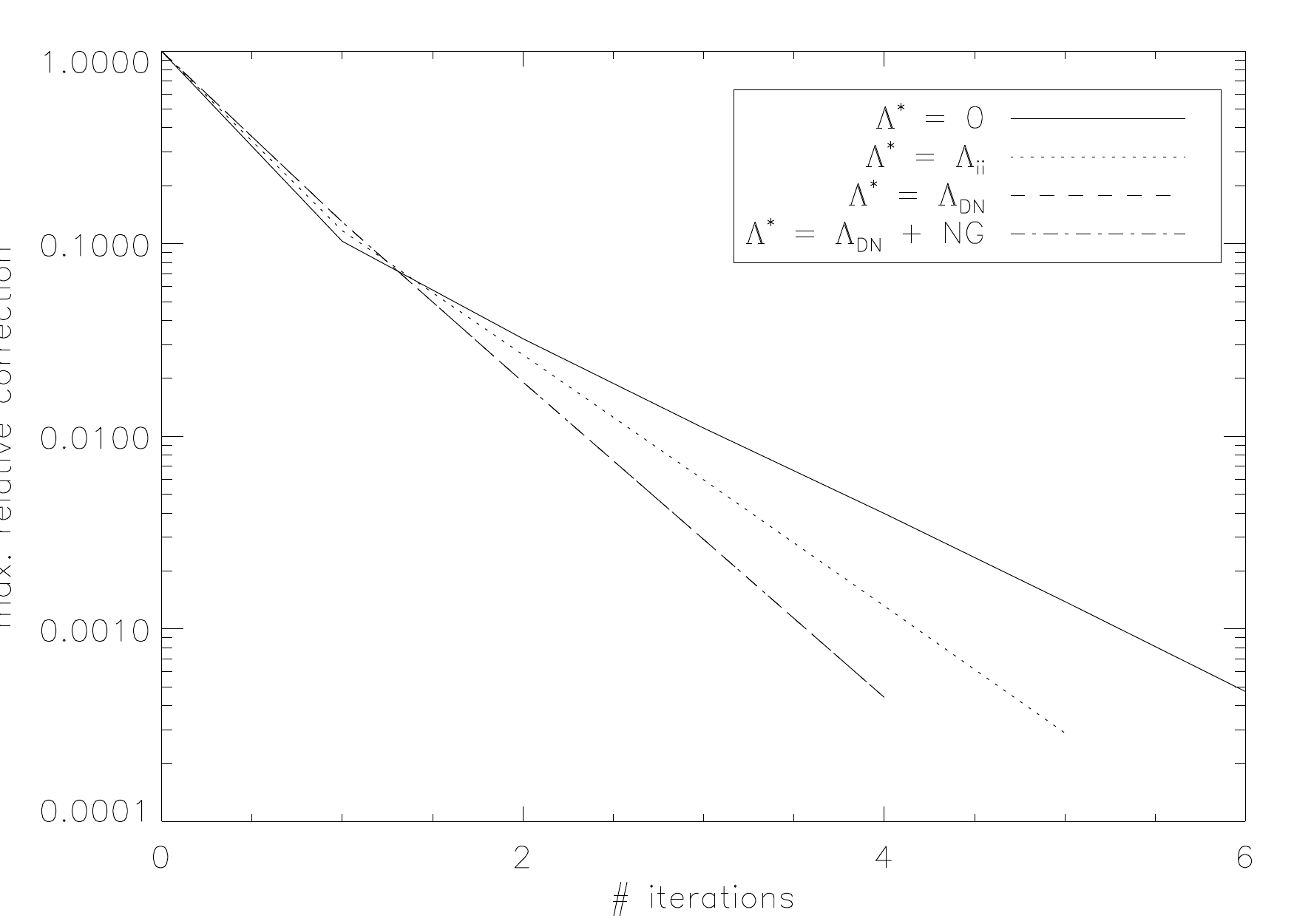}}
   \end{minipage}
}
\\
\resizebox{\hsize}{!}{
   \begin{minipage}{0.5\hsize}
      \resizebox{\hsize}{!}{\includegraphics{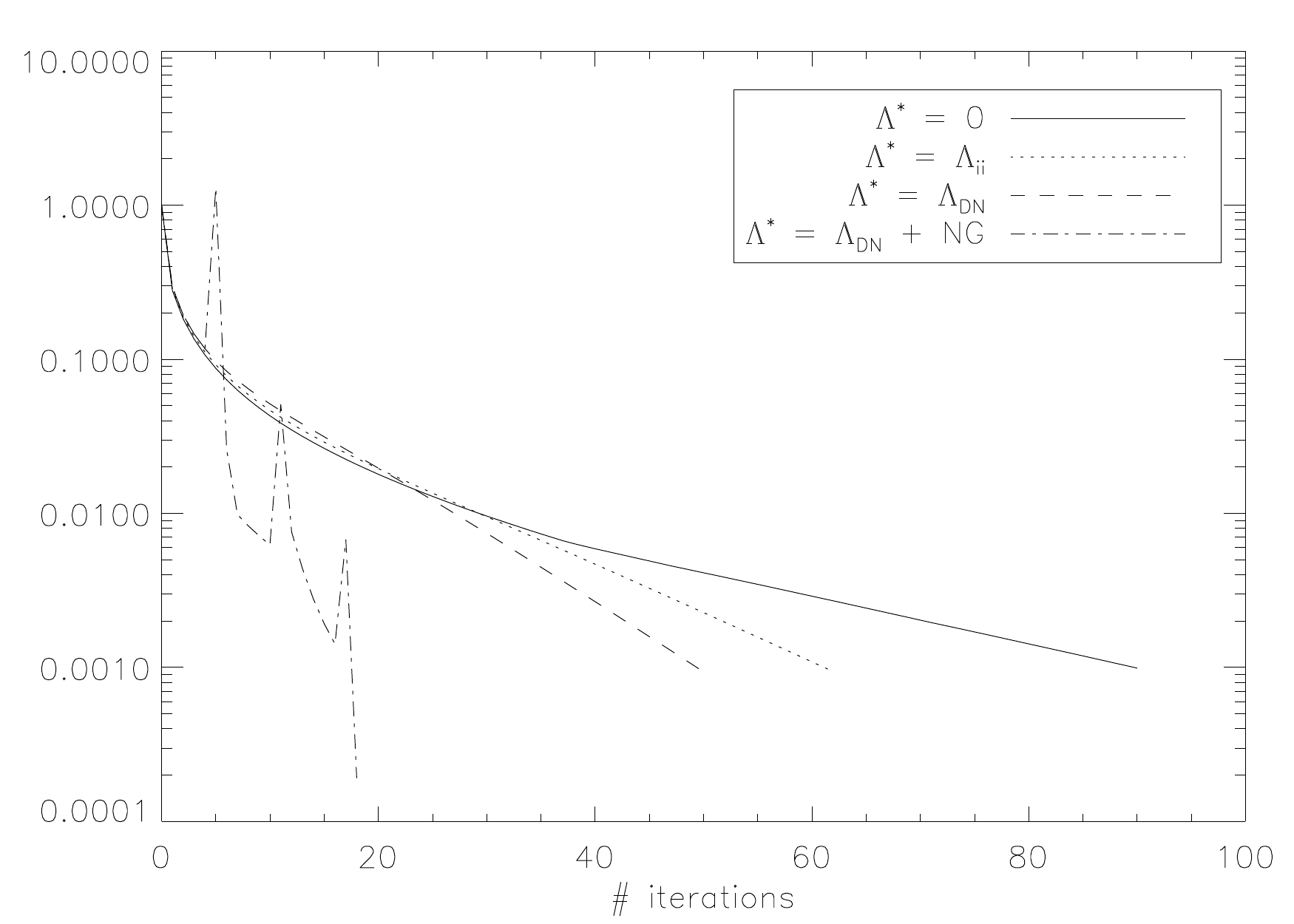}}
   \end{minipage}
   \begin{minipage}{0.5\hsize}
      \resizebox{\hsize}{!}{\includegraphics{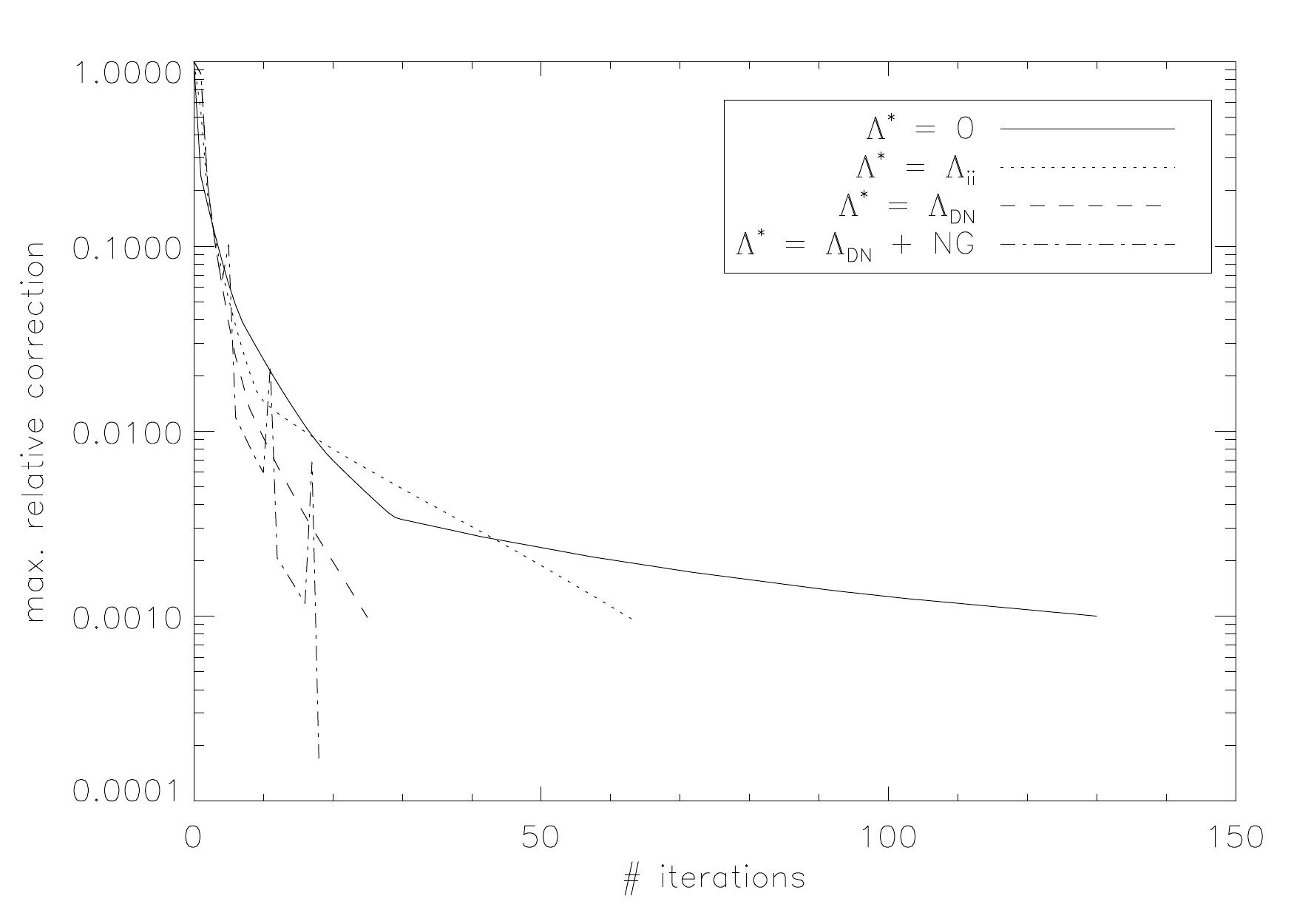}}
   \end{minipage}
}
%
\caption{Convergence behaviour of the continuum transfer with $\epsc=10^{-6}$
  (\textit{left}), and of the line transfer with $\epsl=10^{-6}$
  (\textit{right}), for different acceleration techniques applied to the RT,
  and for a spherical model atmosphere. The optical depth scale varies
  according to $\kcont=10$, $\kline=10$ (\textit{top}) and $\kcont=100$,
  $\kline=10^5$ (\textit{bottom}). See text.}
\label{fig:convergence}
\end{figure*}
To test the convergence behaviour of our ALI implementation with corresponding
ALO, we only considered scattering dominated atmospheres by setting \epsc ~and
\epsl ~to $10^{-6}$. We discuss the continuum transfer in the absence of a
line, and the line transfer assuming an optically thin continuum.  In both
cases, we applied the classical $\Lambda$-iteration, and compare to ALI schemes
using a diagonal or DN-ALO, with Ng-extrapolation switched on or off.
\paragraph{Pure continuum.}
The left panel of Fig.~\ref{fig:convergence} shows the maximum relative
corrections of the mean intensity after each iteration step, for an
intermediate grid resolution of \nx~=~\ny~=~\nz~=~93 spatial points and
\ntheta\nphi~=~968 angular points. Two different continua are shown, referring
to an optically thick ($\kcont=100$), and marginally optically thick
($\kcont=10$) case.  Obviously, the classical $\Lambda$-iteration converges
only very slowly (if at all) in the high optical-depth regime, requiring
$N_{\rm conv}^{\rm (classical)}=90$ iterations until the convergence criterion
of maximum relative corrections being less than $10^{-3}$ is fulfilled. We
emphasize that a (steep) gradient of the `convergence curve' is required to
achieve actual convergence, rather than this (arbitrarily chosen) number. For
the optically thick problem, even the diagonal and also the DN-ALO have severe
convergence problems, with $N_{\rm conv}^{\rm (diag)}=62$ and $N_{\rm
  conv}^{\rm (DN)}=50$, respectively. To ensure fast convergence, the
Ng-acceleration is urgently needed, since it boosts the relative corrections
significantly, reducing the number of required iterations for the optically
thick problem to $N_{\rm conv}^{\rm (DN+NG)}=18$ (lower left panel of
Fig.~\ref{fig:convergence}).
\paragraph{Line case.}
With the same set-up as above, we display the maximum relative corrections of
the scattering integral after each iteration step in the right panel of
Fig.~\ref{fig:convergence}. We applied two different line-strength parameters,
$\kline=10$, $10^5$, which correspond to a weak and strong line, respectively.
Since the line transport is restricted to the finite widths of the resonance
zones, and is therefore intrinsically much more local than the continuum, the
convergence behaviour is accelerated significantly already by the diagonal and
DN-ALO. For the strong line, the required number of iterations until
convergence is reduced from $N_{\rm conv}^{\rm (classical)}=130$ to $N_{\rm
  conv}^{\rm (diag)}=63$, $N_{\rm conv}^{\rm (DN)}=25$ and $N_{\rm conv}^{\rm
  (DN+NG)}=18$, for the diagonal and DN-ALO, excluding or including the
Ng-acceleration, respectively (lower right panel of
Fig.~\ref{fig:convergence}).
%
%
\subsection{Boundary conditions and zero-opacity models}
\label{subsec:numdiff}
Two important points still to be explained refer to the incorporation of
boundary conditions and to numerical diffusion. The latter can be best
understood by considering continuum models with an opacity set to zero.
\paragraph{Boundary conditions.}
\begin{figure}[t]
\resizebox{\hsize}{!}
{\includegraphics{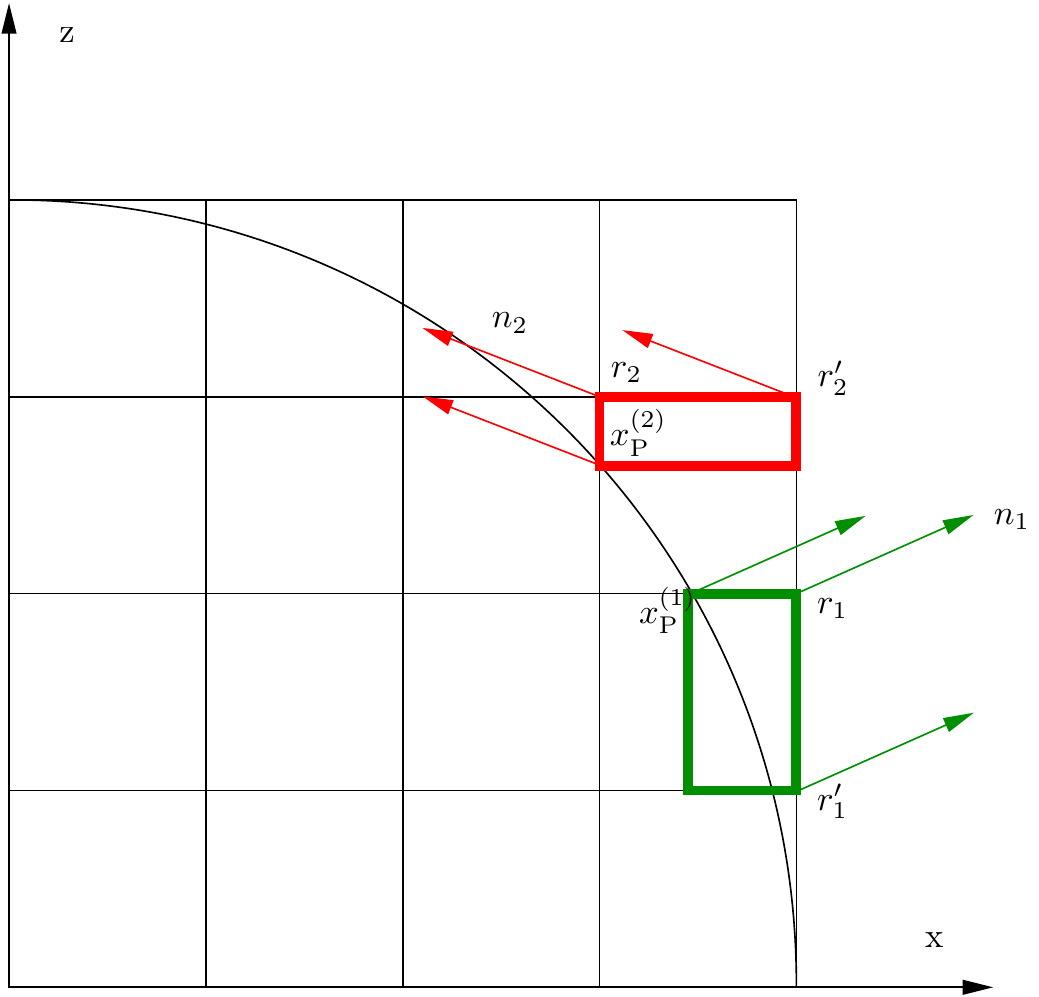}}
\caption{Boundary conditions for two different points, $r_1$, $r_2$, and
  different directions, $n_1$, $n_2$. \textit{Green:} A ray originating from
  the stellar photosphere. To calculate the intensity at $r_1$ in direction
  $n_1$, the intensities at the neighbouring grid points, $x_{\rm P}^{(1)}$
  and $r_1'$, need to be known. A boundary condition is required for grid
  point $x_{\rm P}^{(1)}$, while the intensity at $r_1'$ results within the
  `normal' RT-scheme.  \textit{Red:} A ray originating from outside the
  photosphere. For the grid point $r_2$, a boundary condition has to be
  specified at the corresponding phantom point, $x_{\rm P}^{(2)}$.}
\label{fig:boundary}
\end{figure}
At the outer boundary, the intensities coming from outside are set to zero,
whereas those coming from inside are calculated within the RT scheme, and need
not to be specified. Close to the star, the situation is more complicated: The
inner grid points are generally located close to the boundary, but not
directly on it, due to the difference of Cartesian and spherical grids. The
intensities at those grid points are calculated by the standard FVM-RT
(Eq. \ref{eq:eqrt_disc}), but using different grid cells, defined by the
intersection(s) of the original grid cell with the (spherical) photosphere
(see Fig. \ref{fig:boundary}). The intensity for certain directions needs to
be specified at those so-called phantom points. Figure \ref{fig:boundary}
displays phantom points corresponding to two distinct grid points, $r_1$,
$r_2$, and different ray directions, $n_1$, $n_2$. Radiation originating from
the stellar surface (direction $n_1$ in Fig. \ref{fig:boundary}) is set to
$B_\nu(\Teff)$\footnote{We note here that we are considering a core-halo
  approximation. For future applications, we will include the diffusion limit
  at the lower boundary, of course.} at the corresponding phantom point,
$x_{\rm P}^{(1)}$. For some grid points and ray directions, however
(\eg~direction $n_2$ in Fig. \ref{fig:boundary}), even intensities incident
onto the photosphere need to be specified at the corresponding phantom
point.

Within the core-halo approximation used here, inwards directed intensities
should, in principle, be set to zero at the lower boundary, that is $I_{\rm
  phantom-point}=0$. If, on the other hand, the lower boundary is located at
significant optical depths (as for the majority of the cases considered here),
a specification within a simplified diffusion approximation, $I_{\rm
  phantom-point} \approx B_\nu(\Teff)$, is more appropriate. Although the
diffusion approximation is no longer justified when concentrating on purely
scattering atmospheres, backscattering of photons in such environments mimicks
a similar effect, at least if the optical depths are not too low.

We have tested this issue by considering an optically thin model as the most
extreme test-bed, using both alternative descriptions for those critical
directions. The mean relative errors for both alternatives are of the same
order, and do not significantly differ from those arising under more physical
conditions discussed later (with larger optical depths at the lower boundary,
see Sect.~\ref{subsec:varodepth} and Table~\ref{tab:mint_odepth}). Thus, we
apply $B_{\nu}(\Teff)$ as the inner boundary condition for those critical rays
as well, also because this procedure is less time-consuming, since it avoids
conditional clauses in the innermost loop of the code.
\paragraph{Zero-opacity models.} 
\begin{figure}[t]
\begin{minipage}{9cm}
   \resizebox{\hsize}{!}{\includegraphics{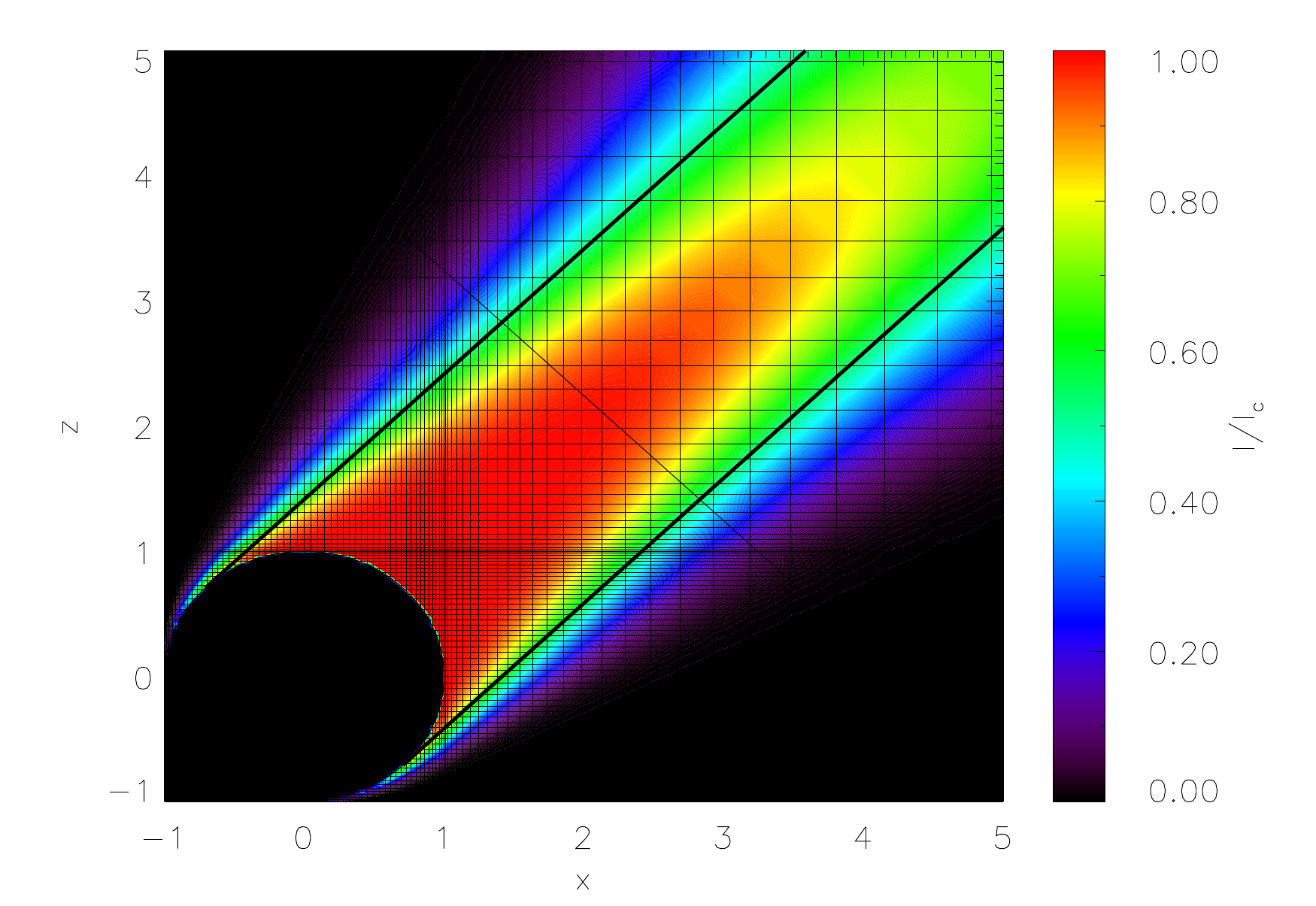}} 
\end{minipage}
\begin{minipage}{9cm}
   \resizebox{\hsize}{!}{\includegraphics{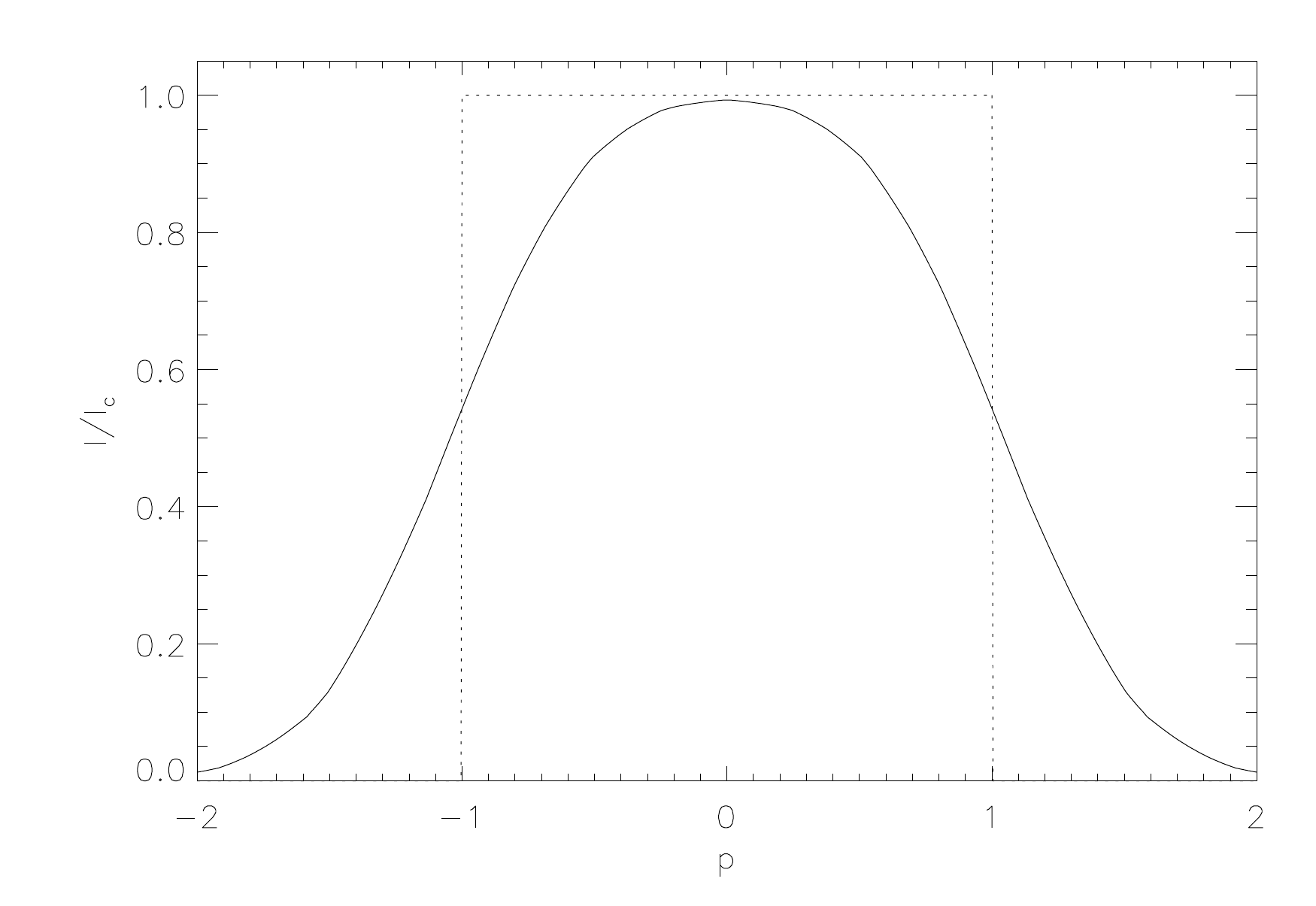}}
\end{minipage}
\begin{minipage}{9cm}
   \resizebox{\hsize}{!}{\includegraphics{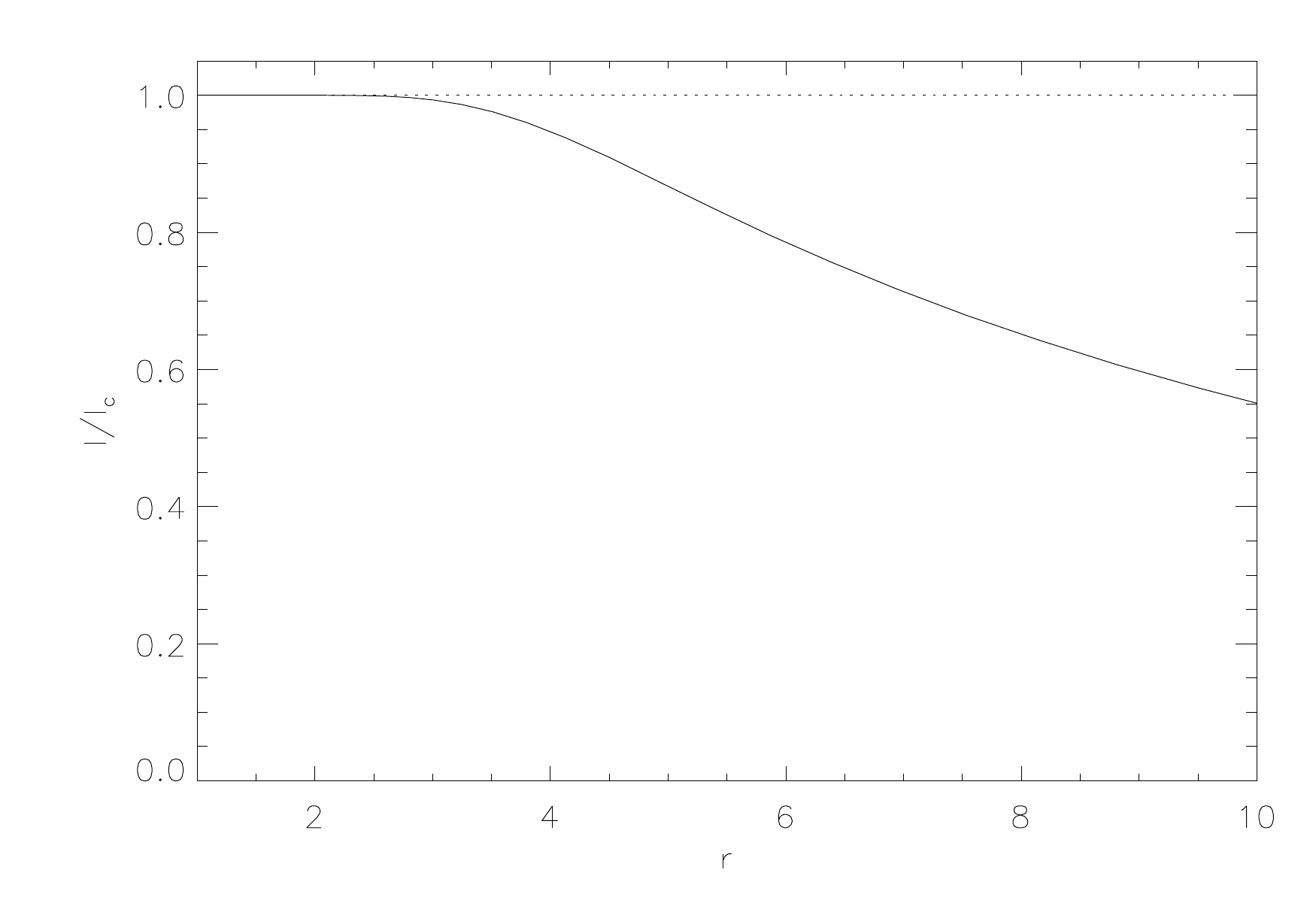}}
\end{minipage}
%
\caption{\textit{Top:} Specific intensities for direction $\vecown{n}=(1,0,1)$
  in the x-z-plane, for a zero-opacity model. Overplotted is the spatial grid
  with a typical size of $\nx=\ny=\nz=133$ grid points. The thick lines
  indicate the theoretical boundary of the beam. \textit{Middle:} Projected
  intensity profile through the area indicated by a straight line in the top
  panel, covering all rays with direction $\vecown{n}$ and impact parameter $p
  = \pm2\, \Rstar$, and theoretical intensity profile denoted by dotted lines.
  The central distance from the origin is 3~\Rstar. \textit{Bottom:} Specific
  intensity along $\vecown{n}$, compared to the theoretical value indicated by
  the dotted line.}
\label{fig:searchlight}
\end{figure}
\begin{figure}[t]
\resizebox{\hsize}{!}{\includegraphics{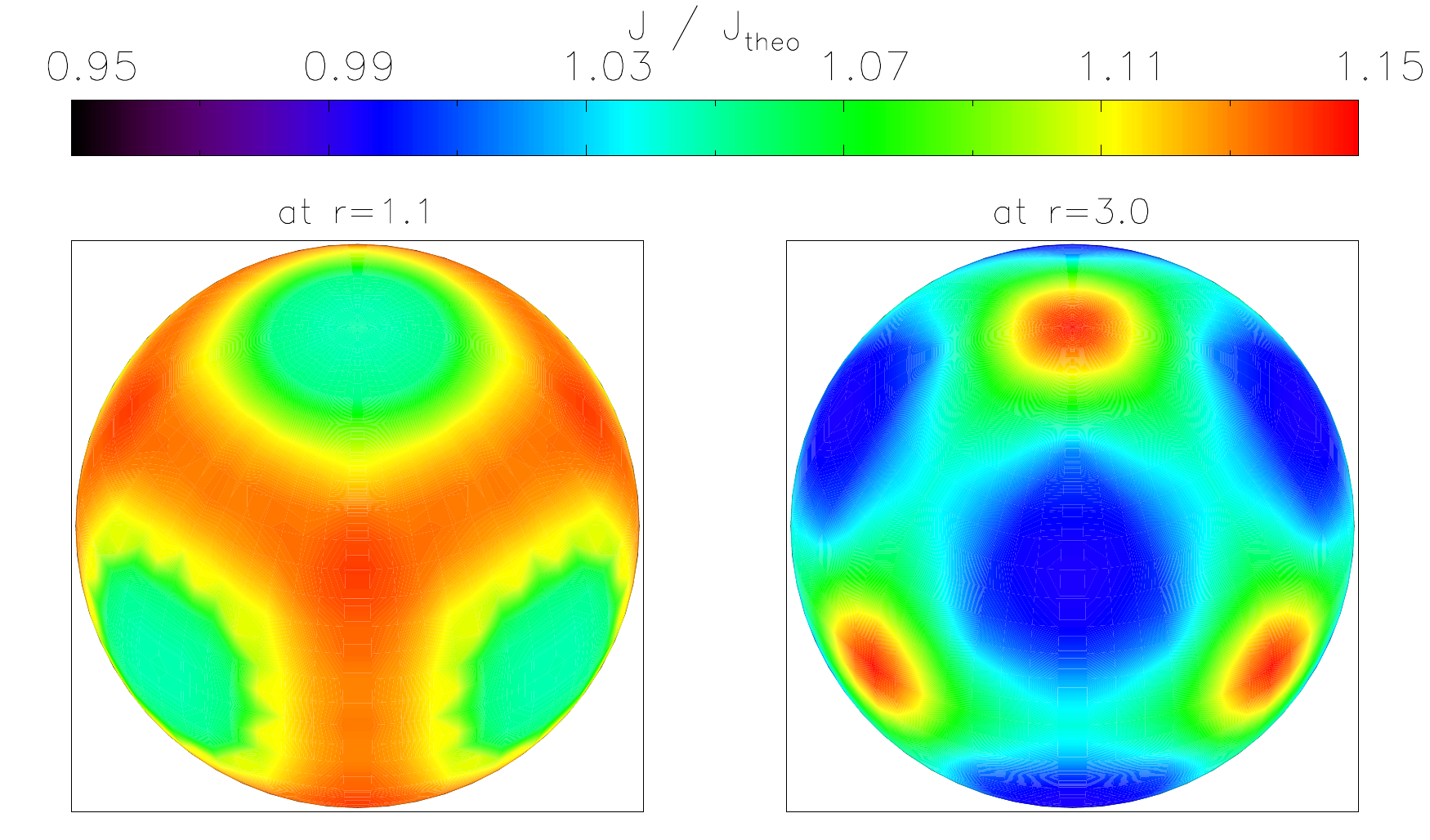}}
%
\caption{Contours of the mean intensity for a zero-opacity model, normalized
  by its theoretical value, on spherical surfaces, $r=1.1\, \Rstar$ (left) and
  $r=3\, \Rstar$ (right). The line of sight is along the vector
  $\vecown{n}=(1,1,1)$. $\nx=\ny=\nz=133$ grid points have been used.}
\label{fig:mint_thin2}
\end{figure}
Numerical diffusion is a major error source within the FVM (it also occurs in
the SC method, though to a lesser extent, see \citealt{Kunasz88} and
\citealt{Ibgui13} for an analysis of this effect in the context of multi-D SC
methods). Quite generally, rays propagating parallel to the grid axes are
nearly undisturbed by this effect, whereas those propagating at different
angles are effectively widened, due to the diffusion of intensity into
neighbouring cells. This diffusion results from the finite size of the grid
cells, and the competition between the $\Delta x / \Delta y$, $\Delta x /
\Delta z$ and $\Delta y / \Delta z$ terms in the discretized EQRT
(Eq.~\ref{eq:eqrt_disc}), for any given direction, and can only be minimized
by increasing the grid resolution.

To obtain a qualitative measure of numerical diffusion errors, we performed a
searchlight beam test with ray direction $\vecown{n}=(1,0,1)$, by setting the
opacity to zero. For consistency, the critical boundary conditions have been
set to zero.

The top panel of Fig.~\ref{fig:searchlight} shows the contours of the specific
intensity in the x-z-plane, and the middle panel displays the intensity
through an aperture perpendicular to the ray direction. In our projection, the
aperture appears as a straight line in the x-z-plane, specified by its
distance from the origin and the impact parameter. Finally, the bottom panel
of Fig.~\ref{fig:searchlight} shows the intensity along the considered
direction, at the centre of the beam.

Evidently, numerical diffusion plays a crucial role. We expect corresponding
errors to be most severe when photons propagate over large distances, for
instance, for optically thin continua, or, in the line case, before they hit
the resonance zones. Since the resonance zones are mostly quite narrow, while
the path-length of freely propagating line photons is usually quite large (at
least if the continuum is comparatively weak, as in most realistic
conditions), numerical diffusion errors are of major significance for the line
transfer.

To address the impact of numerical diffusion on the final solution, we
consider the mean intensity for zero-opacity models. Figure
\ref{fig:mint_thin2} shows the mean intensity (scaled by its theoretical value
obtained from the dilution factor) for such models on spherical surfaces at
two distinct radii, $r = 1.1\, \Rstar$ and $r = 3\, \Rstar$.  We find a clear
pattern of the shape of the mean intensities: Close to the star, the mean
intensities are reasonably accurate on the axes, in contrast to the regions
away from the axes, where they become overestimated. Far from the star, the
situation reverses, with reasonable results away from the axes, and an
overestimate on the axes. This behaviour is explained in the following.

For a given grid point on or close to a coordinate axis, and far from the
stellar surface, core-rays, that means those originating from the stellar
surface, remain nearly undisturbed by numerical diffusion.  Without diffusion,
only such core-rays would contribute to the mean intensity. Due to numerical
diffusion, however, also non-core rays contribute, that means those which form
a certain angle \wrt~the considered grid axis, since they have been fed with
intensity by corresponding core-rays propagating in the same direction
(widening of the effective aperture, see above). Consequently, the mean
intensity becomes overestimated.

For grid points far from the star, and away from the major axes,
core-rays and non-core rays are both affected by numerical diffusion,
resulting in an under- and over-estimation of the intensity,
respectively. Consequently, there is a significant cancellation of both
effects, and the mean intensity remains close to its expected value.

At points located on the grid axes close to the star, numerical
diffusion plays only a minor role, mostly because the contributing non-core
rays are propagating almost perpendicular to the considered axis, which means
parallel to one of the other axes, with negligible diffusion errors.

Away from the axes, and close to the star, contributing non-core rays are
significantly inclined \wrt~the coordinate axes, and thus strongly fed by
diffusion effects. Thus, the mean intensities become overestimated.

Due to the different effects for different ray directions and for different
regions in the atmosphere, any symmetry will be broken. This error cannot be
avoided, and is minimized only for higher grid resolutions.  As the important
part of the radiative transfer is mainly located near to the star (where the
densities are largest), and the numerical diffusion errors are not too large
in this regime (up to a radius of $r\lesssim 3-4 \, \Rstar$), the 3D solution
scheme should deliver at least qualitatively correct results.
%
\subsection{Variation of optical depth}
\label{subsec:varodepth}
\begin{figure*}[t]
\resizebox{\hsize}{!}{
   \begin{minipage}{0.33\hsize}
      \resizebox{\hsize}{!}{\includegraphics{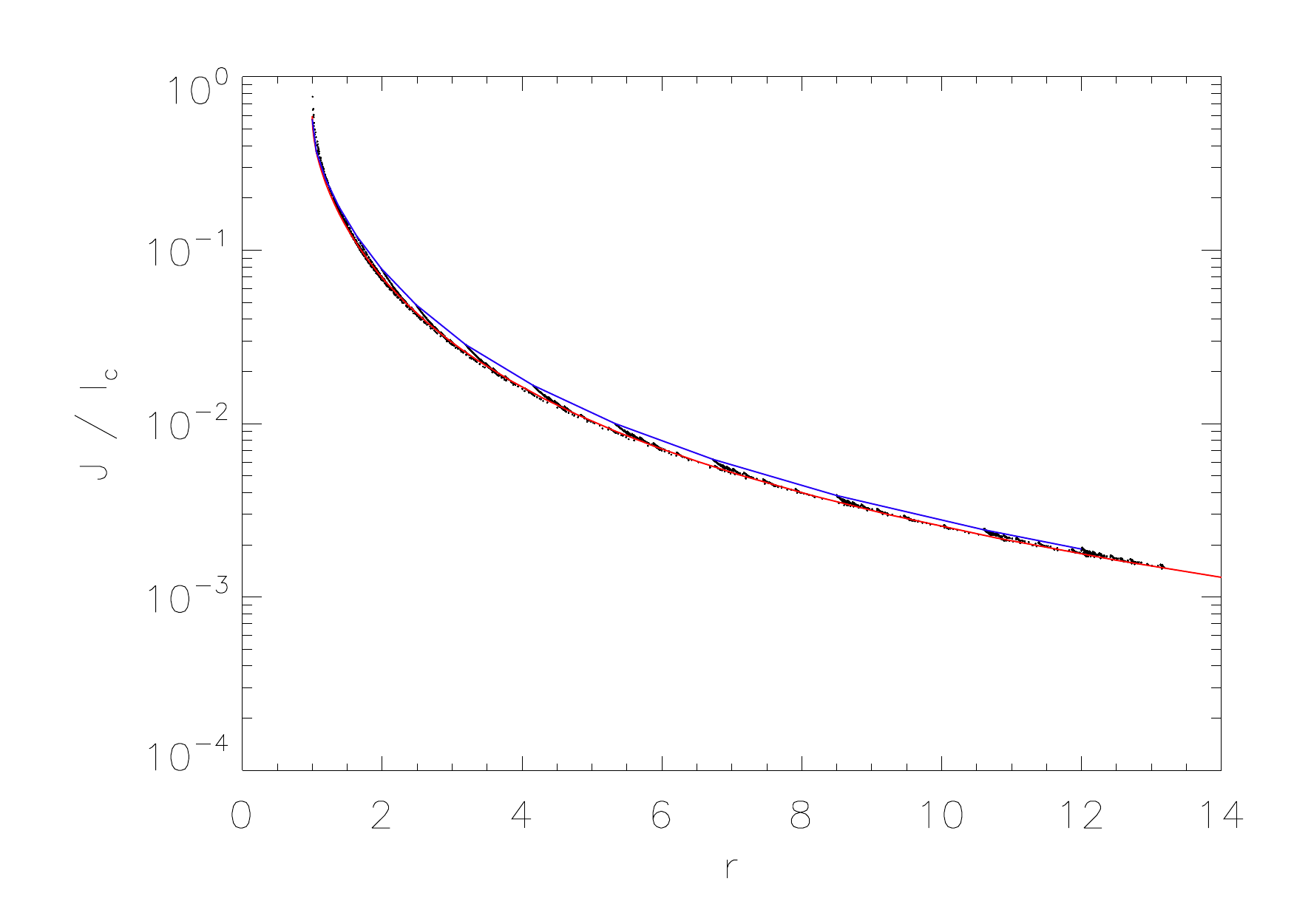}}
   \end{minipage}
   \begin{minipage}{0.33\hsize}
      \resizebox{\hsize}{!}{\includegraphics{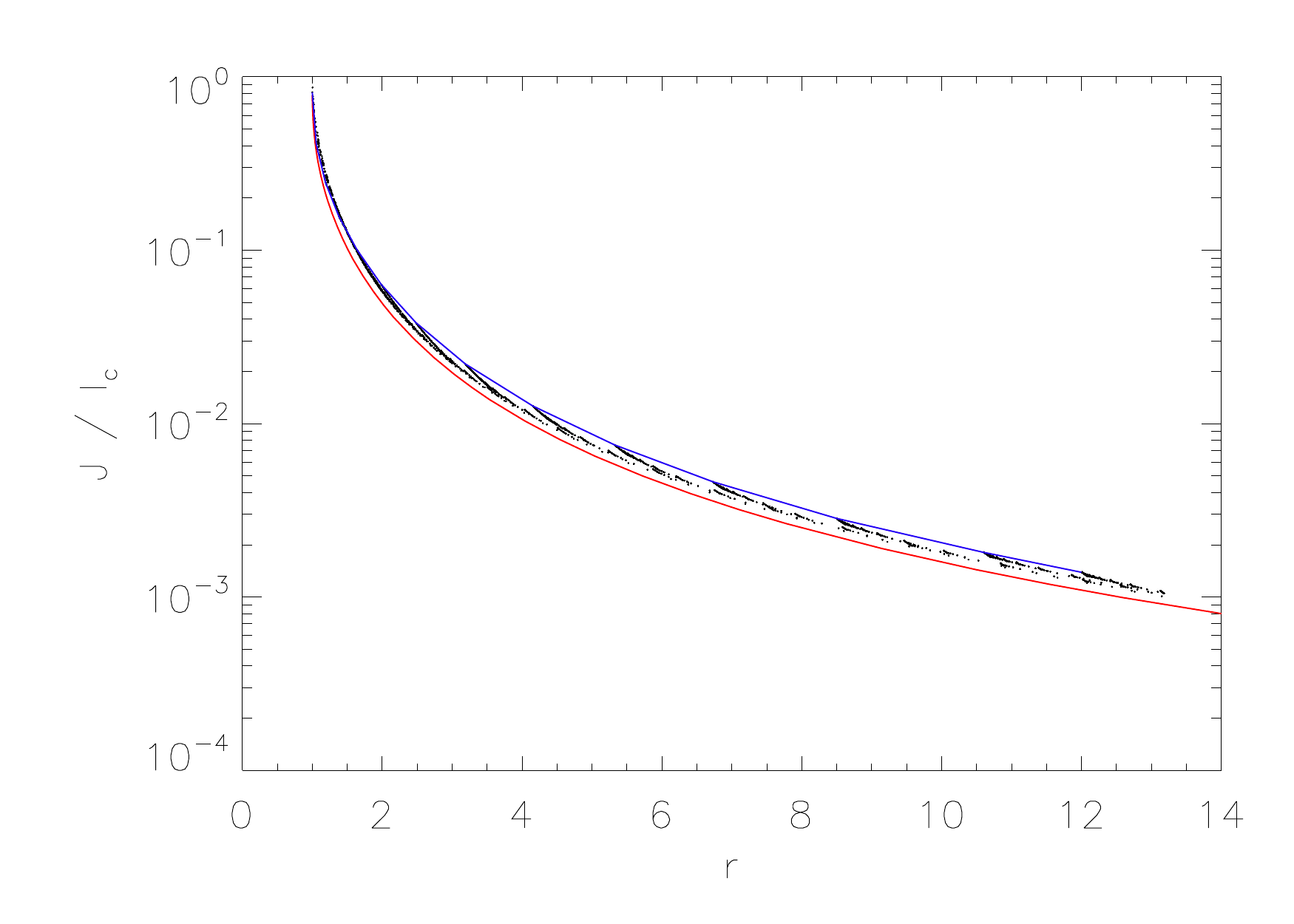}} 
   \end{minipage}
   \begin{minipage}{0.33\hsize}
     \resizebox{\hsize}{!}{\includegraphics{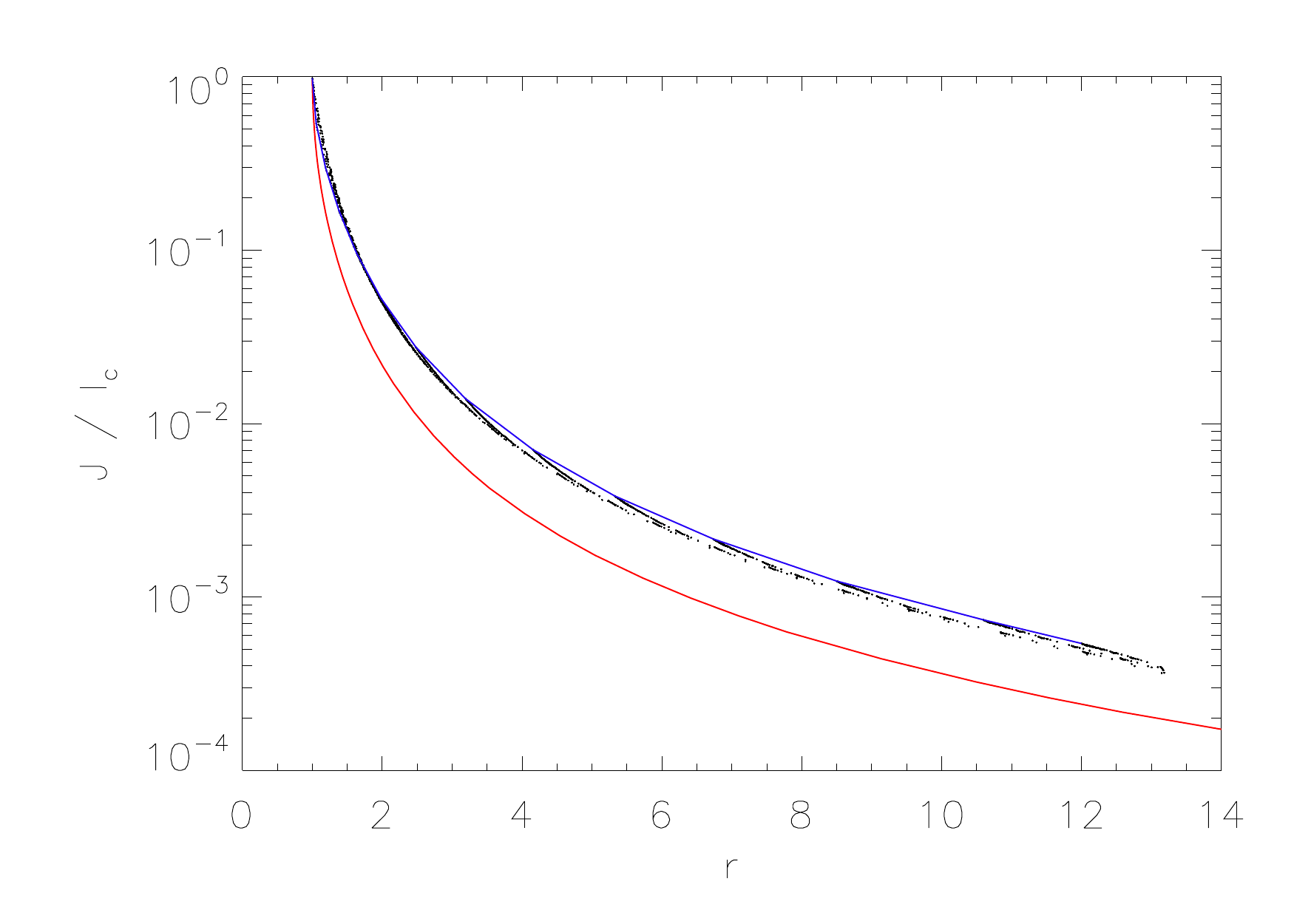}} 
   \end{minipage}
}
\\
\resizebox{\hsize}{!}{
   \begin{minipage}{0.33\hsize}
      \resizebox{\hsize}{!}{\includegraphics{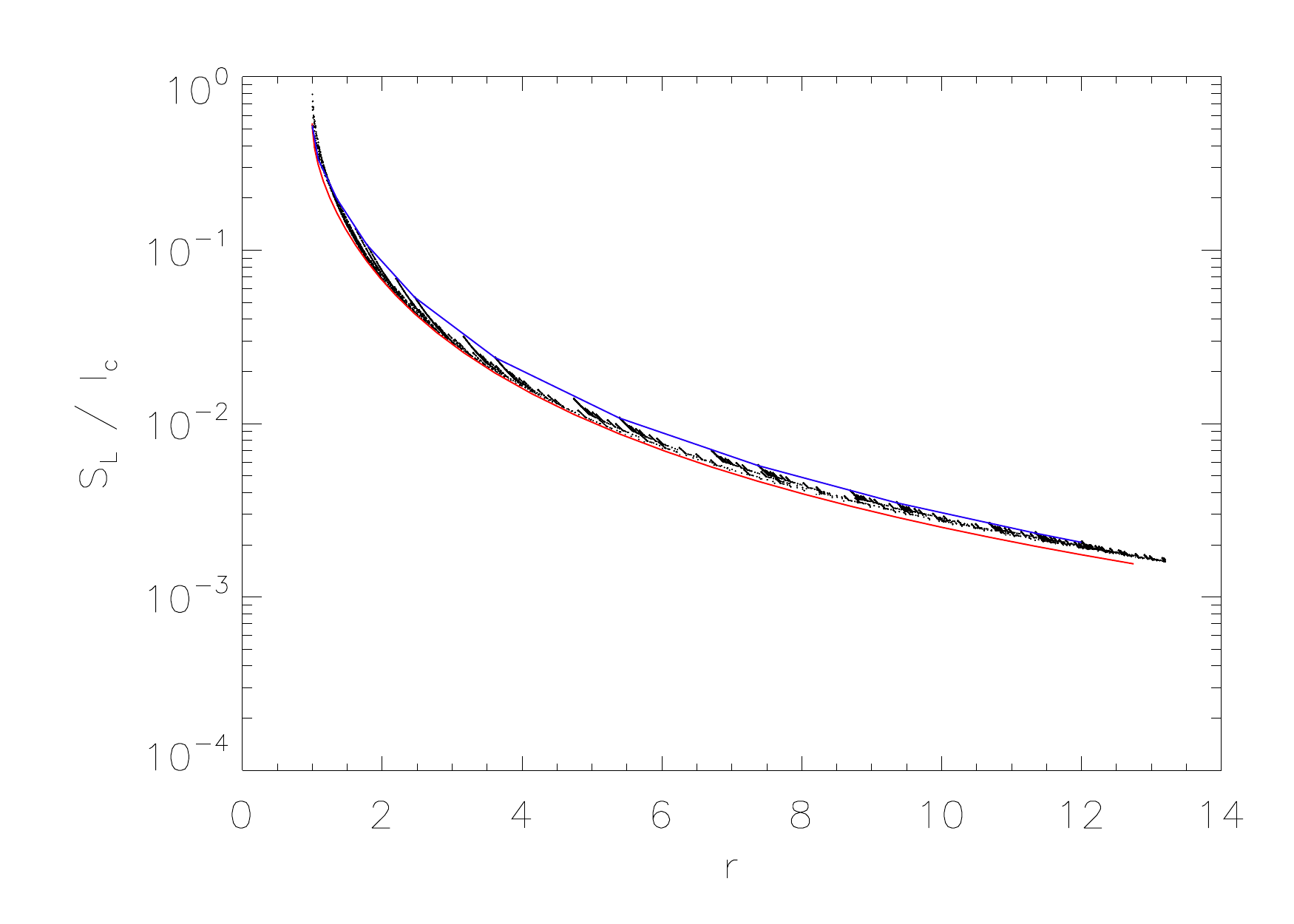}} 
   \end{minipage}
   \begin{minipage}{0.33\hsize}
      \resizebox{\hsize}{!}{\includegraphics{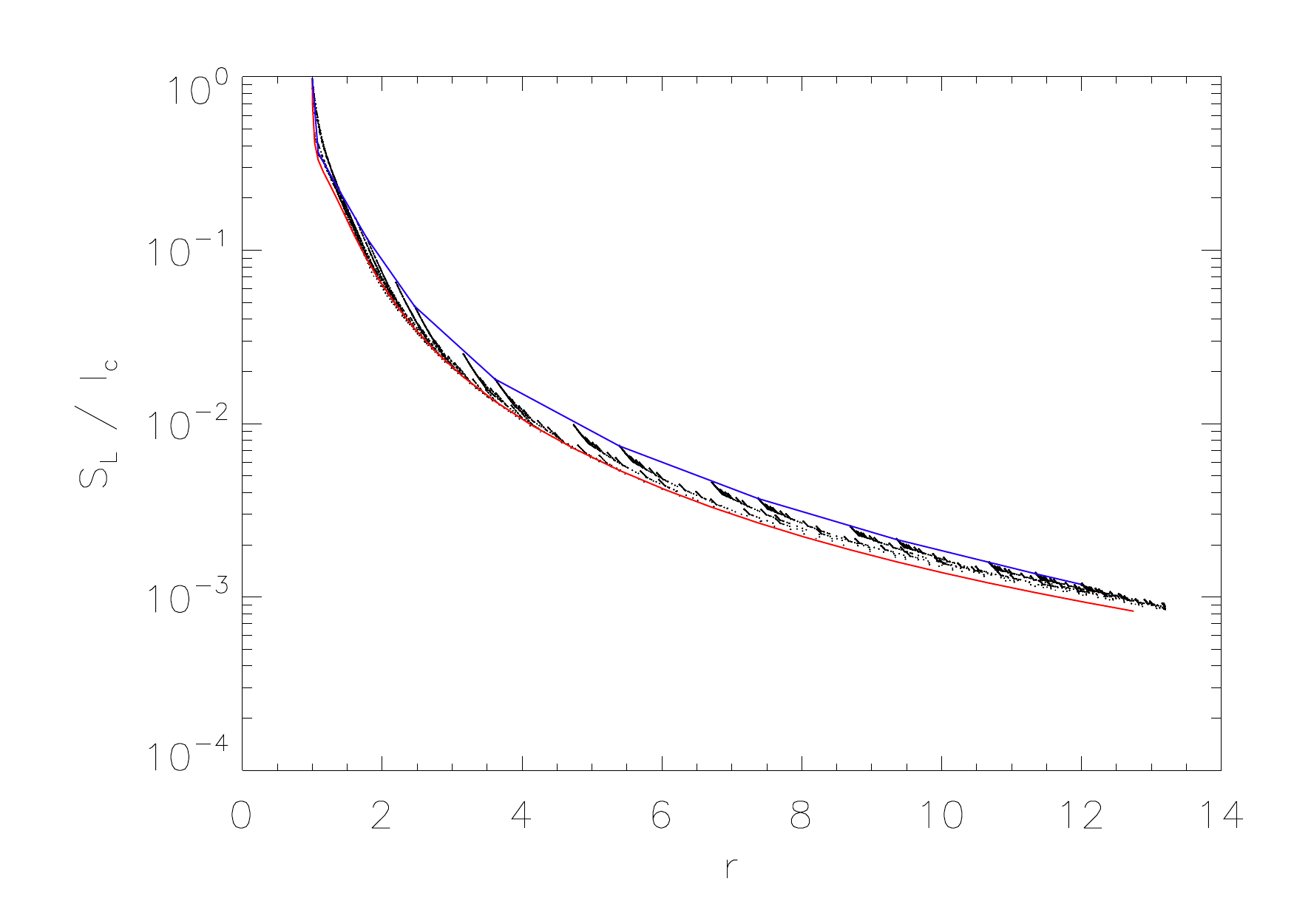}} 
   \end{minipage}
   \begin{minipage}{0.33\hsize}
      \resizebox{\hsize}{!}{\includegraphics{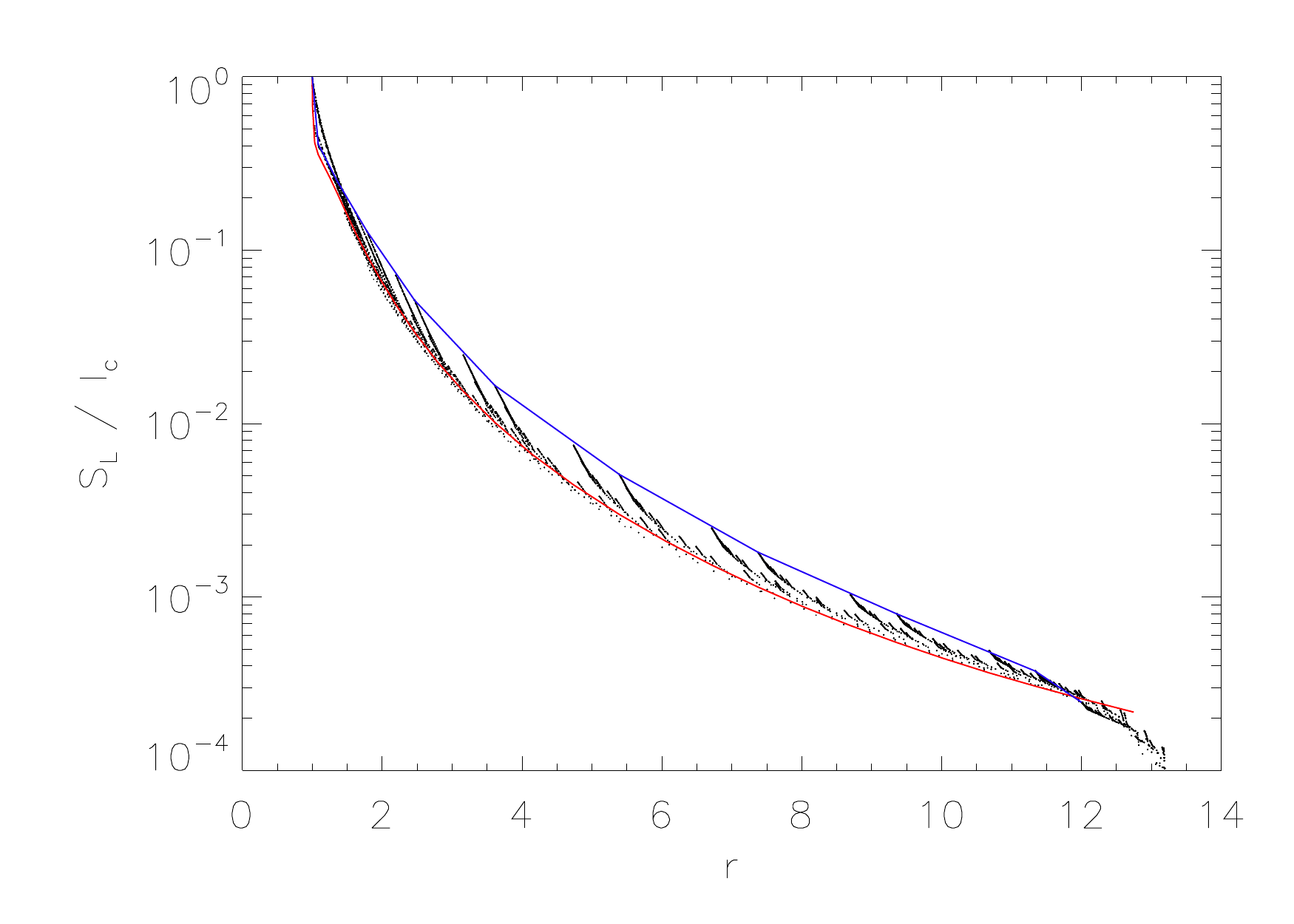}}
   \end{minipage}
}
%
\caption{\textit{Top:} Mean intensities (scaled by the emitted intensity from
  the stellar core, $I_{\rm c}$) for the continuum transport, with
  $\epsc=10^{-6}$, and $\kcont=(10^0, 10^1, 10^2)$ corresponding to $\tau_r =
  (0.17, 1.7, 17.0)$, from left to right. \textit{Bottom:} Line source
  functions (scaled by $I_{\rm c}$) assuming an optically thin continuum, and
  with $\epsl=10^{-6}$, and $\kline=(10^0, 10^3, 10^5)$ from left to
  right. The red line corresponds to the accurate 1D solution, the blue line
  is the solution along the x-axis, and the dots (mainly located in between
  the red and the blue line) correspond to the solution of (arbitrarily)
  selected grid points (only a subset of all grid points is shown to compress
  the images).}
\label{fig:mint_odepth}
\end{figure*}
\begin{table}
\begin{center}
\caption{Mean and maximum relative errors of our 3D solution scheme, when
  compared to an `accurate' 1D solution. The top panel shows the errors of
  the mean intensity for a pure continuum with different \kcont ~parameters,
  while the bottom panel shows the errors of the line source function with
  different line strengths \kline ~and an optically thin background
  continuum. The thermal contribution, $\epsc, \epsl$, has been set to
  $10^{-6}$ in both cases.}
\label{tab:mint_odepth}
\begin{tabular}{cccc}

\kcont & $10^0$ & $10^1$ & $10^2$  \\
\hline
\noalign{\vskip 0.5mm}
$\tau_r$ & $0.17$ & $1.7$ & $17$  \\
\hline
\noalign{\vskip 0.5mm}
$\bar{\Delta J} [\%]$ & 3.35 & 20.29 & 127.91 \\
\hline
\noalign{\vskip 0.5mm}
$\Delta J_{\rm max} [\%]$ & 49.28 & 42.18 & 151.34 \\
\hline\hline
\noalign{\vskip 0.5mm}
\kline & $10^0$ & $10^3$ & $10^5$ \\
\hline
\noalign{\vskip 0.5mm}
$\bar{\Delta \sline} [\%]$ & 12.38 & 18.47 & 23.00 \\
\hline
\noalign{\vskip 0.5mm}
$\Delta S_{\rm L, ~max} [\%]$ & 61.95 & 82.50 & 94.13

\end{tabular}
\end{center}
\end{table}
We finally compare the results from our 3D solutions for a spherically
symmetric wind with corresponding ones from 1D solutions using spherical
coordinates, as a function of optical depth.  Within the continuum transport,
the optical depth has been varied by increasing the linear scaling factor,
$\kcont=(10^0, 10^1, 10^2)$, which defines a radial optical-depth scale,
$\tau_r = (0.17, 1.7, 17.0)$, at the lowermost point.  The solutions for the
mean intensities, together with the 1D solution, are shown in
Fig.~\ref{fig:mint_odepth}, top panel. The line transitions have been
calculated for the same model, assuming an optically thin continuum in order
to extract the error from the line transport alone. The adopted line-strength
parameters, $\kline=(10^0, 10^3, 10^5)$, describe a weak, intermediate and
strong line, respectively. The corresponding solutions are shown in the bottom
panel of Fig.~\ref{fig:mint_odepth}.  To ensure convergence, we have used
the DN-ALO together with the Ng-acceleration (see
Sect.~\ref{subsec:convergence}) for all test calculations. Thus, the
differences between the 1D and 3D solution originate from the formal solution
scheme alone, and not from a false convergence. The mean and maximum relative
errors are summarized in Table \ref{tab:mint_odepth}.

The mean errors are increasing together with the optical thickness, because
the FVM becomes a first-order scheme for increasing optical-depth
steps\footnote{A first-order scheme is sufficiently accurate if $\exp (-\Delta
  \tau ) \approx 1-\Delta \tau$, \ie~if the optical-depth steps, $\Delta
  \tau$, are small.}. The errors in the line case do not exceed roughly 25\%,
and are generally lower than those for the continuum, because the RT is much
more local (see above)\footnote{We emphasize that for the line transfer, the
  ratio of the photon destruction probability, \epsl, to the photon escape
  probability is less than unity for all our models, indicating that the
  line-transfer is, in principal, non-local. However, for the considered
  spherically symmetric problems, no multiple resonances arise, and the line
  is formed within a single, well-localized resonance region.}, and thus, the
error is not being propagated through the complete grid. On the other hand,
numerical diffusion plays a larger role (see Sect.~\ref{subsec:numdiff}),
resulting in minimum errors of $\gtrsim 10\%$ even for very weak lines.

For optically thick continua (with $\tau_r \gtrsim 20$), the continuum
transfer breaks down, giving mean errors larger than $100
\%$. Figure \ref{fig:mint_odepth} shows that the errors for the continuum and
line transport are mostly due to an overestimation of the mean intensities and
scattering integrals, respectively. With respect to the local distribution of
the errors, we find a similar behaviour as for the zero-opacity models (on
vs. away from the axes for different radii). The maximum errors are mostly
found at the same locations as in Fig. \ref{fig:mint_thin2}.

Finally, we conclude that the overall error is a combination of errors
introduced by numerical diffusion, and the first-order scheme for optically
thick environments. The line transport is generally more reliable than the
continuum transport, which should be treated with higher spatial resolution.
\subsection{Emergent flux profile}
\begin{figure*}[t]
\resizebox{\hsize}{!}{
   \begin{minipage}{0.33\hsize}
      \resizebox{\hsize}{!}{\includegraphics{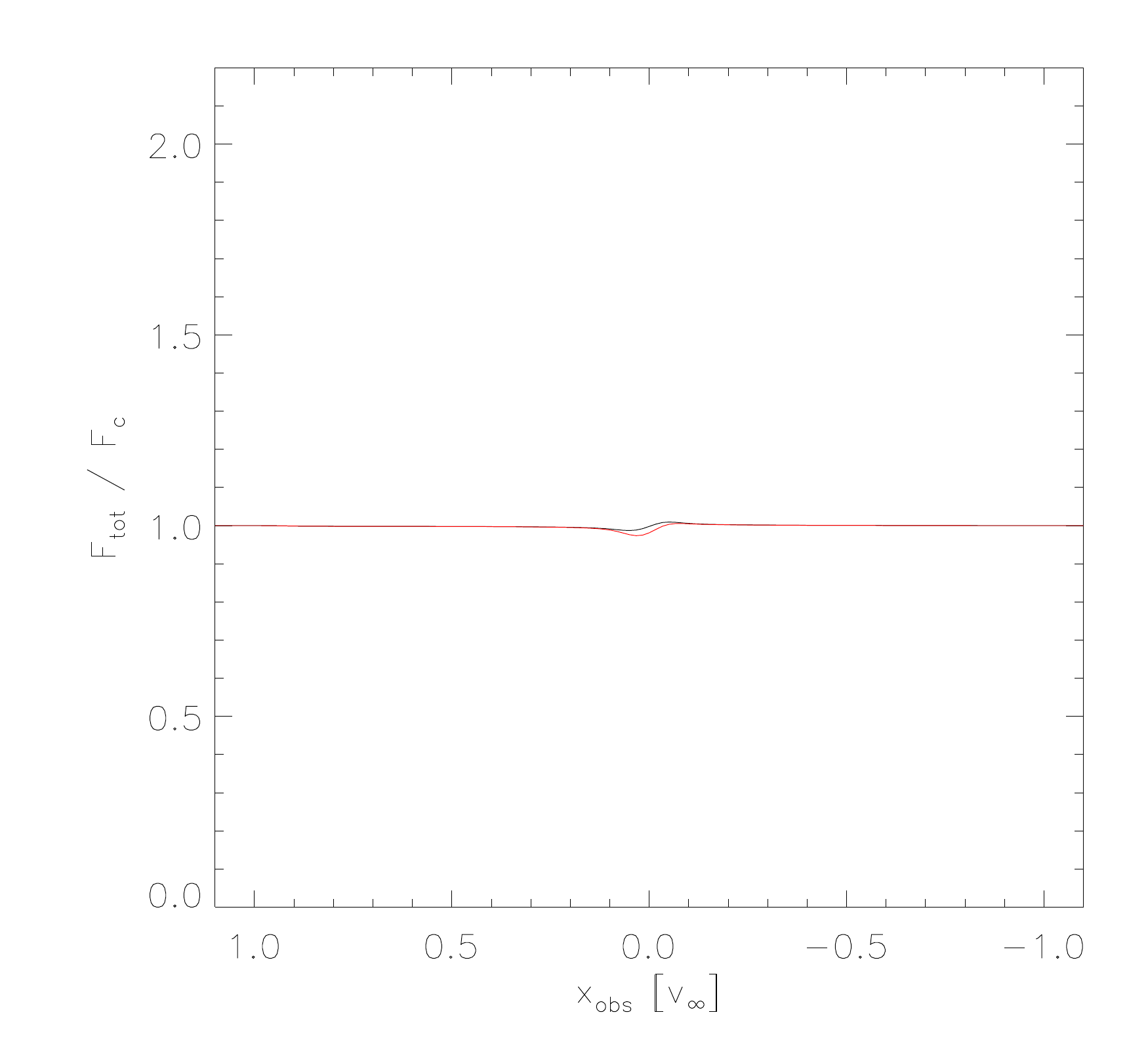}}
   \end{minipage}
   \begin{minipage}{0.33\hsize}
      \resizebox{\hsize}{!}{\includegraphics{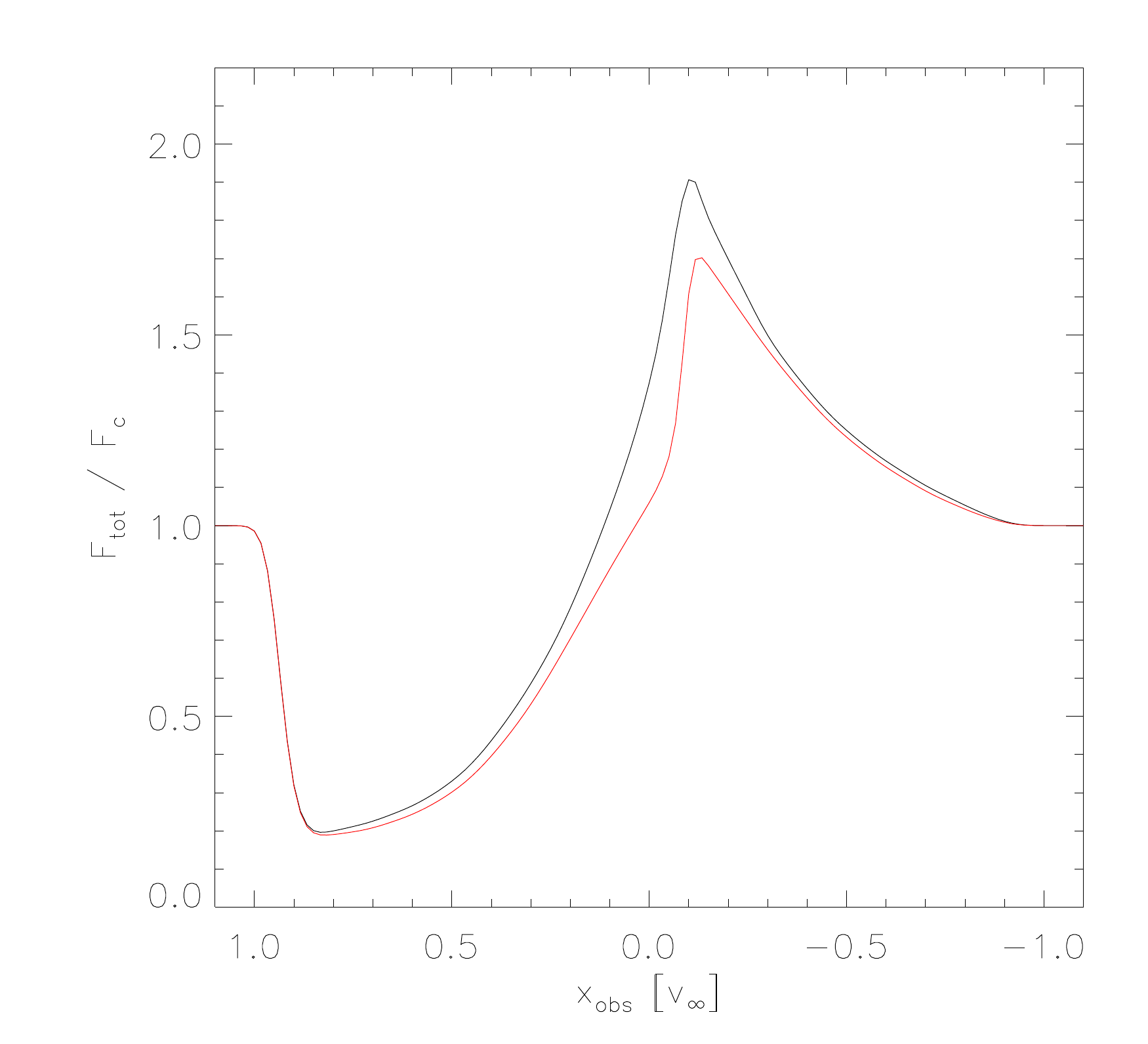}}
   \end{minipage}
   \begin{minipage}{0.33\hsize}
      \resizebox{\hsize}{!}{\includegraphics{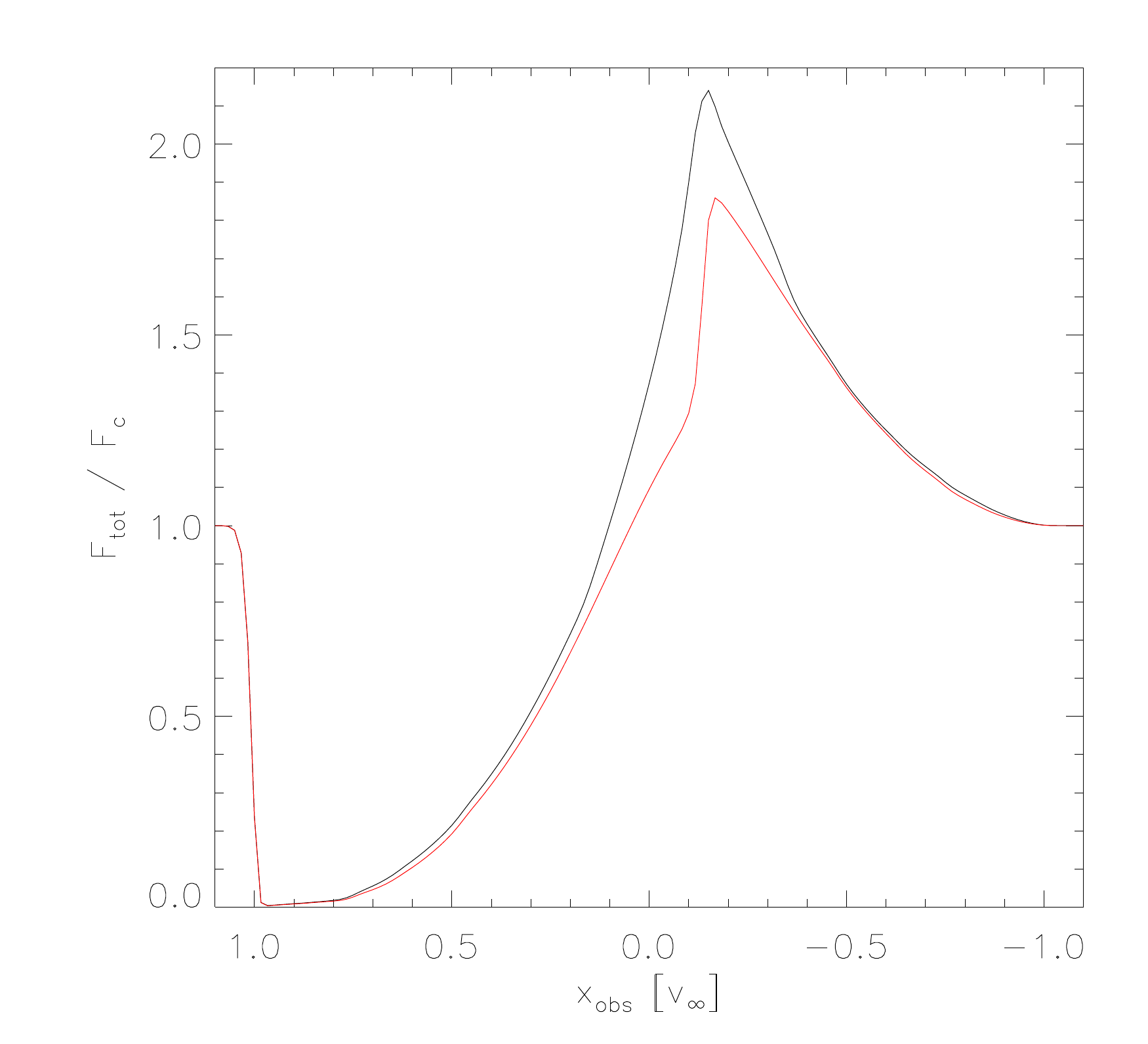}}
   \end{minipage}
}
%
\caption{Emergent line profiles for spherically symmetric models with
  different line-strengths, $\kline=(10^0, 10^3, 10^5)$, from left to
  right. The red line corresponds to the accurate 1D solution, and the black
  line is the solution corresponding to our 3D calculations.}
\label{fig:line_profiles}
\end{figure*}
To calculate the line profiles, we implemented a postprocessing LC
method. Within this method, the converged source function is used to derive
the formal solution along characteristics in a cylindrical coordinate system,
with the z-axis aligned with the line of sight. The emergent intensity is
finally integrated over the projected stellar disc, which gives the flux (see,
\eg~\citealt{Lamersetal87}, \citealt{Busche2005}, \citealt{Sundqvist12c}, for
more details).  To obtain the source functions at each position of a given
characteristics, a 3D tri-linear interpolation is performed in
log-space. Though this could be applied to the velocity components and
opacities as well, we calculate those properties from analytic expressions
whenever possible. This way, we avoid errors introduced by the interpolation
scheme, which have been found to be of the order of only few percent for a
$\beta$-velocity law, but have rather strong influence for the ADM models, where
the interpolation of shear-velocities is not a simple task.

The emergent profiles for the same models as above are shown in
Fig.~\ref{fig:line_profiles}, with differences between the 1D and 3D models
originating from the line source function alone. Generally, when compared to
the 1D solution, the line profiles from our 3D code overestimate the emission
part due to the larger source functions. The absorption edge is slightly
red-shifted from its theoretical value at $x_{\rm obs}\approx 1 + \vth/\vinf$,
because the calculation volume extends only up to $13.2\, \Rstar$ (where
$v(13.2\, \Rstar) = 0.92\, \vinf$). This issue, however, could be improved by
enlarging the size of the calculation volume.

Overall, despite the slightly enhanced emission peak, we find that the line
profile can be reproduced by our FVM, in combination with an accurate
postprocessing routine, for a wide range of line-strength parameters.  Thus,
our method allows for a qualitative interpretation of line profiles even for
the most extreme test cases, that is for strong scattering lines.
\section{Wind-ablation} 
\label{sec:wind_ablation}
\begin{figure}[t]
\resizebox{\hsize}{!}
{\includegraphics{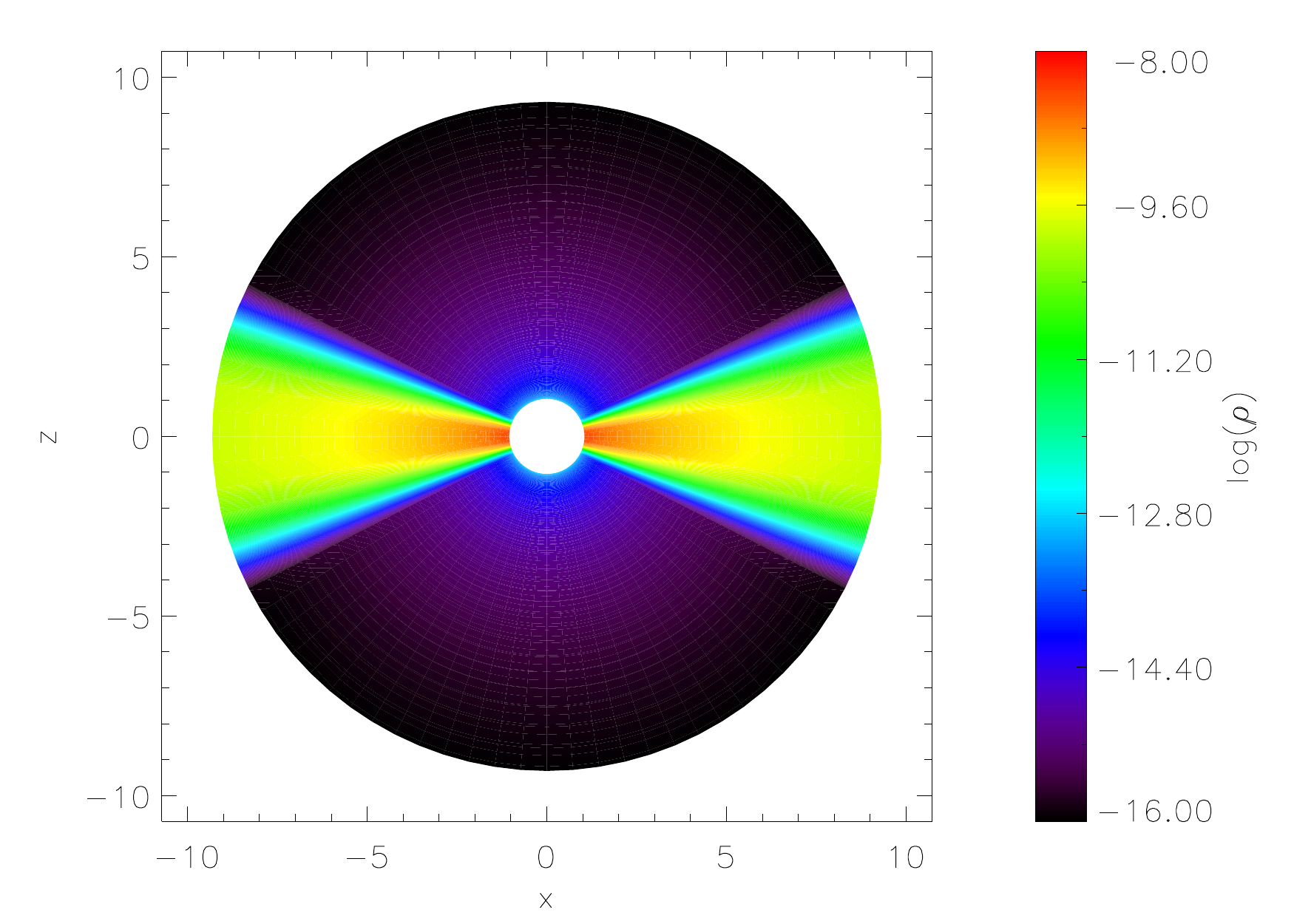}}
%
\caption{Density structure for the wind-ablation model with $\tau_{\rm disc} =
  1400$, in the x-z-plane.}
\label{fig:density_abl}
\end{figure}
\begin{figure}[t]
\resizebox{\hsize}{!}
{\includegraphics{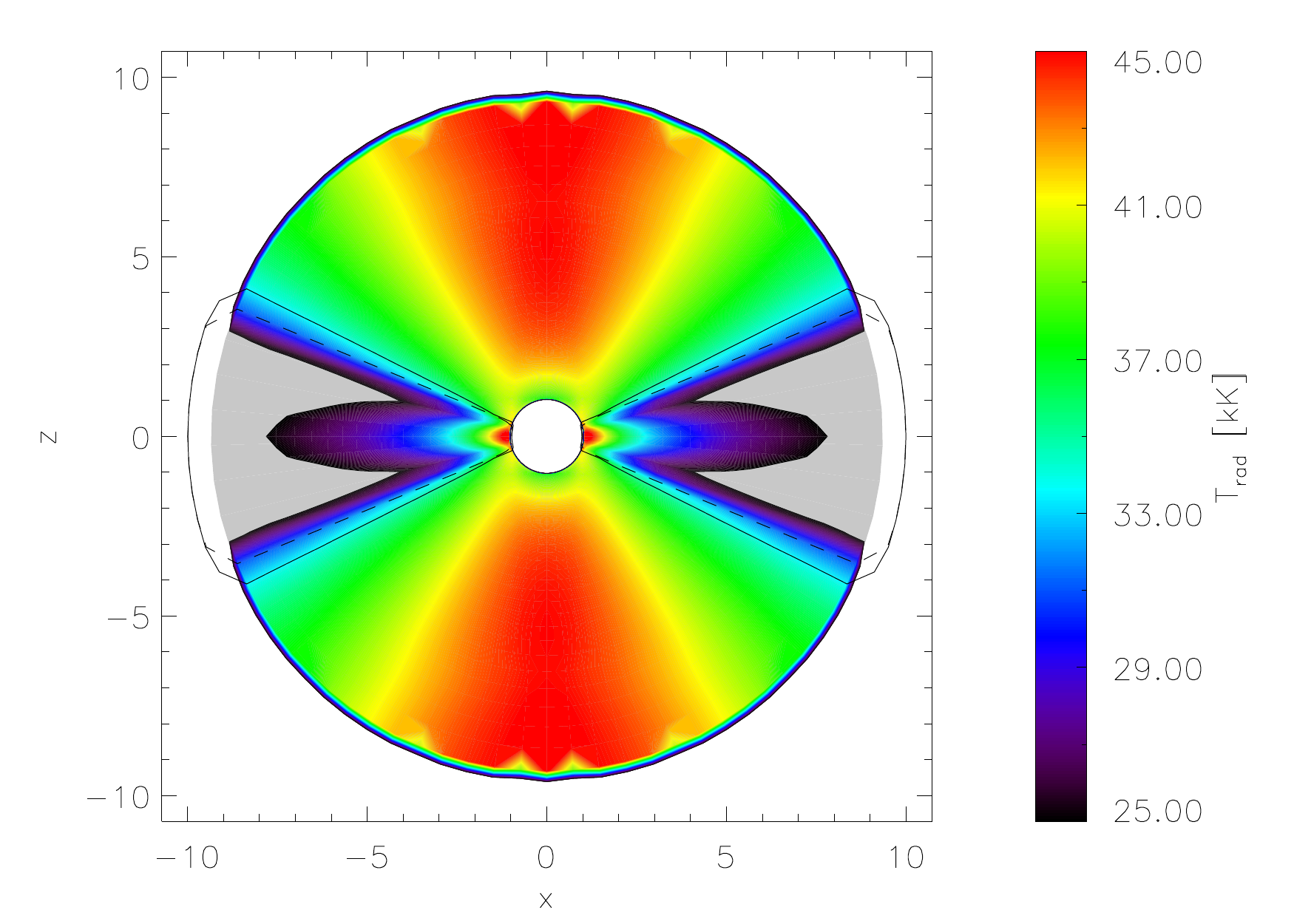}}
%
\caption{As Fig.~\ref{fig:density_abl}, but for the radiation
  temperature. Additionally, the density contours corresponding to a decrease
  in line force by $f = 10$ (solid) and $f = 100$ (dashed) are displayed. Both
  contours indicate the transition region from wind to disc (see text). The
  grey colour corresponds to radiation temperatures less than the colour-coded
  minimum value.}
\label{fig:trad_abl}
\end{figure}
\begin{table}
\begin{center}
\caption{Stellar and wind parameters for the wind-ablation model. The
  line-strength parameters have been set to $\alpha=0.66$, $\bar{Q}=2500$, and
  $Q_0=2200$.}
\label{tab:paramABLATION}
\begin{tabular}{c|c|c|c}

\multicolumn{1}{c}{\Teff [kK]} & 
\multicolumn{1}{c}{\Rstar [\rsun]} &
\multicolumn{1}{c}{\vmin [\kms]} &
\multicolumn{1}{c}{$\mdot_{\rm wind}$ [\msunyr]}
\\
\multicolumn{1}{c}{36} & 
\multicolumn{1}{c}{9.4} &
\multicolumn{1}{c}{22} &
\multicolumn{1}{c}{$1.5 \cdot 10^{-7}$}

\\\hline\hline
\multicolumn{1}{c}{\logg} & 
\multicolumn{1}{c}{$\beta$} &
\multicolumn{1}{c}{\vinf [\kms]} &
\multicolumn{1}{c}{}
\\
\multicolumn{1}{c}{3.9} &
\multicolumn{1}{c}{1} &
\multicolumn{1}{c}{2200} &
\multicolumn{1}{c}{}
\end{tabular}
\end{center}
\end{table}
Using 3D radiation-hydrodynamic simulations, \cite{Kee15} and \cite{Kee16}
modelled the ablation of circumstellar discs around massive stars, due to
radiative line driving. They showed that a significant line-force arises due
to the coupling of non-radially streaming photons to the non-radial velocity
field of circumstellar discs (see also \citealt[their Figure 1]{Kee16}). The
line-force has been calculated within a Sobolev-approach, by means of
line-strength distribution functions.  Contrasted to the original formulation
by \citet{CAK}, they followed the parameterization by \cite{Gayley95}. The
full 3D line acceleration can then be written as\footnote{Strictly speaking,
  Eq.  \eqref{eq:lforce} holds only when the strongest line is optically
  thick. See \cite{Kee15}, Chapter 2, for a complete derivation and
  discussion.}
\beq
\label{eq:lforce}
\vecown{g}_{\rm lines} \approx \dfrac{\kappa_{\rm e} \bar{Q}}{(1-\alpha) c
  (Q_0 \kappa_{\rm e} c \rho)^\alpha} \int \bigl( \vecown{n} \cdot
\vecown{\nabla} \cdot (\vecown{n} \cdot \vecown{v})\bigr)^\alpha I(\vecown{n})
\vecown{n} \dd \Omega \,,
\eeq
where $\vecown{n}$ and $\vecown{v}$ describe the direction of the considered
ray and the velocity-vector, respectively. $\rho$ is the density, $\bar{Q}$,
$Q_0$, and $\alpha$ describe the line-strength distribution, and were taken
from the calibration of \citet[their Table 2]{Puls00}, for the considered
\Teff.

For optically thin continua (\eg~in classical Be stars), the incident
intensity $I(\vecown{n})$ can be directly replaced by the intensity
originating from the stellar core, $I_{\rm c}$, and Eq.  \eqref{eq:lforce} can
be solved by quadrature, for a given density and velocity structure. For
accreting high-mass stars (see Sect.~\ref{sec:intro}), that means for massive
objects in their late formation phases, however, the circumstellar discs are
optically thick, and at least two major problems arise:

Firstly, due to absorption and scattering processes, the incident intensity at
a considered point needs to be calculated by a global solution of the
radiation field, which is very time consuming in hydrodynamic
simulations. \cite{Kee15} developed an efficient method to delimit the
contribution by either calculating the absorption part alone (giving a lower
limit of the incident intensity), or assuming the disc to be optically thin
(giving an upper limit). A comparison between the irradiation obtained from
their method to the irradiation obtained from our 3D code (including
scattering of photons) will be presented in a forthcoming paper, and shall not
be discussed here.

In this paper, we only consider the second problem of optically thick
environments: Since the disc partly blocks the irradiation from the star, the
radiation field might become considerably reduced. However, it is this
(ionizing) radiation field, which mainly determines the ionization stages of
the considered plasma, and consequently might influence the line-strength
distribution function. To address this issue, we proceed as follows: First,
the radiation field for a specific hydrodynamic structure (see below) is
computed by our 3D code, and the resulting mean intensity is translated to a
corresponding radiation temperature (using $\Jnu=: W \cdot \Bnu(\Trad)$ with
dilution factor $W$; the derived radiation temperature would correspond to
\Teff\, if the star had an optically thin, spherically symmetric
atmosphere). Corresponding line-strength parameters could again be obtained
from \citet{Puls00}. In regions where the local radiation temperature is
similar to the effective one, one can safely assume that the parameters of the
line-strength distribution remain at their original input values, and indeed
can be used to calculate the line force throughout all following
hydro-timesteps.  If, on the other hand, \Trad\ differed significantly from
\Teff, this would imply that these parameters would need to be consistently
adapted within the hydrodynamical evolution.

For our analysis, we used a wind+Keplerian-disc model similar to the initial
conditions for the accreting O7-star system as considered by
\cite{Kee16}. This model describes such objects as already defined in the
introduction as accreting high-mass stars (see \citealt{Hosokawa2010},
\citealt{Kuiper2016}). The wind and stellar parameters are given in Table
\ref{tab:paramABLATION}, following \cite{Kee16}.
A radial optical depth of the disc, $\tau_{\rm disc}=1400$, has been
adopted. We approximated the continuum by pure Thomson-scattering, $\epsc=0$,
to ensure frequency independence. This is a fair assumption for the 500-2000
\AA ~range, where the majority of line-driving happens
(\eg~\citealt{Puls00}). Of course, we would expect thermalization in the
disc's deeper layers. Due to the dominating $\rho^{-\alpha}$-dependence of the
line force (Eq. \ref{eq:lforce}) and the large densities inside the disc,
however, most of the wind-ablation occurs at the surface layers, and we do not
need to care about the details in the inner parts. This fact is even more
important, since it allows us to apply our 3D FVM method, although being aware
of the large errors of the continuum transfer for optically thick media. To
ensure that the transition region from the wind to the disc is not subject to
(larger) numerical uncertainties, we have performed a test calculation with
doubled grid resolution ($\nx^{\rm test}=\ny^{\rm test}=\nz^{\rm
  test}=265$). Although we found, as expected, differences in the inner part
of the disc, our results for the outer part and the wind region are (almost)
identical. We can, therefore, safely assume that the obtained solutions, at
least in the aforementioned regions, are only mildly affected by numerical
artefacts.

The density structure and radiation temperature (the latter computed by our
code) are shown in Fig.~\ref{fig:density_abl} and \ref{fig:trad_abl},
respectively. The radiation temperature in the wind (here: along the z-axis)
exceeds the effective temperature by a factor of roughly 1.25. In order to
ensure that this is not a numerical effect, we have checked this issue by
calculating the same wind model, however applying an optically thin disc with
$\tau_{\rm disc} = 1.4 \cdot 10^{-3}$. For such a model, \Trad ~and \Teff
~turned out to be fairly identical. We thus conclude that the enhancement of
radiation temperature in our original model is due to additional irradiation
of the wind from the disc, by scattering off photons from the disc. Most
likely, this effect will induce latitudinal line-force components (also to be
addressed in a forthcoming paper).

Wind-ablation dominates in the transition region between wind and disc. We
define this region by calculating the decrease in line-force by a certain
factor, $f$, due to density effects alone, that means assuming the same
ionization stages and the same velocity structure. Such a reduction of
line-force or line-acceleration is easily cast into an enhancement of density
via Eq. \eqref{eq:lforce},
\beq
\dfrac{g_{\rm lines}^{\rm (disc)}(r,\Theta)}{g_{\rm lines}^{\rm (wind)}(r)} <
\frac{1}{f} \quad
\leftrightarrow \quad \Biggl(\dfrac{\rho_{\rm disc}(r,\Theta)}{\rho_{\rm
    wind}(r)}\Biggr)^\alpha > f \,, 
\eeq
where the radius-dependent quantities from the wind can be measured along the
z-axis. In this picture, $f$ should be chosen such that the corresponding
decrease in line-force represents the border from the wind region to the
region where the line-force is negligible (\ie~inside the disc). As a first
guess, we adopted $f = 10$, and display the corresponding density contour in
Fig.~\ref{fig:trad_abl}.  Since a factor $f = 10$ seems to be somewhat
artificial, we additionally display the density contour corresponding to $f =
100$.

From our simulations, we then find that both contours are located within a
range of \Trad\ between roughly 31 and 33~kK, which is of the same order as
the effective temperature, \Teff\ = 36 kK.  We thus conclude that the
ionization stages at the disc surface are not changing too much, when compared
to the ionization stages in the wind, and that the line-strength
parameterization of the wind can also be used to calculate the line-force at
the surface of such optically thick circumstellar discs. Due to significant
scattering of photons off the disc, a multi-D description of the radiative
transfer might need to be incorporated into the hydrodynamic simulations, to
account for all force-components.
\section{Dynamical magnetospheres: HD191612} 
\label{sec:dm} 
As a first application to line transitions, we modelled UV resonance lines in
dynamical magnetospheres, that means atmospheres which form in slowly rotating
magnetic OB-stars (in contrast to the so-called centrifugal magnetospheres,
which form in fast rotating magnetic OB-stars). As a prototypical case, we
considered the Of?p star \hdname{HD191612}, which has a negligible equatorial
rotation speed of $\vrot \approx 1.4\, \kms$ (\citealt{Howarth07},
\citealt{Sundqvist12c}). \cite{Marcolino13} already calculated corresponding
resonance lines for this star, by extending the 3D formal solver developed by
\cite{Sundqvist12c} to a `3D Sobolev with exact integration' method (SEI,
\citealt{Lamersetal87}), and applying this method to a set of 100
two-dimensional MHD-simulation snapshots, equidistantly distributed over the
azimuth-angle to enable a 3D description of the atmosphere.  At least for the
\Ha ~line (where the source function is taken from prototypical 1D
NLTE-calculations), such a patching-technique produces quite similar results
as full 3D MHD simulations (see \citealt{udDoula2013}).  In
Sect.~\ref{subsec:mhd}, we use the same simulations as a benchmark for our 3D
code, and compare the obtained line profiles to those from
\cite{Marcolino13}. In Sect.~\ref{subsec:adm}, we calculate analogous line
profiles for the ADM model developed by \cite{Owocki16}, to investigate in how
far their simplified description of the magnetosphere can be used as a
reasonable substitute for elaborate MHD simulations. We already note here,
that such a simplified approach would be favourable to MHD simulations,
because it provides (within the applied approximations) an average, steady
state solution for the magnetospheric structure, and avoids time-consuming
hydrodynamic simulations.
\subsection{MHD models}
\label{subsec:mhd}
\begin{figure}[t]
\resizebox{\hsize}{!}{
   \begin{minipage}{\hsize}
      \resizebox{\hsize}{!}{\includegraphics{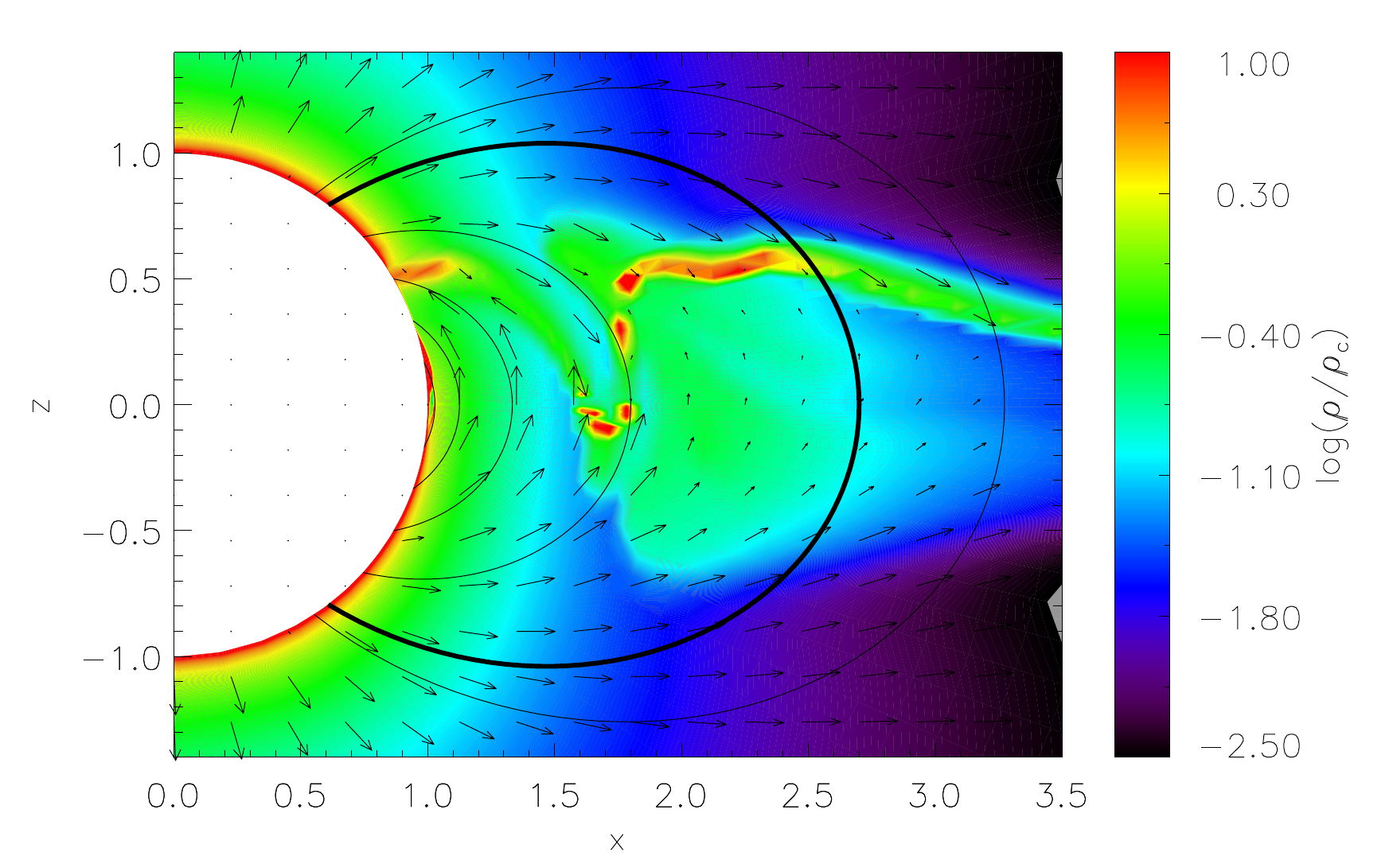}}
   \end{minipage}
}
\\
\resizebox{\hsize}{!}{
   \begin{minipage}{\hsize}
      \resizebox{\hsize}{!}{\includegraphics{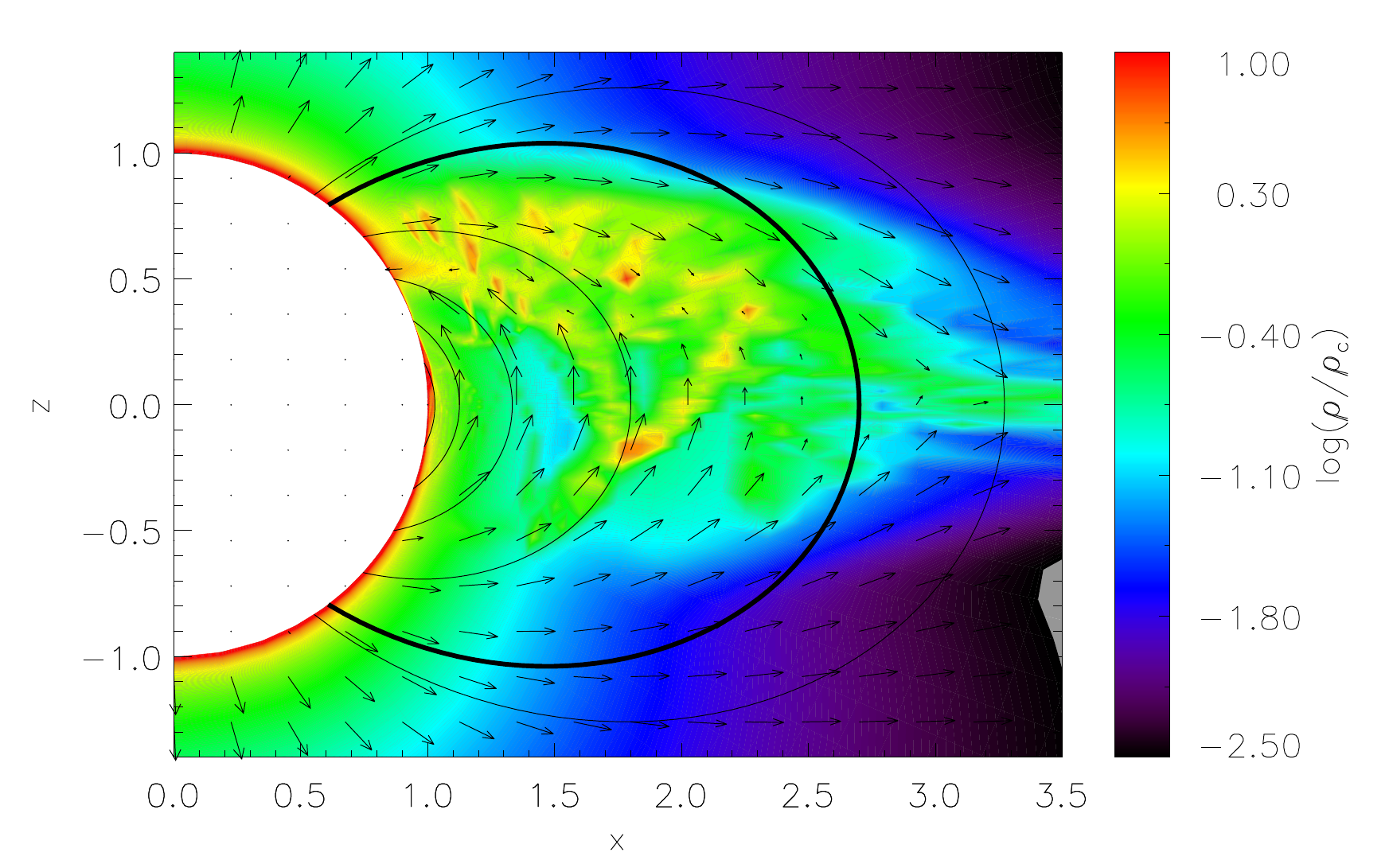}}
   \end{minipage}
}
%
\caption{Upper panel: density structure for an example snapshot from the MHD
  simulations for \hdname{HD191612}, as performed by
  \cite{Sundqvist12c}. Lower panel: azimuthal average of the MHD
  simulations. In both figures, the density has been normalized by a typical
  downflow density, $\rho_{\rm c}:=\mdotz/\bigl(4\pi\Rstar^2\vesc\bigr)$, with
  \mdotz ~from Table \ref{tab:hd191612}, and $\vesc \approx 800 \,\kms$ the
  photospheric escape velocity. The velocity field is displayed by arrows,
  with the length of the velocity vectors limited to 0.5 \vesc. We
  additionally show the dipole magnetic field of the ADM models used in
  Sect. \ref{subsec:adm} (solid lines, and thick solid line for $\ralf = 2.7
  \, \Rstar$). The corresponding magnetic axis is aligned with the z-axis. The
  grey colour corresponds to densities outside the range indicated on the
  right.}
\label{fig:density_mhd}
\end{figure}
\begin{figure}
\resizebox{\hsize}{!}{
   \begin{minipage}{\hsize}
      \resizebox{\hsize}{!}{\includegraphics{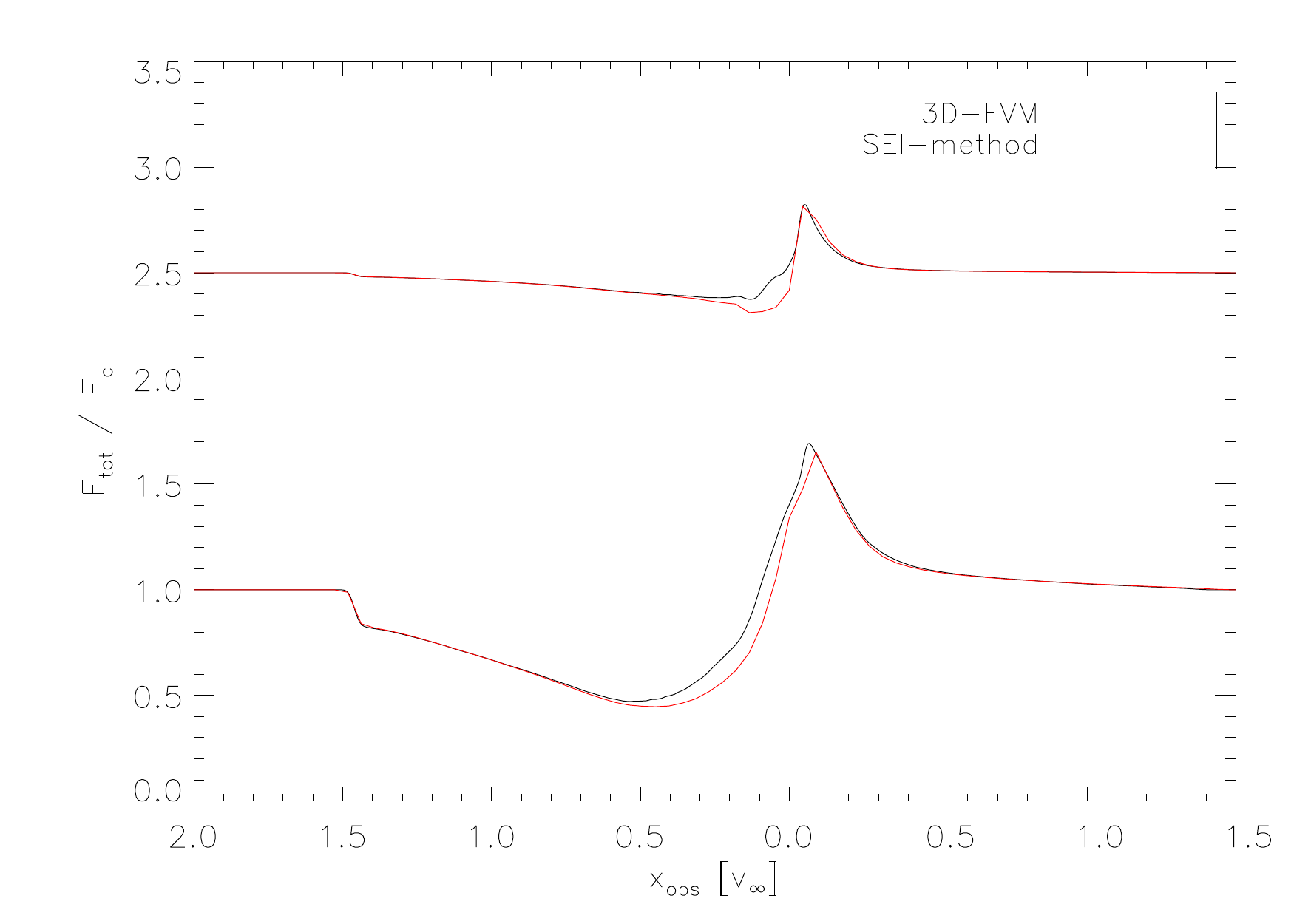}}
   \end{minipage}
}
\\
\resizebox{\hsize}{!}{
   \begin{minipage}{\hsize}
      \resizebox{\hsize}{!}{\includegraphics{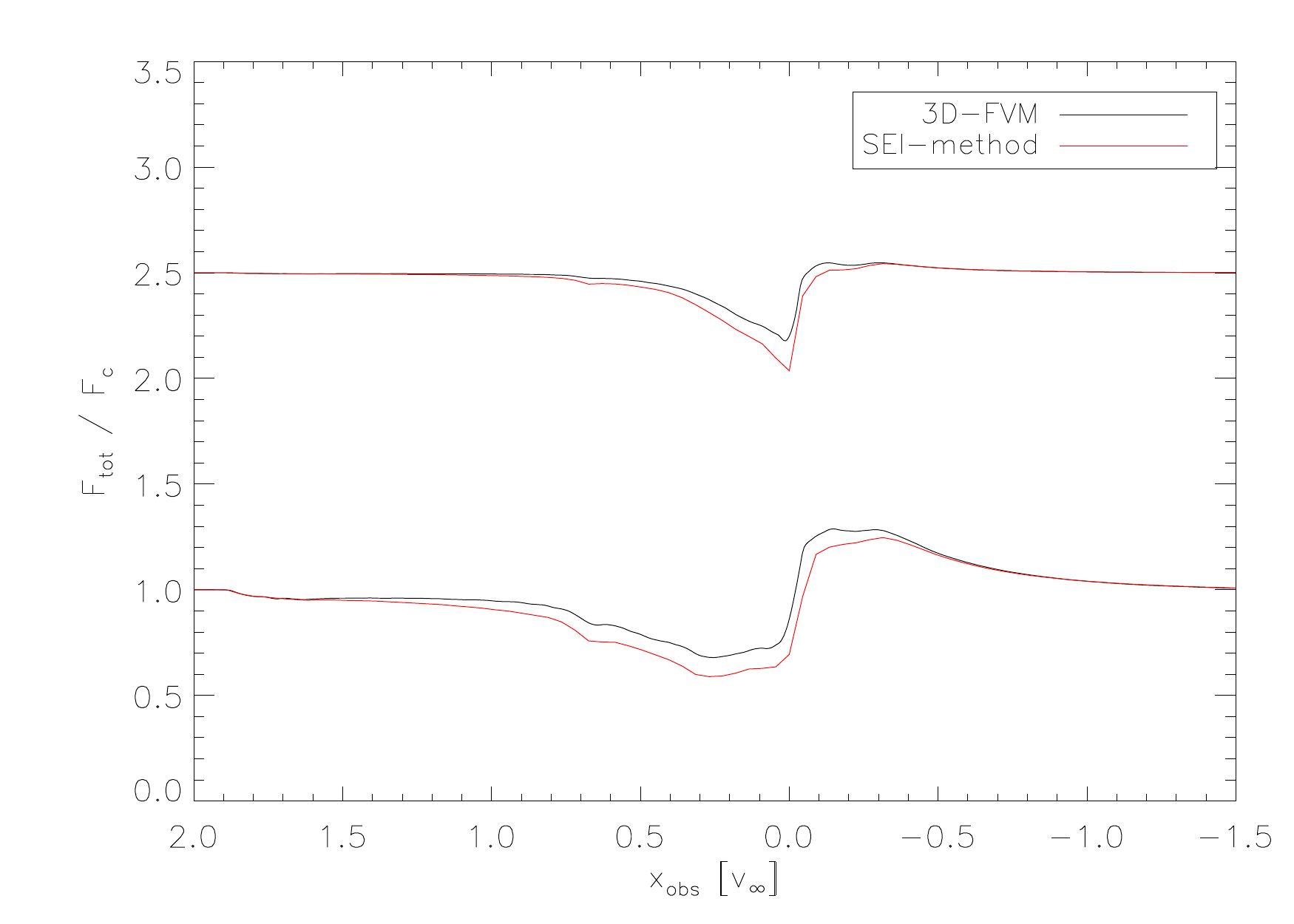}}
   \end{minipage}
}
%
\caption{UV resonance-line profiles for the MHD models, as obtained from our
  3D-code (black) and from the SEI method by \cite{Marcolino13} (red). Two
  different line-strength parameters, \kapline\, = ~0.1 and ~1.0, have been
  used. For convenience, the line profiles for \kapline\, = ~0.1 have been
  shifted vertically by 1.5. The upper and lower panels show the synthetic
  line profiles for pole-on and equator-on observers, respectively. The
  abscissa has been scaled to \vinf\, = 2700 ~\kms, the `observed' 1D value
  applied by \cite{Marcolino13}.}
\label{fig:profile_mhd}
\end{figure}
To understand the behaviour of the line profiles presented below, we first
explain the basic characteristics of (non-rotating) magnetic winds.  For a
more detailed discussion, we refer the reader to the seminal work by
\cite{udDoula02} and \cite{udDoula08}. These authors introduced a magnetic
dipole field as an initial condition, and evolved the (initially spherical)
stellar wind according to the MHD equations. Within ideal MHD\footnote{Ideal
  MHD is a fair approximation in hot star winds, due to the high
  conductivity.}, the material follows the (closed) magnetic-field lines in
regions where the magnetic energy exceeds the kinetic energy of the wind
(close to the star), whereas, in the opposite case, the field lines follow the
(almost radial) mass flow (far from the star). The border of both regions can
be roughly described by the Alfv\'en-Radius, $\ralf\approx\etastar^{1/4}$,
with wind-confinement parameter $\etastar := \bigl(\Bpole^2
\Rstar^2\bigr)/\bigl(4 \mdotz \vinf\bigr)$, and \Bpole ~the polar
magnetic-field strength evaluated at the stellar radius (see
\citealt{udDoula02}).

Within closed-field regions, material originating from opposite footpoints
shocks (and accumulates) in the equatorial plane. Due to the
$1/\rho^\alpha$-dependence of the line-force (see Eq. \ref{eq:lforce}), the
net-force becomes dominated by gravity, and produces an inflow along the
magnetic field lines in a `snake-like' pattern.

In the open-field regions, the presence of the magnetic field, together with a
frozen-in mass flow, results in a density decrease when compared with
spherically symmetric models, due to the faster-than-radial expansion of the
flow-tube area (see Figure 7 in \citealt{udDoula02}). Consequently, the
line-force becomes increased, resulting in higher terminal velocities than in
1D non-magnetic models. A single snapshot and an azimuthal average of the
applied MHD simulations are shown in Fig. \ref{fig:density_mhd}.

Based on such MHD simulations, \cite{Sundqvist12c} calculated corresponding
\Ha-line profiles, while \cite{Marcolino13} investigated the UV resonance line
formation. To remain consistent with the calculations by \cite{Marcolino13},
we apply \epsl\, = 0, and use their description of the line-strength
parameter, \kapline, originally introduced by \cite{Hamann80}. \kapline ~is
related to the line-strength parameter from Eq. \eqref{eq:kline} by
\beq
\label{eq:kapline}
\kapline=\dfrac{1}{4\pi m_{\rm p}} \dfrac{\mdot}{\Rstar\vinf^2} \dfrac{1 +
  \hei \YHe}{1+4\YHe}\sigmae \vthfid \kline \,,
\eeq
with $\hei=2$ and $\YHe=N_{\rm He}/N_{\rm H} = 0.1$, the number of free
electrons per helium atom, and helium abundance by number, respectively.

Although we use a micro-turbulent velocity of \vturb\, = 100 ~\kms ~for the
determination of the source function, we calculate the final line profile
(somewhat inconsistently) for \vturb\, = 50 ~\kms, as done by
\cite{Marcolino13}. The line profiles obtained from our 3D code and the SEI
line profiles from \cite{Marcolino13} are displayed in
Fig. \ref{fig:profile_mhd}, for two different line-strength parameters,
\kapline\, = 0.1 and 1.0, respectively.

The agreement between the two methods is excellent. The minor differences in
the emission part are related to two effects: Firstly, the methods for
determining the source functions (SEI implying very narrow resonance lines
vs. FVM accounting for much broader ones, due to \vturb\, = 100 ~\kms) are
quite different, and a certain deviation must be present. Secondly, the
(general) overestimation of the scattering integrals and thus source functions
due to the FVM might play a role as well. Also the absorption parts of the
line profiles observed equator-on (lower panel of Fig.~\ref{fig:profile_mhd})
are not perfectly matched. This (small) effect is most likely simply due to
different formulations of the numerical solvers.

A comparison of these line profiles with those from corresponding spherically
symmetric models has already been performed by \cite{Marcolino13}, and we
summarize only the most important characteristics: (1) The absorption trough
for pole-on and equator-on observers extends beyond the 1D terminal velocity,
as expected from the MHD atmospheric structure. We note that such a large
extension has not been observed for \hdname{HD191612}.  (2) The emission for
equator-on observers is reduced (compared to the 1D case), due to the lower
densities in the emission plane (\eg~the polar plane for line-centre
frequencies with $x_{\rm OBS}=0$).  (3) The particular form of the line
profiles is determined by the different mapping of projected velocities for
different observer directions.

Given the overall agreement of the two different methods, we conclude that the
SEI and our 3D FVM solutions are consistent. Additionally, we are highly
confident that the line formation is described correctly (at least
qualitatively), since both methods are completely independent. Under this
assumption we are able to study the UV line formation within the ADM model.
\subsection{ADM models}
\label{subsec:adm}
\begin{table}
\tabcolsep0.38mm
\begin{center}
\caption{Stellar and wind parameters of \hdname{HD191612} (left panel). \Teff,
  \logg, \Rstar, \mdot, \vinf ~have been derived by \cite{Howarth07}, and
  \Bpole ~is adopted from \cite{Wade11}. For the ADM model, we adapted the
  mass-loss rate and terminal velocity at the pole (right panel), to be
  consistent with the MHD simulations from Fig.~\ref{fig:density_mhd}.}
\label{tab:hd191612}
\begin{tabular}{c|c|c|c}
\multicolumn{1}{l}{\Teff [kK]} & \multicolumn{1}{l||}{35} &
\multicolumn{1}{l}{\hspace{1ex}} & \multicolumn{1}{l}{} \\
\multicolumn{1}{l}{\logg} & \multicolumn{1}{l||}{3.5} &
\multicolumn{1}{l}{\hspace{1ex}} & \multicolumn{1}{l}{} \\
\multicolumn{1}{l}{\Rstar [\rsun]} & \multicolumn{1}{l||}{14.5} &
\multicolumn{1}{l}{\hspace{1ex}} & \multicolumn{1}{l}{} \\
\multicolumn{1}{l}{\vinf [\kms]} & \multicolumn{1}{l||}{2700} &
\multicolumn{1}{l}{\hspace{1ex}$\vinf^{\rm (pole)}$~[\kms]}
  & \multicolumn{1}{l}{3963} \\
\multicolumn{1}{l}{\mdot [\msunyr]} & \multicolumn{1}{l||}{$1.6 \cdot
  10^{-6}$} & \multicolumn{1}{l}{\hspace{1ex}\mdotz [\msunyr]} &
\multicolumn{1}{l}{$1.1 \cdot 10^{-6}$} \\ 
\multicolumn{1}{l}{\Bpole [G]} & \multicolumn{1}{l||}{2450} &
\multicolumn{1}{l}{\hspace{1ex}} & \multicolumn{1}{l}{} \\
\multicolumn{1}{l}{} &
\multicolumn{1}{l||}{} & \multicolumn{1}{l}{\hspace{1ex} $\Rightarrow$ \ralf [\Rstar]} &
\multicolumn{1}{l}{2.7} 
\end{tabular}
\end{center}
\end{table}
\begin{figure}
\resizebox{\hsize}{!}{
   \begin{minipage}{\hsize}
      \resizebox{\hsize}{!}{\includegraphics{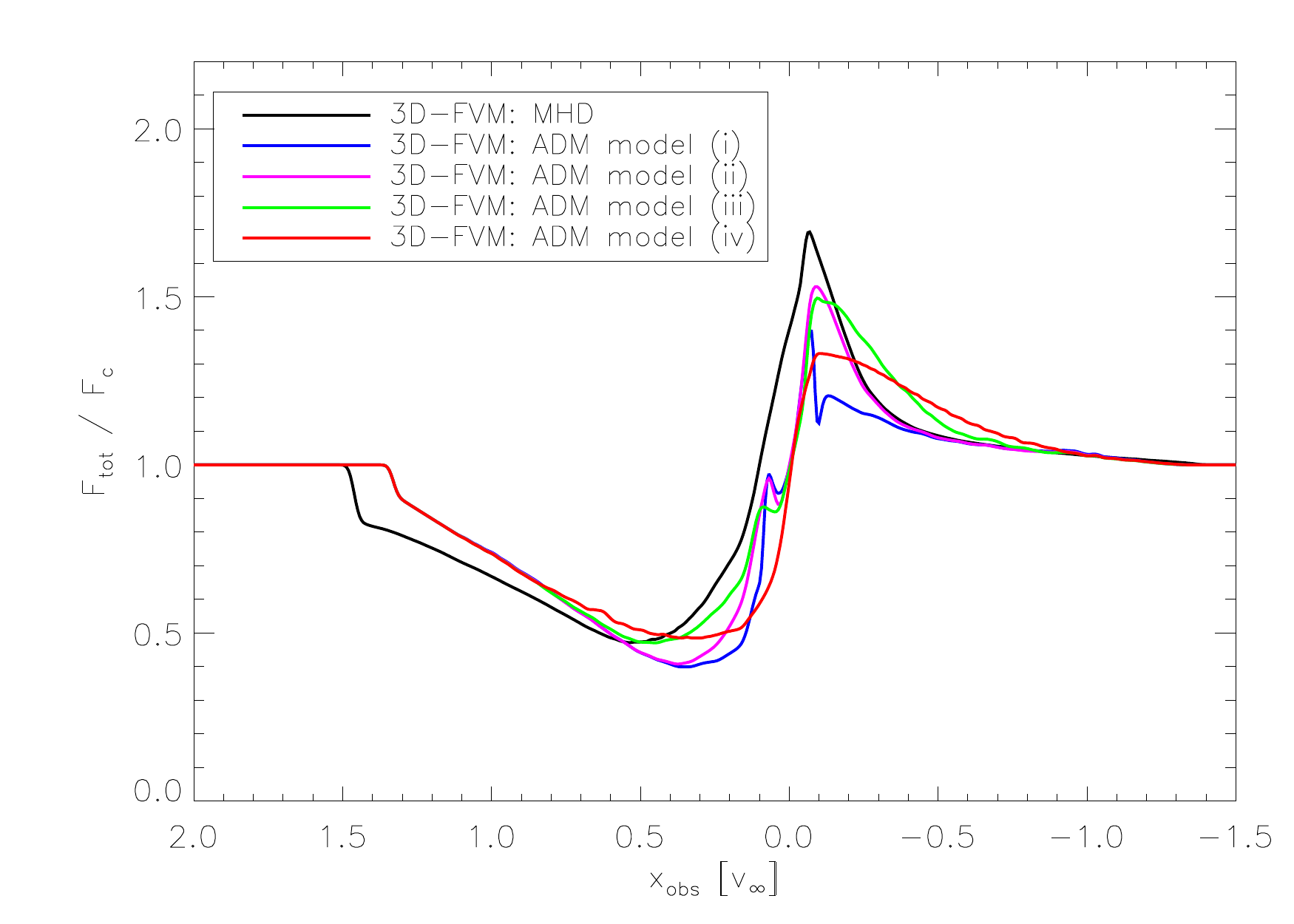}}
   \end{minipage}
}
\\
\resizebox{\hsize}{!}{
   \begin{minipage}{\hsize}
      \resizebox{\hsize}{!}{\includegraphics{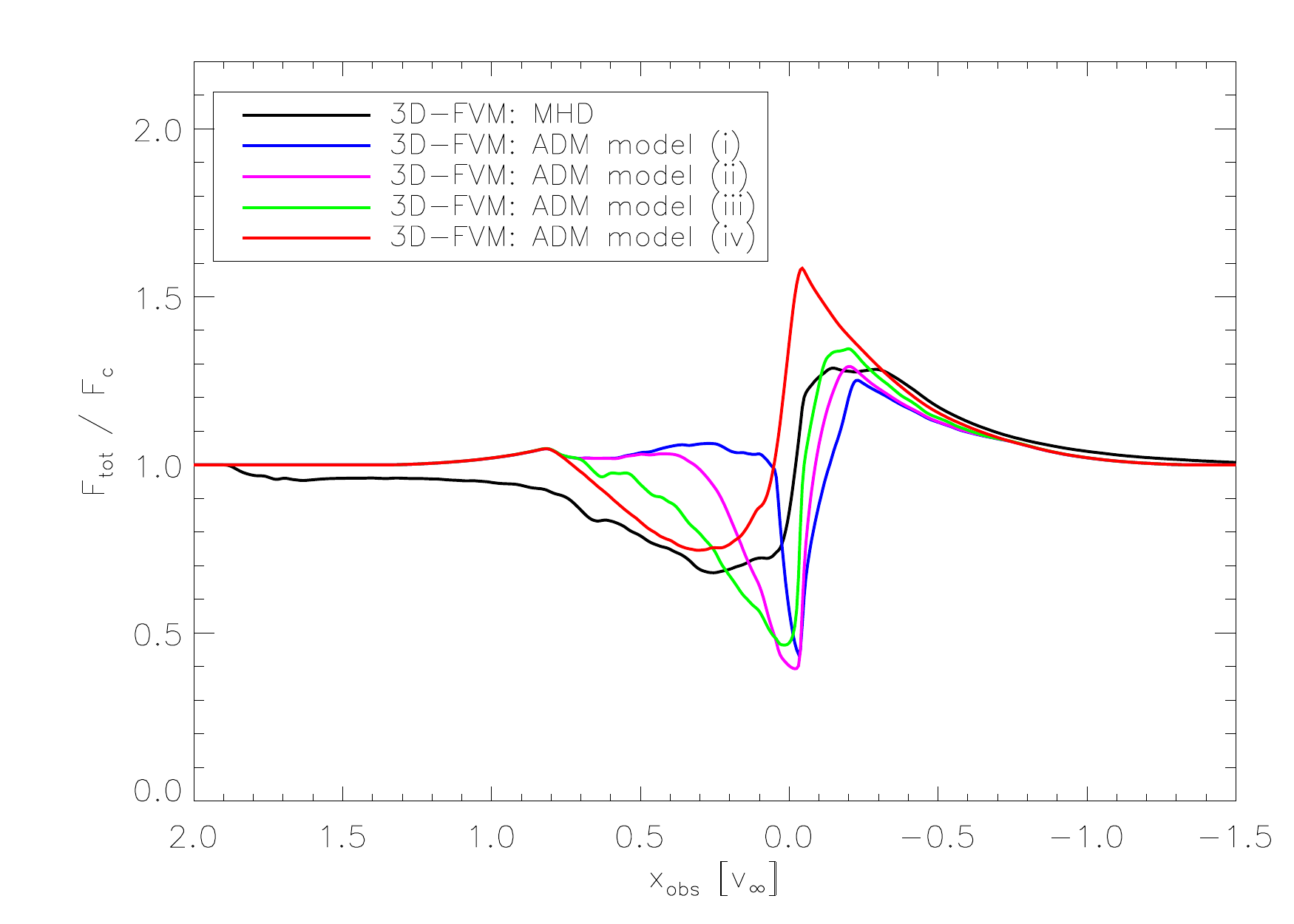}}
   \end{minipage}
}
%
\caption{UV resonance-line profiles obtained from our 3D code, for the MHD
  simulation (black, as Fig. \ref{fig:profile_mhd}), and the different ADM
  models, (i) to (iv) (see text). The line-strength parameter has been set to
  \kapline\ = 1. The upper and lower panels show the synthetic profiles for
  pole-on and equator-on view, respectively. As in Fig.~\ref{fig:profile_mhd},
  the abscissa has been scaled to \vinf\, = 2700 ~\kms, the `observed' 1D
  value applied by \cite{Marcolino13}.}
\label{fig:profile_adm}
\end{figure}
\cite{Owocki16} developed an analytic description of dynamical magnetospheres,
in order to set a framework similar to the $\beta$-velocity-field prescription
for spherically symmetric winds. This ADM formalism provides a
time-independent, steady-state solution for dynamical magnetospheres, which is
comparable to the average of several MHD-simulation snapshots, and has been
corroborated by a comparison of synthetic \Ha ~lines with observations. The
formation of resonance lines within the ADM framework, however, has not been
analysed yet, and is the focus of this section. For that purpose, we aim at
modelling the MHD atmospheric structure from above with the ADM method, and
compare the resulting line profiles.

Within the ADM method, \cite{Owocki16} divide the atmosphere into two major
zones. The border between both regions is given by the condition
$\rapex=\ralf$, where the apex-radius, $\rapex$, is defined as the distance
between the origin and the intersection of magnetic equator and closed dipole
magnetic-field line attached to a considered point (see left panel of
Fig. \ref{fig:profiles_density_adm} for clarification). In the following, we
call these two regions the `closed-field region' ($\rapex < \ralf$) and the
`outer wind' ($\rapex > \ralf$). The closed-field region consists of three
different components:
\begin{itemize}
\item wind-upflow component: The magnetic loops are fed with material ejected
  from the stellar surface. The matter flow follows the dipole magnetic-field
  lines, with absolute velocities calculated from a $\beta$-velocity law,
  using $\beta=1$.
\item post-shock component: The collision of outflows following the B-field
  lines from opposite foot points leads to a shock at the magnetic equator,
  resulting in a hot and dense post-shock region. The extent of this region is
  controlled by a (dimensionless) cooling parameter, \chiinf, where
  $1/\chiinf$ describes the efficiency of radiative cooling by X-ray emission
  (see \citealt{udDoula14} for details). In test calculations, however, this
  component turned out to have only very small influence on the UV line
  formation. Thus, and to keep the model as simple as possible, we neglect the
  post-shock component in this work.
\item cooled-downflow component: As the post-shock gas cools, its density
  increases, and the line-force decreases. Thus, the cooled and compressed gas
  is pulled back onto the stellar surface by gravity, resulting in a downflow
  starting at the magnetic equator. The gas is accelerated from zero velocity
  along the B-field lines to the escape speed at the stellar surface.
\end{itemize}
For their \Ha ~analysis, \cite{Owocki16} only considered the cooled downflow,
because of the mostly larger densities of this component. Since the infall
occurs in episodic infall events, the closed-field region is actually highly
structured, and the authors found rather large clumping factors $\langle
\rho^2 \rangle / \langle \rho \rangle$ (of the order of several tens).  Under
the assumption of clumps that are optically thin then, this clumping factor
can be used to translate the actual (structured) density-distribution to the
mean opacities and emissivities of recombination lines ($\rho^2$-processes,
see \eg~\citealt{puls08b}). For the UV resonance line formation (linear in
$\rho$), micro-clumping has no direct impact on the mean opacities. Therefore,
and because of the different densities and velocities within the upflow and
downflow components, an explicit description of the structured medium is
required when considering UV resonance lines. As it is not a priori clear how
to treat the combination of the above mentioned components, we consider four
different approaches, and model the closed-field region by:
\begin{enumerate}
\item[(i)] Applying only the cooled-downflow component.
\item[(ii)] Introducing a statistical treatment, where the probabilities of
  using either the wind-upflow or the cooled downflow-component when
  calculating the radiative transfer are here defined as
   \beq
   \nonumber
   P_{\rm w} := \dfrac{\rho_{\rm w}}{\rho_{\rm w} + \rho_{\rm c}} \,, \qquad
   P_{\rm c} := \dfrac{\rho_{\rm c}}{\rho_{\rm w} + \rho_{\rm c}} = 1 - P_{\rm w} \,.
   \eeq
  This approach preferentially chooses the component with higher density and
  lower velocity\footnote{Both quantities are connected by the continuity
    equation.}, in other words that component with the larger timescale for
  the matter flow.
\item[(iii)] Introducing flux-tubes that alternating consist of the downflow
  and upflow component.
\item[(iv)] Applying only the wind-upflow component.
\end{enumerate}
The models are ordered such that the contribution of the wind-upflow component
is increasing from model (i) to (iv).

As a zeroth-order approximation, \cite{Owocki16} model the outer wind (at
$\rapex > \ralf$) by the wind-upflow component, that means by a flow following
closed magnetic-field lines even in that region. This is a fair assumption for
modelling the polar regions, since it accounts for the faster than radial
decline of the density (see also \citealt{OuD04}). Moreover, the magnetic
field lines are nearly radial in these regions, thus resulting in a nearly
radial outflow similar to the MHD simulations. On the other hand, the velocity
vectors near the equatorial regions are modelled with a large latitudinal
component, whereas they are radially directed within the (more realistic) MHD
simulations. Thus, a match of the ADM and MHD magnetospheric structure in the
equatorial region cannot be achieved within the standard formulation. With
respect to UV line formation, this is the major drawback of the ADM formalism,
and will influence the line formation (see below).

To set the base density, \cite{Owocki16} introduced the mass-loss rate of the
star if it had no magnetic field, \mdotz, which determines the loop-feeding
rate. With the input parameters from Table \ref{tab:hd191612}, the dynamical
magnetosphere can be modelled according to the recipe from
\cite{Owocki16}\footnote{We have increased the wind-upflow and cooled-downflow
  densities by a factor of two, which is missing in their original
  equations.}.  We used the values of \mdotz ~and $\vinf^{\rm (pole)}$ ~in
Table \ref{tab:hd191612}, right panel, to adapt the ADM model to the MHD
simulations. For our model parameters, the Alfv\'en-Radius, $\ralf = 2.7
~\Rstar$, has been calculated from the mass-loss rate, terminal velocity and
magnetic-field strength. We stress that the adopted mass-loss rate is not
necessarily the `true' one, nor the mass-loss rate the star would have if no
magnetic field was present. For the applied ADM models, the resulting density
stratification, magnetic-field lines and velocity vectors in the xz-plane are
shown in the left panel of Fig.~\ref{fig:profiles_density_adm}. Here and in
the following, the equatorial plane coincides with the plane of the magnetic
equator, since we assume the magnetic axis to be aligned with the z-axis.

Compared to the MHD structure (see Fig.~\ref{fig:density_mhd}), the ADM
densities in the closed-field region are best represented by model (ii) and
(iii), that is by a combination of downflow and upflow component. In the outer
wind near the equator, the densities are underestimated due to the
aforementioned different description of the velocity field (see
Fig.~\ref{fig:profiles_density_adm}, left panel).

Once again we apply $\epsl=0$, and compare the corresponding line profiles
with line-strength parameter, \kapline\ = 1, for equator-on and pole-on
observers, with the line profiles obtained from the MHD simulations (see
Fig. \ref{fig:profile_adm}). For clarification,
Fig. \ref{fig:profiles_density_adm} additionally displays all line profiles
with their emission and absorption parts. The differences between the
profile-sets can be explained as follows.

\emph{For pole-on observers:} With increasing contribution from the upflow
component, the emission peak becomes broader, because the emitting volume at
intermediate to high velocities increases. Simultaneously, the cooled downflow
component with only low absolute velocities decreases, resulting in a lower
emission peak near the line centre.  An exception is model (i), for which the
emission peak at low frequency shifts is lowest, because the emission of the
upflow component near the star (with high densities and low velocities) is
missing. When compared to the line profiles from the MHD simulations, the best
result is obtained for model (ii), that is for the statistical description of
upflow and downflow component in the closed-field region. Even for this model,
however, we only get a relatively poor match with the MHD profiles. Since our
ADM models cover a large range of combinations of upflow and downflow
component (including the most extreme cases of a pure upflow and a pure
downflow), this finding suggests that the outer wind region is inadequately
modelled.  Indeed, the major differences of the line profiles can be explained
(at least qualitatively) by the different description of the outer wind: (1)
In the ADM models, the emission peaks close to line centre (\ie~at $x_{\rm
  OBS} \approx 0$, with corresponding resonance zones at projected velocities
$\vecown{n}\cdot\vecown{v} \approx 0$) are underestimated compared to the MHD
model, since within all ADM models also the mass flow in the outer wind is
adopted to follow closed magnetic field lines. This assumption becomes
problematic in equatorial regions, since here the ADM wind flows almost
perpendicular to the plane, whereas it is almost radial in the MHD
case. Consequently, when viewed pole on, only large projected velocities are
present in the corresponding area of the equatorial plane, where the latter
creates a large part of low-velocity emission in the MHD model. This part is
now missing in the ADM models, and the emitting area is almost limited to the
downflow component (with generally low projected velocities). Thus, the
profile becomes shallower than in the MHD case.  (2) Within the blue
absorption trough, the absorption column in front of the star is slightly
decreased, because the velocity vectors are once again following the magnetic
field lines, and do not perfectly match the MHD simulations. The differences
of the line profiles can thus be explained by the different description of the
outer wind region alone.

\emph{For equator-on observers}, the emission peak of the ADM models becomes
stronger and shifted to the blue side with increasing contribution of the
upflow component. Additionally, the absorption part on the blue side
increases, while it decreases on the red side. This behaviour is readily
explained: As the upflow contribution in front of the star (with projected
velocities directed towards the observer, thus affecting the blue side of the
profile) grows, the downwind contribution (with projected velocities directed
away from the observer, and affecting the red side of the profile) is
diminished. Consequently, the absorption in front of the star increases on the
blue side, and decreases on the red one. Again, when compared to the MHD
model, none of the obtained profiles provides a good agreement. While model
(iv) reproduces the absorption part on the blue side relatively well, the
red-sided absorption part is highly underestimated. On the other hand, a
better model for the red-sided absorption (\eg~model iii) underestimates the
absorption part on the blue side. In fact, it is not possible to
simultaneously model the blue and red absorption by only tuning the
composition of the closed-field region, suggesting that (at least) the outer
wind needs to be treated differently. For instance, assuming a radial outflow
in the equatorial plane of the outer wind region would increase the blue-sided
absorption (and emission), while preserving the rather good behaviour of model
(iii) on the red side.

Taking all this evidence together, we conclude that the (present) ADM model
needs to be improved for the modelling of UV resonance lines, at least in the
outer wind. Such a re-formulation then needs to include a consistent
description of the actual velocity and density stratification, accounting for
the delicate interplay between B-field and wind.
\section{Summary and conclusions} 
\label{conclusions} 
In this paper, we introduced a newly developed 3D FVM code, which solves the
equation of radiative transfer for continuum- and line-scattering problems
(the latter approximated by a two-level-atom in the present version). An
observer's-frame formulation allows us to consider arbitrary velocity fields
(and density structures).

Within the ALI scheme, the code iterates the source functions to convergence,
using a non-local approximate $\Lambda$-operator, and extrapolating subsequent
iterates by the Ng-formalism. For the most challenging problems of optically
thick, scattering dominated atmospheres, we obtained a satisfying convergence
behaviour, with relative corrections between subsequent iterates of less than
$10^{-3}$ within 20 iteration steps. Due to this convergence behaviour, we
were able to analyse the performance of the 3D FVM for such optically thick,
scattering dominated atmospheres.

A comparison of spherically symmetric problems calculated with our 3D code and
an accurate 1D solver shows that the FVM requires a relatively high spatial
resolution. Continuum errors are of the order of $\lesssim 20 \%$ for
marginally optically thick atmospheres, that is for typical O-star conditions,
however increase rapidly for larger optical depths, due to the first-order
scheme. These errors can only be reduced by applying either a higher grid
resolution, or when using more accurate solution schemes (\eg~the SC method
with appropriate interpolations). To analyse rapidly expanding stellar winds
with large continuum optical depths, the development of a 3D SC method is
planned for the future in our group.

The line transfer, on the other hand, performs much better, with relative
errors less than $25 \%$ even for strong lines. The resulting profiles as
obtained from a postprocessing LC method are reasonably accurate.

Due to significant numerical diffusion, intrinsic to the FVM, we found a
minimum error (for optically thin continua and weak lines) of roughly
10\%. Additionally, any symmetry of a considered problem is broken, due to the
distinct behaviour of numerical diffusion for different ray directions and in
different regions of the atmosphere.  Numerical diffusion errors, however,
could be minimized by increasing the grid resolution in the outer parts of the
atmosphere.  Accounting for the sound reproduction of line profiles for
spherically symmetric models, we conclude that our code is ready to be used
also for arbitrary 3D atmospheric structures, at least if the continuum
displays an only moderate optical depth.

As a first application to continuum-scattering problems, we estimated the
radiation temperatures of wind-ablation models, focusing on the transition
region between a line-driven wind and the optically thick circumstellar disc
(as present during the late phases of massive star formation in accreting high
mass stars). We found a reduction of radiation temperatures by only few
percent, which indicates that the ionization stages in this region are
(almost) the same as in the wind.  Thus, a line-distribution formalism with
the same set of line-strength parameters as used in the wind can be applied to
obtain the line acceleration that finally ablates the disc.  Because of the
numerical inaccuracies of the FVM, our findings must be taken with caution,
and possibly rechecked with more elaborate methods or a much higher grid
resolution. To analyse the complete evolution of optically thick circumstellar
discs, the impact of continuum scattering on latitudinal forces still needs to
be investigated, and is left to future studies.

As a benchmark for our code regarding the line transfer in non-spherical
models, we considered the same MHD simulations of dynamical magnetospheres as
used by \cite{Marcolino13}, and compared the resulting UV resonance-line
profiles to those obtained from their 3D-SEI analysis. The profiles as viewed
both polar-on and equator-on are in excellent agreement, indicating that our
3D FVM performs well also for such atmospheric models. We additionally applied
the analytic dynamical magnetosphere framework (ADM, \citealt{Owocki16}), and
modelled the corresponding atmospheric structure by adopting four different
descriptions of the closed-field regions. A comparison between the obtained
line profiles and those for the MHD simulations from above showed significant
differences. These were explained by the (somewhat insufficient) description
of the outer wind region within the (present) ADM formulation, primarily in
the equatorial plane. An improvement of the underlying assumptions is planned
for future work.
\begin{acknowledgements}
We thank our anonymous referee for valuable comments and suggestions. Many
thanks to V. Petit and S. Owocki for fruitful discussions about the ADM.  LH
gratefully acknowledges support from the German Research Foundation, DFG,
under grant PU 117/9-1. NDK acknowledges support from the German DFG under
grant KU 2849/3-1, which funds the Emmy Noether research group on ``Accretion
Flows and Feedback in Realistic Models of Massive Star Formation''. JOS
acknowledges funding from the European Union’s Horizon 2020 research and
innovation programme under the Marie Sklodowska-Curie grant agreement
no. 656725.
\end{acknowledgements}

\bibliographystyle{aa}
\bibliography{bib_levin}

\begin{thebibliography}{80}
\expandafter\ifx\csname natexlab\endcsname\relax\def\natexlab#1{#1}\fi

\bibitem[{{Abbott} {et~al.}(2016){Abbott}, {Abbott}, {Abbott}, {Abernathy},
  {Acernese}, {Ackley}, {Adams}, {Adams}, {Addesso}, {Adhikari}, \&
  et~al.}]{Abbott16}
{Abbott}, B.~P., {Abbott}, R., {Abbott}, T.~D., {et~al.} 2016, Physical Review
  Letters, 116, 061102

\bibitem[{{Adam}(1990)}]{Adam90}
{Adam}, J. 1990, \aap, 240, 541

\bibitem[{{Alecian} {et~al.}(2013){Alecian}, {Wade}, {Catala}, {Grunhut},
  {Landstreet}, {Bagnulo}, {B{\"o}hm}, {Folsom}, {Marsden}, \&
  {Waite}}]{Alecian13}
{Alecian}, E., {Wade}, G.~A., {Catala}, C., {et~al.} 2013, \mnras, 429, 1001

\bibitem[{{Amarsi} {et~al.}(2016){Amarsi}, {Asplund}, {Collet}, \&
  {Leenaarts}}]{Amarsi16}
{Amarsi}, A.~M., {Asplund}, M., {Collet}, R., \& {Leenaarts}, J. 2016, \mnras,
  455, 3735

\bibitem[{{Belczynski} {et~al.}(2016){Belczynski}, {Holz}, {Bulik}, \&
  {O'Shaughnessy}}]{Belczynski16}
{Belczynski}, K., {Holz}, D.~E., {Bulik}, T., \& {O'Shaughnessy}, R. 2016,
  \nat, 534, 512

\bibitem[{{Busche} \& {Hillier}(2005)}]{Busche2005}
{Busche}, J.~R. \& {Hillier}, D.~J. 2005, \aj, 129, 454

\bibitem[{{Cannon}(1973)}]{Cannon73}
{Cannon}, C.~J. 1973, \apj, 185, 621

\bibitem[{{Castor} {et~al.}(1975){Castor}, {Abbott}, \& {Klein}}]{CAK}
{Castor}, J.~I., {Abbott}, D.~C., \& {Klein}, R.~I. 1975, \apj, 195, 157

\bibitem[{{Cherepashchuk}(1976)}]{Cherepa1976}
{Cherepashchuk}, A.~M. 1976, Soviet Astronomy Letters, 2, 138

\bibitem[{{Cranmer} \& {Owocki}(1996)}]{CO96}
{Cranmer}, S.~R. \& {Owocki}, S.~P. 1996, \apj, 462, 469

\bibitem[{{de Mink} {et~al.}(2013){de Mink}, {Langer}, {Izzard}, {Sana}, \& {de
  Koter}}]{deMink13}
{de Mink}, S.~E., {Langer}, N., {Izzard}, R.~G., {Sana}, H., \& {de Koter}, A.
  2013, \apj, 764, 166

\bibitem[{{Dufton} {et~al.}(2011){Dufton}, {Dunstall}, {Evans}, {Brott},
  {Cantiello}, {de Koter}, {de Mink}, {Fraser}, {H{\'e}nault-Brunet},
  {Howarth}, {Langer}, {Lennon}, {Markova}, {Sana}, \& {Taylor}}]{Dufton11}
{Dufton}, P.~L., {Dunstall}, P.~R., {Evans}, C.~J., {et~al.} 2011, \apjl, 743,
  L22

\bibitem[{{Fabiani Bendicho} {et~al.}(1997){Fabiani Bendicho}, {Trujillo
  Bueno}, \& {Auer}}]{FB97}
{Fabiani Bendicho}, P., {Trujillo Bueno}, J., \& {Auer}, L. 1997, \aap, 324,
  161

\bibitem[{{Gayley}(1995)}]{Gayley95}
{Gayley}, K.~G. 1995, \apj, 454, 410

\bibitem[{{Georgiev} {et~al.}(2006){Georgiev}, {Hillier}, \&
  {Zsarg{\'o}}}]{Georgiev06}
{Georgiev}, L.~N., {Hillier}, D.~J., \& {Zsarg{\'o}}, J. 2006, \aap, 458, 597

\bibitem[{{Gr{\"a}fener} {et~al.}(2002){Gr{\"a}fener}, {Koesterke}, \&
  {Hamann}}]{Graf02}
{Gr{\"a}fener}, G., {Koesterke}, L., \& {Hamann}, W.-R. 2002, \aap, 387, 244

\bibitem[{{Hamann}(1980)}]{Hamann80}
{Hamann}, W.-R. 1980, \aap, 84, 342

\bibitem[{{Hamann}(1981)}]{Hamann81a}
{Hamann}, W.-R. 1981, \aap, 93, 353

\bibitem[{{Hauschildt}(1992)}]{Haus92}
{Hauschildt}, P.~H. 1992, \jqsrt, 47, 433

\bibitem[{{Hauschildt} \& {Baron}(2006)}]{Haus06}
{Hauschildt}, P.~H. \& {Baron}, E. 2006, \aap, 451, 273

\bibitem[{{Hillier} \& {Miller}(1998)}]{hilliermiller98}
{Hillier}, D.~J. \& {Miller}, D.~L. 1998, \apj, 496, 407

\bibitem[{{Hosokawa} {et~al.}(2010){Hosokawa}, {Yorke}, \&
  {Omukai}}]{Hosokawa2010}
{Hosokawa}, T., {Yorke}, H.~W., \& {Omukai}, K. 2010, \apj, 721, 478

\bibitem[{{Howarth} {et~al.}(2007){Howarth}, {Walborn}, {Lennon}, {Puls},
  {Naz{\'e}}, {Annuk}, {Antokhin}, {Bohlender}, {Bond}, {Donati}, {Georgiev},
  {Gies}, {Harmer}, {Herrero}, {Kolka}, {McDavid}, {Morel}, {Negueruela},
  {Rauw}, \& {Reig}}]{Howarth07}
{Howarth}, I.~D., {Walborn}, N.~R., {Lennon}, D.~J., {et~al.} 2007, \mnras,
  381, 433

\bibitem[{{Ibgui} {et~al.}(2013{\natexlab{a}}){Ibgui}, {Hubeny}, {Lanz}, \&
  {Stehl{\'e}}}]{Ibgui13}
{Ibgui}, L., {Hubeny}, I., {Lanz}, T., \& {Stehl{\'e}}, C. 2013{\natexlab{a}},
  \aap, 549, A126

\bibitem[{{Ibgui} {et~al.}(2013{\natexlab{b}}){Ibgui}, {Hubeny}, {Lanz},
  {Stehl{\'e}}, {Gonz{\'a}lez}, \& {Chi{\`e}ze}}]{Ibgui13a}
{Ibgui}, L., {Hubeny}, I., {Lanz}, T., {et~al.} 2013{\natexlab{b}}, in
  Astronomical Society of the Pacific Conference Series, Vol. 474, Numerical
  Modeling of Space Plasma Flows (ASTRONUM2012), ed. N.~V. {Pogorelov},
  E.~{Audit}, \& G.~P. {Zank}, 66

\bibitem[{{Jones}(1973)}]{Jones73b}
{Jones}, H.~P. 1973, \apj, 185, 183

\bibitem[{{Jones} \& {Skumanich}(1973)}]{Jones73a}
{Jones}, H.~P. \& {Skumanich}, A. 1973, \apj, 185, 167

\bibitem[{{Kee}(2015)}]{Kee15}
{Kee}, N.~D. 2015, PhD thesis, University of Delaware

\bibitem[{{Kee} {et~al.}(2016){Kee}, {Owocki}, \& {Sundqvist}}]{Kee16}
{Kee}, N.~D., {Owocki}, S., \& {Sundqvist}, J.~O. 2016, \mnras, 458, 2323

\bibitem[{{Kuiper} {et~al.}(2016){Kuiper}, {Turner}, \& {Yorke}}]{Kuiper2016}
{Kuiper}, R., {Turner}, N.~J., \& {Yorke}, H.~W. 2016, \apj, 832, 40

\bibitem[{{Kunasz} \& {Auer}(1988)}]{Kunasz88}
{Kunasz}, P. \& {Auer}, L.~H. 1988, \jqsrt, 39, 67

\bibitem[{{Kunasz} \& {Olson}(1988)}]{Kunasz88b}
{Kunasz}, P.~B. \& {Olson}, G.~L. 1988, \jqsrt, 39, 1

\bibitem[{{Lamers} {et~al.}(1987){Lamers}, {Cerruti-Sola}, \&
  {Perinotto}}]{Lamersetal87}
{Lamers}, H.~J.~G.~L.~M., {Cerruti-Sola}, M., \& {Perinotto}, M. 1987, \apj,
  314, 726

\bibitem[{{Leenaarts} \& {Carlsson}(2009)}]{Leenaarts09}
{Leenaarts}, J. \& {Carlsson}, M. 2009, in Astronomical Society of the Pacific
  Conference Series, Vol. 415, The Second Hinode Science Meeting: Beyond
  Discovery-Toward Understanding, ed. B.~{Lites}, M.~{Cheung}, T.~{Magara},
  J.~{Mariska}, \& K.~{Reeves}, 87

\bibitem[{{Lobel} \& {Blomme}(2008)}]{Lobel08}
{Lobel}, A. \& {Blomme}, R. 2008, \apj, 678, 408

\bibitem[{{Lucy}(1983)}]{Lucy83}
{Lucy}, L.~B. 1983, \apj, 274, 372

\bibitem[{{Maeder}(1999)}]{MaederIV}
{Maeder}, A. 1999, \aap, 347, 185

\bibitem[{{Maeder} \& {Meynet}(2000)}]{MaederVI}
{Maeder}, A. \& {Meynet}, G. 2000, \aap, 361, 159

\bibitem[{{Marcolino} {et~al.}(2013){Marcolino}, {Bouret}, {Sundqvist},
  {Walborn}, {Fullerton}, {Howarth}, {Wade}, \& {ud-Doula}}]{Marcolino13}
{Marcolino}, W.~L.~F., {Bouret}, J.-C., {Sundqvist}, J.~O., {et~al.} 2013,
  \mnras, 431, 2253

\bibitem[{{Mihalas}(1978)}]{mihalasbook78}
{Mihalas}, D. 1978, {Stellar atmospheres (2nd edition)} (San Francisco:
  W.~H.~Freeman and Co., 1978)

\bibitem[{{Mullan}(1984)}]{Mullan84}
{Mullan}, D.~J. 1984, \apj, 283, 303

\bibitem[{{Ng}(1974)}]{ng74}
{Ng}, K.-C. 1974, \jcp, 61, 2680

\bibitem[{{Olson} {et~al.}(1986){Olson}, {Auer}, \& {Buchler}}]{OAB86}
{Olson}, G.~L., {Auer}, L.~H., \& {Buchler}, J.~R. 1986, \jqsrt, 35, 431

\bibitem[{{Olson} \& {Kunasz}(1987)}]{Olson87}
{Olson}, G.~L. \& {Kunasz}, P.~B. 1987, \jqsrt, 38, 325

\bibitem[{{Owocki} \& {ud-Doula}(2004)}]{OuD04}
{Owocki}, S.~P. \& {ud-Doula}, A. 2004, \apj, 600, 1004

\bibitem[{{Owocki} {et~al.}(2016){Owocki}, {ud-Doula}, {Sundqvist}, {Petit},
  {Cohen}, \& {Townsend}}]{Owocki16}
{Owocki}, S.~P., {ud-Doula}, A., {Sundqvist}, J.~O., {et~al.} 2016, \mnras,
  462, 3830

\bibitem[{{Patankar}(1980)}]{patankarbook80}
{Patankar}, S.~V. 1980, Numerical heat transfer and fluid flow, Series on
  Computational Methods in Mechanics and Thermal Science (Hemisphere Publishing
  Corporation, 1980)

\bibitem[{{Pauldrach} {et~al.}(2001){Pauldrach}, {Hoffmann}, \&
  {Lennon}}]{pauldrach01}
{Pauldrach}, A.~W.~A., {Hoffmann}, T.~L., \& {Lennon}, M. 2001, \aap, 375, 161

\bibitem[{{Petit} {et~al.}(2017){Petit}, {Keszthelyi}, {MacInnis}, {Cohen},
  {Townsend}, {Wade}, {Thomas}, {Owocki}, {Puls}, \& {ud-Doula}}]{Petit17}
{Petit}, V., {Keszthelyi}, Z., {MacInnis}, R., {et~al.} 2017, \mnras, 466, 1052

\bibitem[{{Petit} {et~al.}(2013){Petit}, {Owocki}, {Wade}, {Cohen},
  {Sundqvist}, {Gagn{\'e}}, {Ma{\'{\i}}z Apell{\'a}niz}, {Oksala}, {Bohlender},
  {Rivinius}, {Henrichs}, {Alecian}, {Townsend}, {ud-Doula}, \& {MiMeS
  Collaboration}}]{Petit13}
{Petit}, V., {Owocki}, S.~P., {Wade}, G.~A., {et~al.} 2013, \mnras, 429, 398

\bibitem[{{Pittard}(2009)}]{pittard09}
{Pittard}, J.~M. 2009, \mnras, 396, 1743

\bibitem[{{Prilutskii} \& {Usov}(1976)}]{Prilutskii1976}
{Prilutskii}, O.~F. \& {Usov}, V.~V. 1976, \sovast, 20, 2

\bibitem[{{Puls}(1991)}]{Puls91}
{Puls}, J. 1991, \aap, 248, 581

\bibitem[{{Puls} \& {Herrero}(1988)}]{PH88}
{Puls}, J. \& {Herrero}, A. 1988, \aap, 204, 219

\bibitem[{{Puls} {et~al.}(1993){Puls}, {Owocki}, \& {Fullerton}}]{POF93}
{Puls}, J., {Owocki}, S.~P., \& {Fullerton}, A.~W. 1993, \aap, 279, 457

\bibitem[{{Puls} {et~al.}(2000){Puls}, {Springmann}, \& {Lennon}}]{Puls00}
{Puls}, J., {Springmann}, U., \& {Lennon}, M. 2000, \aaps, 141, 23

\bibitem[{{Puls} {et~al.}(2005){Puls}, {Urbaneja}, {Venero}, {Repolust},
  {Springmann}, {Jokuthy}, \& {Mokiem}}]{Puls05}
{Puls}, J., {Urbaneja}, M.~A., {Venero}, R., {et~al.} 2005, \aap, 435, 669

\bibitem[{{Puls} {et~al.}(2008){Puls}, {Vink}, \& {Najarro}}]{puls08b}
{Puls}, J., {Vink}, J.~S., \& {Najarro}, F. 2008, \aapr, 16, 209

\bibitem[{{Ram{\'{\i}}rez-Agudelo} {et~al.}(2013){Ram{\'{\i}}rez-Agudelo},
  {Sim{\'o}n-D{\'{\i}}az}, {Sana}, {de Koter}, {Sab{\'{\i}}n-Sanjul{\'{\i}}an},
  {de Mink}, {Dufton}, {Gr{\"a}fener}, {Evans}, {Herrero}, {Langer}, {Lennon},
  {Ma{\'{\i}}z Apell{\'a}niz}, {Markova}, {Najarro}, {Puls}, {Taylor}, \&
  {Vink}}]{RamirezAgudelo13}
{Ram{\'{\i}}rez-Agudelo}, O.~H., {Sim{\'o}n-D{\'{\i}}az}, S., {Sana}, H.,
  {et~al.} 2013, \aap, 560, A29

\bibitem[{{Rivero Gonz{\'a}lez} {et~al.}(2012){Rivero Gonz{\'a}lez}, {Puls},
  {Najarro}, \& {Brott}}]{rivero12}
{Rivero Gonz{\'a}lez}, J.~G., {Puls}, J., {Najarro}, F., \& {Brott}, I. 2012,
  \aap, 537, A79

\bibitem[{{Sana} {et~al.}(2013){Sana}, {de Koter}, {de Mink}, {Dunstall},
  {Evans}, {H{\'e}nault-Brunet}, {Ma{\'{\i}}z Apell{\'a}niz},
  {Ram{\'{\i}}rez-Agudelo}, {Taylor}, {Walborn}, {Clark}, {Crowther},
  {Herrero}, {Gieles}, {Langer}, {Lennon}, \& {Vink}}]{Sana13}
{Sana}, H., {de Koter}, A., {de Mink}, S.~E., {et~al.} 2013, \aap, 550, A107

\bibitem[{{Sana} \& {Evans}(2011)}]{Sana11}
{Sana}, H. \& {Evans}, C.~J. 2011, in IAU Symposium, Vol. 272, Active OB Stars:
  Structure, Evolution, Mass Loss, and Critical Limits, ed. C.~{Neiner},
  G.~{Wade}, G.~{Meynet}, \& G.~{Peters}, 474--485

\bibitem[{{Sana} {et~al.}(2001){Sana}, {Rauw}, \& {Gosset}}]{Sana01}
{Sana}, H., {Rauw}, G., \& {Gosset}, E. 2001, \aap, 370, 121

\bibitem[{{Seelmann} {et~al.}(2010){Seelmann}, {Hauschildt}, \&
  {Baron}}]{Seelmann10}
{Seelmann}, A.~M., {Hauschildt}, P.~H., \& {Baron}, E. 2010, \aap, 522, A102

\bibitem[{{Stenholm} {et~al.}(1991){Stenholm}, {Stoerzer}, \&
  {Wehrse}}]{Stenholm91}
{Stenholm}, L.~G., {Stoerzer}, H., \& {Wehrse}, R. 1991, \jqsrt, 45, 47

\bibitem[{{Stevens} {et~al.}(1992){Stevens}, {Blondin}, \&
  {Pollock}}]{Stevens92}
{Stevens}, I.~R., {Blondin}, J.~M., \& {Pollock}, A.~M.~T. 1992, \apj, 386, 265

\bibitem[{{Sundqvist} {et~al.}(2012){Sundqvist}, {ud-Doula}, {Owocki},
  {Townsend}, {Howarth}, \& {Wade}}]{Sundqvist12c}
{Sundqvist}, J.~O., {ud-Doula}, A., {Owocki}, S.~P., {et~al.} 2012, \mnras,
  423, L21

\bibitem[{{Tessem}(2013)}]{tessem13}
{Tessem}, T. 2013, Master's thesis, University of Bergen

\bibitem[{{Townsend} \& {Owocki}(2005)}]{Townsend05a}
{Townsend}, R.~H.~D. \& {Owocki}, S.~P. 2005, \mnras, 357, 251

\bibitem[{{Trujillo Bueno} \& {Fabiani Bendicho}(1995)}]{Bueno1995}
{Trujillo Bueno}, J. \& {Fabiani Bendicho}, P. 1995, \apj, 455, 646

\bibitem[{{ud-Doula} {et~al.}(2014){ud-Doula}, {Owocki}, {Townsend}, {Petit},
  \& {Cohen}}]{udDoula14}
{ud-Doula}, A., {Owocki}, S., {Townsend}, R., {Petit}, V., \& {Cohen}, D. 2014,
  \mnras, 441, 3600

\bibitem[{{ud-Doula} \& {Owocki}(2002)}]{udDoula02}
{ud-Doula}, A. \& {Owocki}, S.~P. 2002, \apj, 576, 413

\bibitem[{{ud-Doula} {et~al.}(2008){ud-Doula}, {Owocki}, \&
  {Townsend}}]{udDoula08}
{ud-Doula}, A., {Owocki}, S.~P., \& {Townsend}, R.~H.~D. 2008, \mnras, 385, 97

\bibitem[{{ud-Doula} {et~al.}(2013){ud-Doula}, {Sundqvist}, {Owocki}, {Petit},
  \& {Townsend}}]{udDoula2013}
{ud-Doula}, A., {Sundqvist}, J.~O., {Owocki}, S.~P., {Petit}, V., \&
  {Townsend}, R.~H.~D. 2013, \mnras, 428, 2723

\bibitem[{{von Zeipel}(1924)}]{Zeipel24}
{von Zeipel}, H. 1924, \mnras, 84, 665

\bibitem[{{Wade} {et~al.}(2012){Wade}, {Grunhut}, \& {MiMeS
  Collaboration}}]{Wade12}
{Wade}, G.~A., {Grunhut}, J.~H., \& {MiMeS Collaboration}. 2012, in
  Astronomical Society of the Pacific Conference Series, Vol. 464,
  Circumstellar Dynamics at High Resolution, ed. A.~C. {Carciofi} \&
  T.~{Rivinius}, 405

\bibitem[{{Wade} {et~al.}(2011){Wade}, {Howarth}, {Townsend}, {Grunhut},
  {Shultz}, {Bouret}, {Fullerton}, {Marcolino}, {Martins}, {Naz{\'e}}, {Ud
  Doula}, {Walborn}, \& {Donati}}]{Wade11}
{Wade}, G.~A., {Howarth}, I.~D., {Townsend}, R.~H.~D., {et~al.} 2011, \mnras,
  416, 3160

\bibitem[{{Weber} {et~al.}(2013){Weber}, {Pauldrach}, {Knogl}, \&
  {Hoffmann}}]{Weber2013}
{Weber}, J.~A., {Pauldrach}, A.~W.~A., {Knogl}, J.~S., \& {Hoffmann}, T.~L.
  2013, \aap, 555, A35

\bibitem[{{Zsarg{\'o}} {et~al.}(2006){Zsarg{\'o}}, {Hillier}, \&
  {Georgiev}}]{Zsargo2006}
{Zsarg{\'o}}, J., {Hillier}, D.~J., \& {Georgiev}, L.~N. 2006, \aap, 447, 1093

\bibitem[{{Zsarg{\'o}} {et~al.}(2008){Zsarg{\'o}}, {Hillier}, \&
  {Georgiev}}]{Zsargo08a}
{Zsarg{\'o}}, J., {Hillier}, D.~J., \& {Georgiev}, L.~N. 2008, \aap, 478, 543

\end{thebibliography}

\appendix
\section{The discretized EQRT within the FVM}
\label{app:fvm}
\begin{figure}[t]

\resizebox{\hsize}{!}
{\includegraphics{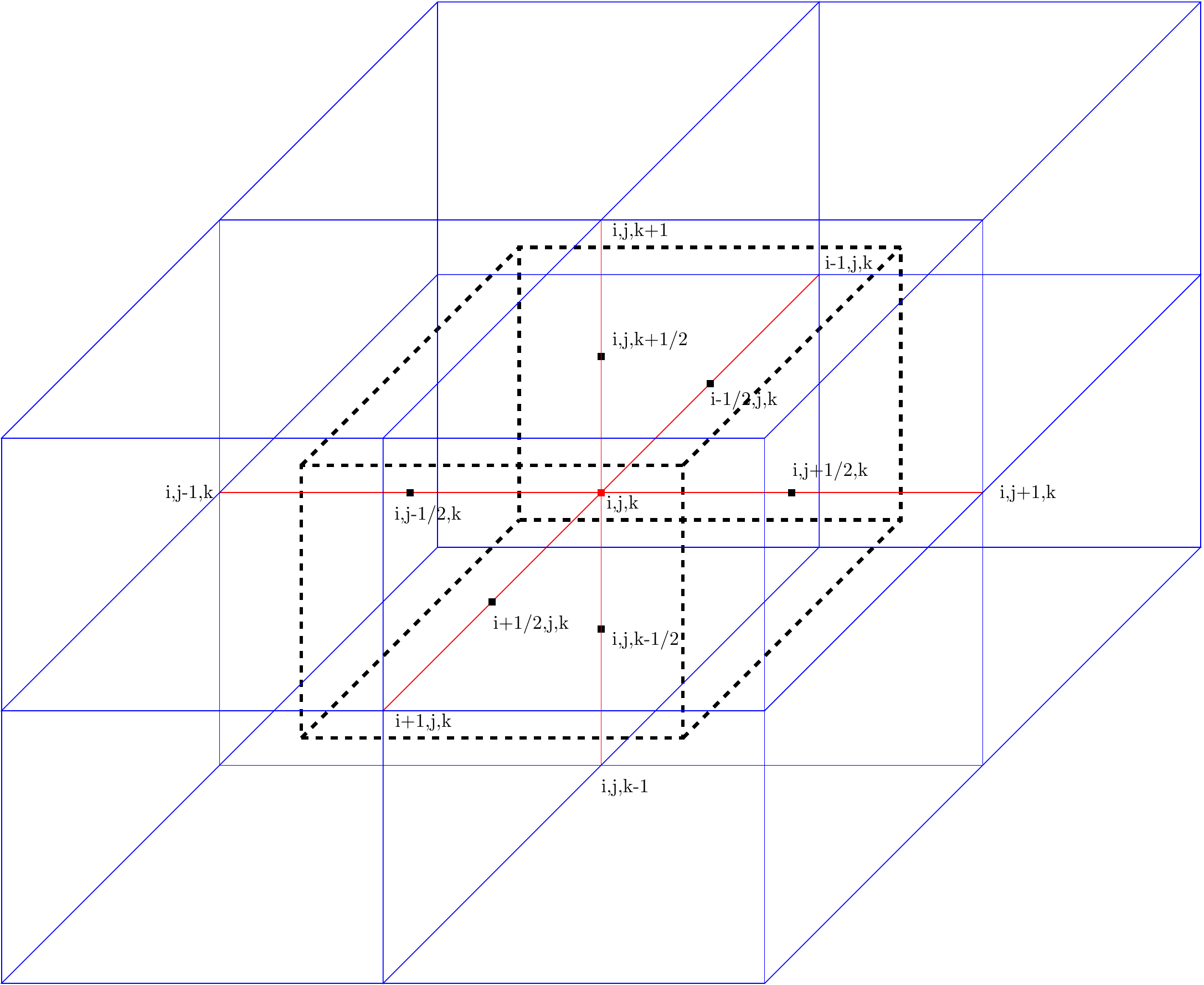}}
\caption{Geometry used within the control-volume approach: The discretized 3D
  spatial grid is shown in blue. The dashed lines indicate the control volume,
  corresponding to a grid-point (i,j,k). The control-volume surfaces are
  located at the centre between the grid-point coordinates.}
\label{fig:fvm_cell}
\end{figure}
Following the ideas of \cite{patankarbook80} and \cite{Adam90}, we here
describe the discretization scheme of the time-independent equation of
radiative transfer, Eq. \eqref{eq:eqrt}. At each grid point,
Eq. \eqref{eq:eqrt} is integrated over a finite control volume (see
Fig. \ref{fig:fvm_cell}). Applying Gauss's theorem, we obtain:
\beq
   \label{eq:eqrt_app}
   \int_{\partial V} I \vecown{n}\cdot\vecown{\dd S} = \int_{V} \chi \bigl(S - I\bigr) ~\dd V \,.
\eeq
Here and in the following, we omit the notation for the explicit frequency
dependence. The left-hand side of Eq. \eqref{eq:eqrt_app} describes the
intensity propagating into and out of the control-volume surfaces, and the
right-hand side corresponds to the grid-cell contribution from sources and
sinks.  Assuming that the variables at the grid points are appropriate mean
values within the corresponding control volume, the right-hand side is easily
integrated, and gives for grid point (i,j,k),
\begin{multline}
   \label{eq:eqrt_rhs}
   \int \chi (S - I) ~\dd V = \\ = \chi_{\rm ijk} (S_{\rm ijk} - I_{\rm ijk})
   (x_{\rm i+1/2}-x_{\rm i-1/2})(y_{\rm j+1/2}-y_{\rm j-1/2})(z_{\rm
     k+1/2}-x_{\rm k-1/2}) \,.
\end{multline}
Since we are using Cartesian coordinates, the integral on the left-hand side
can be readily calculated:
\begin{align}
\label{eq:eqtest}
   \int I & \vecown{n}\cdot\vecown{\dd S} = \nonumber \\
   = & n_x \int_{y_{\rm j-1/2}}^{y_{\rm j+1/2}}
   \int_{z_{\rm k-1/2}}^{z_{\rm k+1/2}} I(x_{\rm i+1/2},y,z) - I(x_{\rm
     i-1/2},y,z) ~\dd y \dd z + \nonumber \\
   + & n_y \int_{x_{\rm i-1/2}}^{x_{\rm i+1/2}}
   \int_{z_{\rm k-1/2}}^{z_{\rm k+1/2}} I(x,y_{\rm j+1/2},z) - I(x,y_{\rm
     j-1/2},z) ~\dd x \dd z + \nonumber \\
   + & n_z \int_{x_{\rm i-1/2}}^{x_{\rm i+1/2}}
   \int_{y_{\rm j-1/2}}^{y_{\rm j+1/2}} I(x,y,z_{\rm k+1/2}) - I(x,y,z_{\rm
     k-1/2}) ~\dd x \dd y \,.
\end{align}
Again, assuming that the intensities at the midpoints of the control-volume
surfaces are representative averages of the corresponding surfaces, we obtain:
\begin{align}
\label{eq:eqrt_lhs}
   \int I & \vecown{n}\cdot\vecown{\dd S} = \nonumber \\
   = & n_x (I_{\rm i+1/2,j,k} - I_{\rm i-1/2,j,k}) (y_{\rm j+1/2}-y_{\rm j-1/2})(z_{\rm
     k+1/2}-x_{\rm k-1/2}) + \nonumber \\ 
   + & n_y (I_{\rm i,j+1/2,k} - I_{\rm i,j-1/2,k}) (x_{\rm i+1/2}-x_{\rm i-1/2})(z_{\rm
     k+1/2}-x_{\rm k-1/2}) + \nonumber \\ 
   + & n_z (I_{\rm i,j,k+1/2} - I_{\rm i,j,k-1/2}) (x_{\rm i+1/2}-x_{\rm i-1/2})(y_{\rm
     j+1/2}-y_{\rm j-1/2}) \,.
\end{align}
Since the control-volume coordinates are positioned at the midpoints of the
grid coordinates, we substitute:
\beqa
\label{eq:app_delx}
x_{\rm i+1/2}-x_{\rm i-1/2} = \dfrac{x_{\rm i+1}-x_{\rm i-1}}{2} \,,\\
\label{eq:app_dely}
y_{\rm j+1/2}-y_{\rm j-1/2} = \dfrac{y_{\rm j+1}-y_{\rm j-1}}{2} \,,\\
\label{eq:app_delz}
z_{\rm k+1/2}-z_{\rm k-1/2} = \dfrac{z_{\rm k+1}-z_{\rm k-1}}{2} \,.
\eeqa
Finally, we use the upwind approximation to replace the (unknown) intensities
at the control-volume surfaces:
\begin{align*}
   &\begin{rcases}
      I_{\rm i+1/2,j,k} &\rightarrow I_{\rm i,j,k}  \\
       I_{\rm i-1/2,j,k} &\rightarrow I_{\rm i-1,j,k} \\
      \alpha &:= 1 
   \end{rcases}
   \makebox[12ex][l]{for $n_x > 0 \,,$}
   \begin{rcases}
      I_{\rm i+1/2,j,k} &\rightarrow I_{\rm i+1,j,k}  \\
       I_{\rm i-1/2,j,k} &\rightarrow I_{\rm i,j,k} \\
      \alpha &:= -1 
   \end{rcases}
   \makebox[12ex][l]{for $n_x < 0 \,,$} \\
   &\begin{rcases}
      I_{\rm i,j+1/2,k} &\rightarrow I_{\rm i,j,k}  \\
       I_{\rm i,j-1/2,k} &\rightarrow I_{\rm i,j-1,k} \\
      \beta &:= 1 
   \end{rcases}
   \makebox[12ex][l]{for $n_y > 0 \,, \quad$}
   \begin{rcases}
      I_{\rm i,j+1/2,k} &\rightarrow I_{\rm i,j+1,k}  \\
       I_{\rm i,j-1/2,k} &\rightarrow I_{\rm i,j,k} \\
      \beta &:= -1 
   \end{rcases}
   \makebox[12ex][l]{for $n_y < 0 \,,$} \\
   &\begin{rcases}
      I_{\rm i,j,k+1/2} &\rightarrow I_{\rm i,j,k}  \\
      I_{\rm i,j,k-1/2} &\rightarrow I_{\rm i,j,k-1} \\
      \gamma &:= 1 
   \end{rcases}
   \makebox[12ex][l]{for $n_z > 0 \,, \quad$}
   \begin{rcases}
      I_{\rm i,j,k+1/2} &\rightarrow I_{\rm i,j,k+1}  \\
      I_{\rm i,j,k-1/2} &\rightarrow I_{\rm i,j,k} \\
      \gamma &:= -1 
   \end{rcases}
   \makebox[12ex][l]{for $n_z < 0 \,.$}
\end{align*}
Combining equations \eqref{eq:eqrt_app}, \eqref{eq:eqrt_rhs},
\eqref{eq:eqrt_lhs}, \eqref{eq:app_delx}-\eqref{eq:app_delz}, and the
definitions of $\alpha, \beta, \gamma$, the discretized \eqrt ~finally reads:
\begin{align}
\label{eq:eqrt_disc0}
 &n_x (I_{\rm ijk} - I_{\rm i-\alpha j k}) \dfrac{y_{\rm j+\beta} - y_{\rm
    j-\beta}}{2}\dfrac{z_{\rm k+\gamma} - z_{\rm k-\gamma}}{2} + \nonumber \\
 + & n_y (I_{\rm ijk} - I_{\rm i j-\beta k}) \dfrac{x_{\rm i+\alpha} - x_{\rm
    i-\alpha}}{2}\dfrac{z_{\rm k+\gamma} - z_{\rm k-\gamma}}{2} +  \nonumber \\
 + & n_z (I_{\rm ijk} - I_{\rm i j k-\gamma}) \dfrac{x_{\rm i+\alpha} - x_{\rm
    i-\alpha}}{2}\dfrac{y_{\rm j+\beta} - y_{\rm j-\beta}}{2} = \nonumber \\
 = &\Bigl[ \chicijk \scontijk + \chibarijk \profile^{\rm ijk} \slineijk -
  (\chicijk + \chibarijk \profile^{\rm ijk}) I_{\rm ijk} \Bigr] \times \nonumber\\
   & \makebox[.8\linewidth][r]{$\times \dfrac{x_{\rm i+\alpha} - x_{\rm i-\alpha}}{2} \dfrac{y_{\rm j+\beta} - y_{\rm
    j-\beta}}{2} \dfrac{z_{\rm k+\gamma} - z_{\rm k-\gamma}}{2} \,,$}
\end{align}
where we already have separated the continuum and line contribution of the
opacity and source function. Collecting terms, and solving for $I_{\rm ijk}$
leads to:
\begin{align}
\label{eq:eqrt_disc_app}
I_{\rm ijk} &= \dfrac{\chicijk}{\chicijk + \chibarijk \profile^{\rm (ijk)}
  +\frac{2 n_x}{x_{\rm i+\alpha} - x_{\rm i-\alpha}} + \frac{2 n_y}{y_{\rm
      j+\beta} - y_{\rm j-\beta}} + \frac{2 n_z}{z_{\rm k+\gamma} - z_{\rm
      k-\gamma}}} \scontijk + \nonumber \\
&+ \dfrac{\chibarijk \profile^{\rm (ijk)}}{\chicijk + \chibarijk \profile^{\rm (ijk)}
  +\frac{2 n_x}{x_{\rm i+\alpha} - x_{\rm i-\alpha}} + \frac{2 n_y}{y_{\rm
      j+\beta} - y_{\rm j-\beta}} + \frac{2 n_z}{z_{\rm k+\gamma} - z_{\rm
      k-\gamma}}} \slineijk + \nonumber\\
&+ \dfrac{\frac{2n_x}{x_{\rm i+\alpha} - x_{\rm i-\alpha}}}{\chicijk + \chibarijk \profile^{\rm (ijk)}
  +\frac{2 n_x}{x_{\rm i+\alpha} - x_{\rm i-\alpha}} + \frac{2 n_y}{y_{\rm
      j+\beta} - y_{\rm j-\beta}} + \frac{2 n_z}{z_{\rm k+\gamma} - z_{\rm
      k-\gamma}}} I_{\rm i-\alpha jk} + \nonumber \\
&+ \dfrac{\frac{2n_y}{y_{\rm j+\beta} - y_{\rm j-\beta}}}{\chicijk + \chibarijk \profile^{\rm (ijk)}
  +\frac{2 n_x}{x_{\rm i+\alpha} - x_{\rm i-\alpha}} + \frac{2 n_y}{y_{\rm
      j+\beta} - y_{\rm j-\beta}} + \frac{2 n_z}{z_{\rm k+\gamma} - z_{\rm
      k-\gamma}}} I_{\rm i j-\beta k}+ \nonumber \\
&+ \dfrac{\frac{2n_z}{z_{\rm k+\gamma} - z_{\rm k-\gamma}}}{\chicijk + \chibarijk \profile^{\rm (ijk)}
  +\frac{2 n_x}{x_{\rm i+\alpha} - x_{\rm i-\alpha}} + \frac{2 n_y}{y_{\rm
      j+\beta} - y_{\rm j-\beta}} + \frac{2 n_z}{z_{\rm k+\gamma} - z_{\rm
      k-\gamma}}} I_{\rm i j k-\gamma} \,.
\end{align}
\section{UV line profiles for different ADM models}
%
\begin{figure*}[h!]
\resizebox{\hsize}{!}{
   \begin{minipage}{0.36\hsize}
      \resizebox{\hsize}{!}{\includegraphics{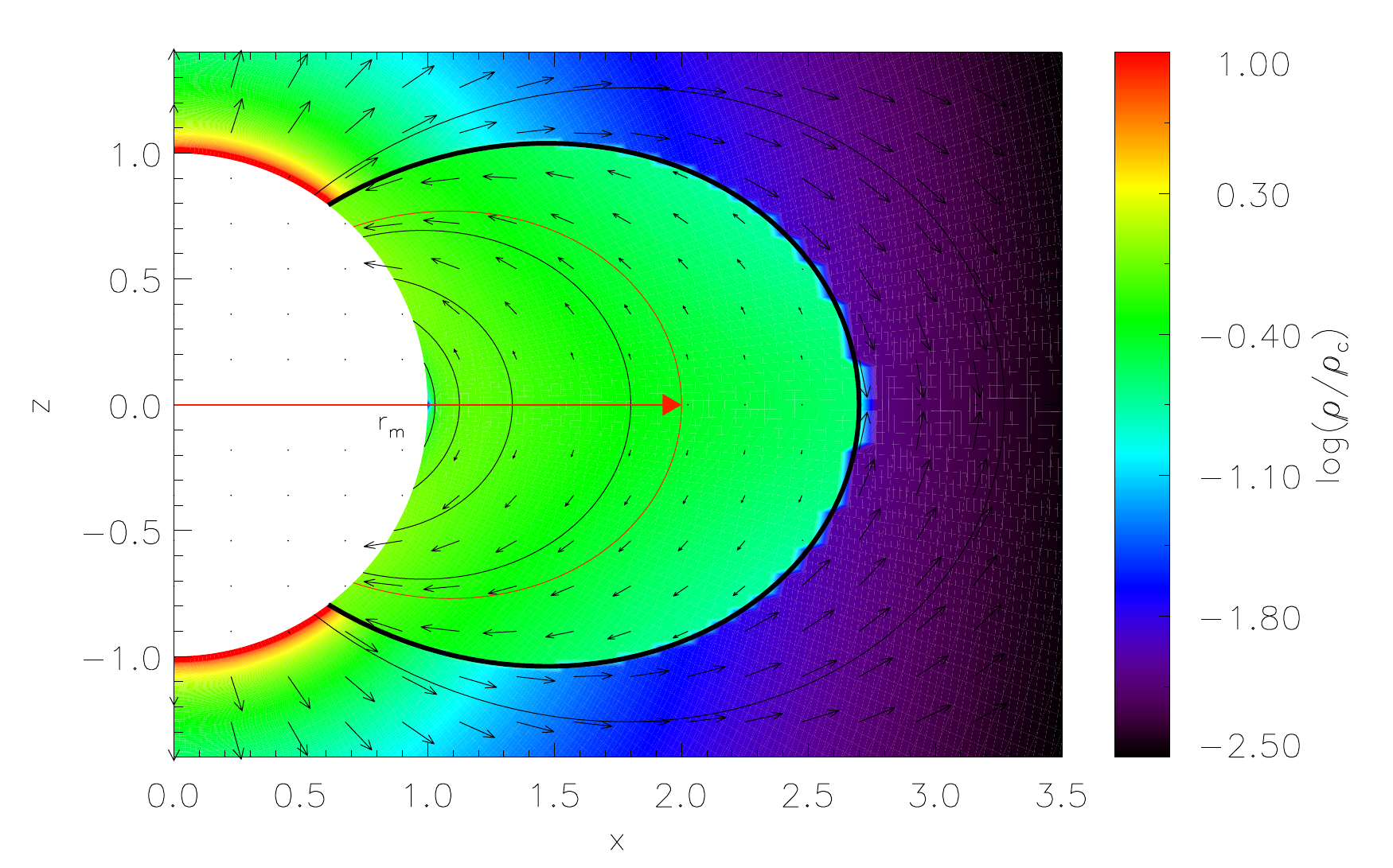}}
   \end{minipage}
   \begin{minipage}{0.32\hsize}
      \resizebox{\hsize}{!}{\includegraphics{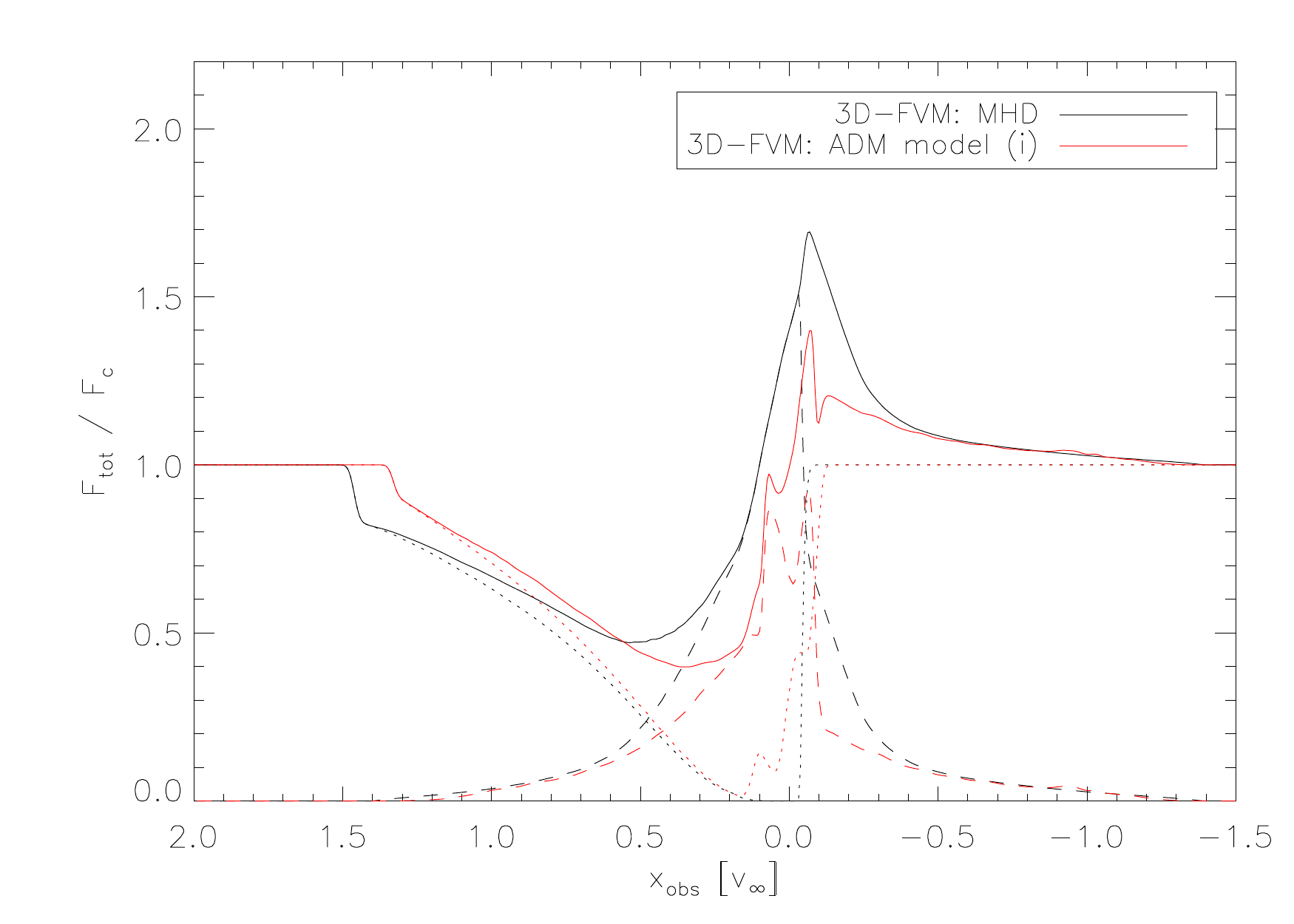}}
   \end{minipage}
   \begin{minipage}{0.32\hsize}
      \resizebox{\hsize}{!}{\includegraphics{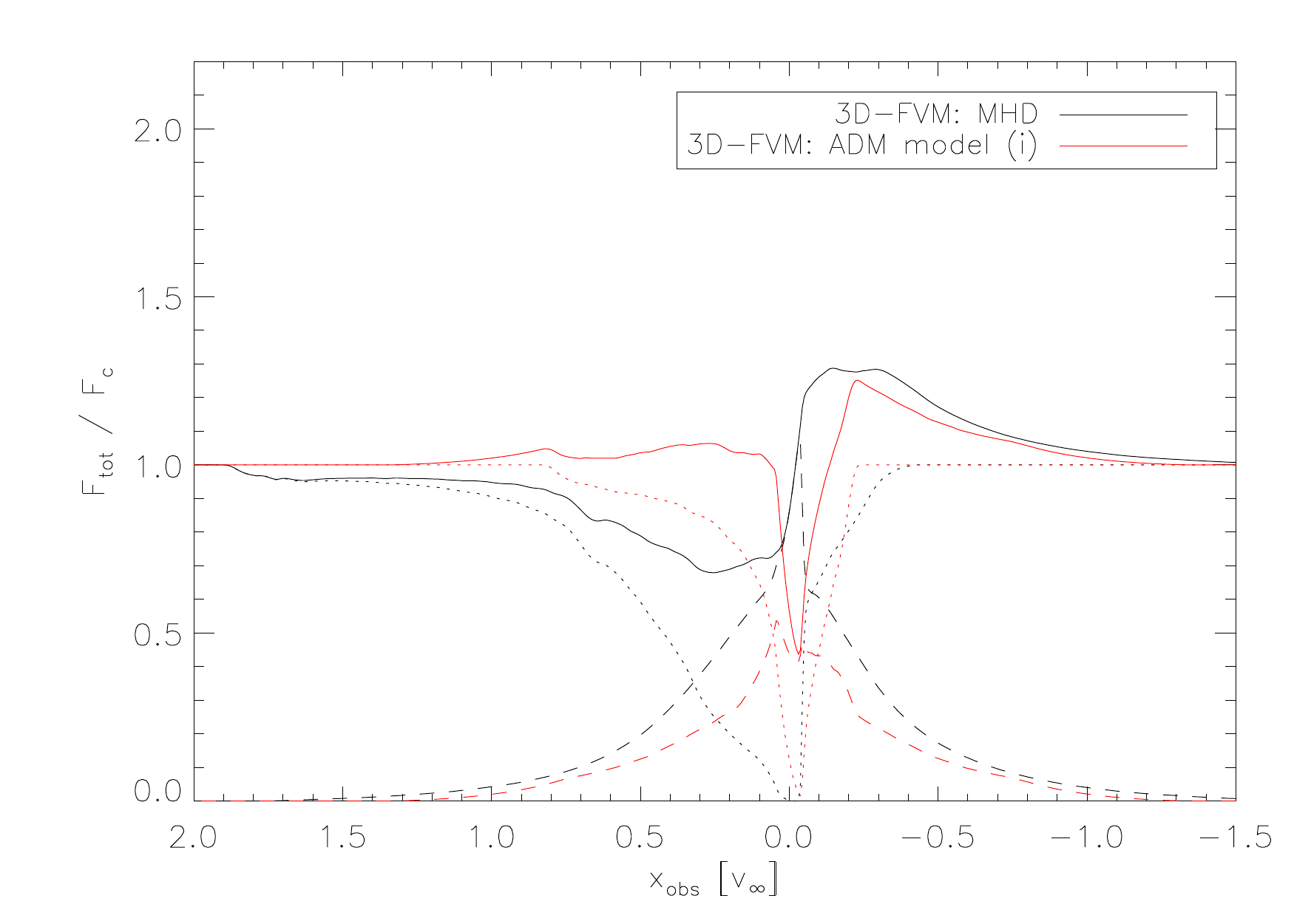}}
   \end{minipage}
}
\\
\resizebox{\hsize}{!}{
   \begin{minipage}{0.36\hsize}
      \resizebox{\hsize}{!}{\includegraphics{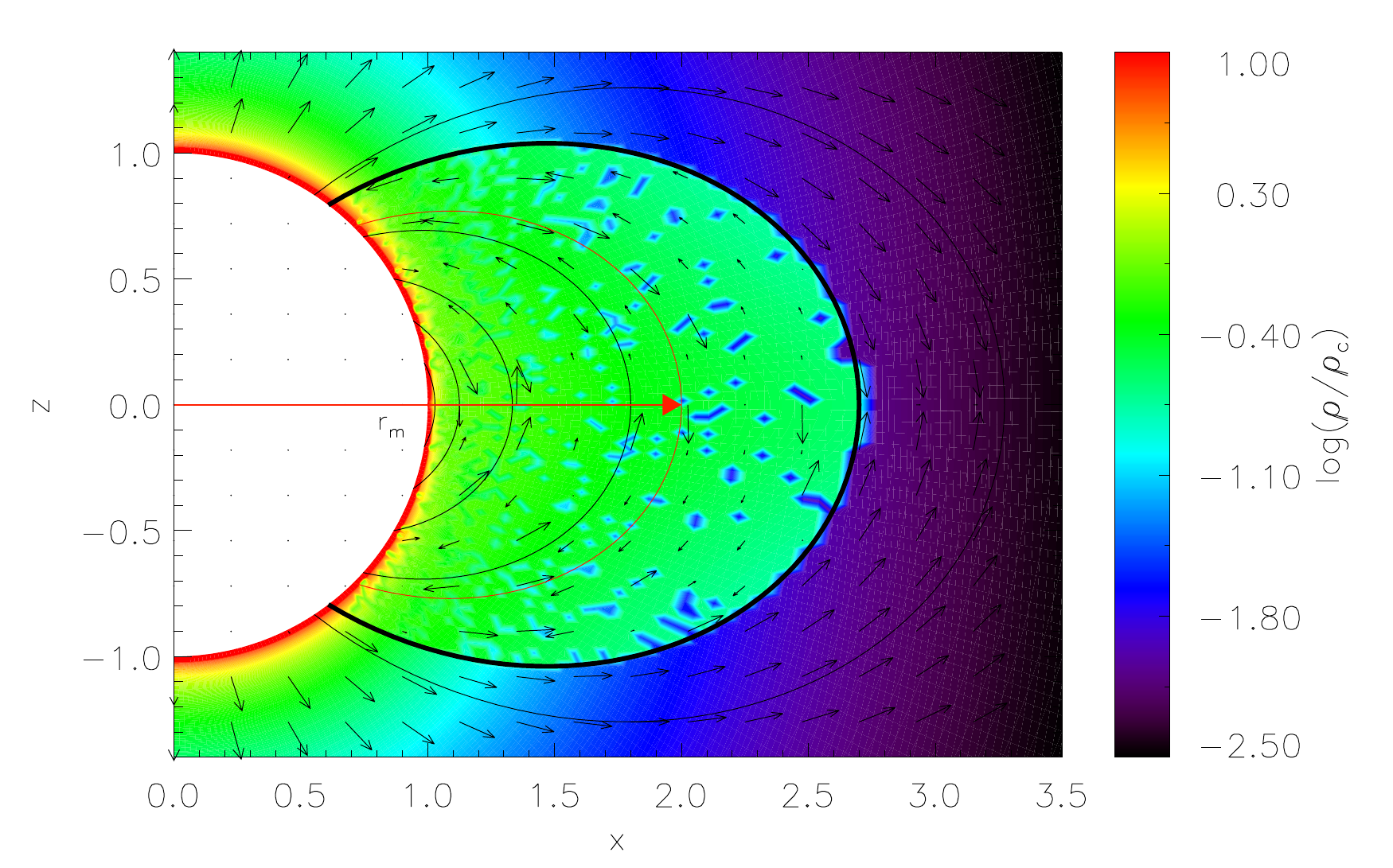}}
   \end{minipage}
   \begin{minipage}{0.32\hsize}
      \resizebox{\hsize}{!}{\includegraphics{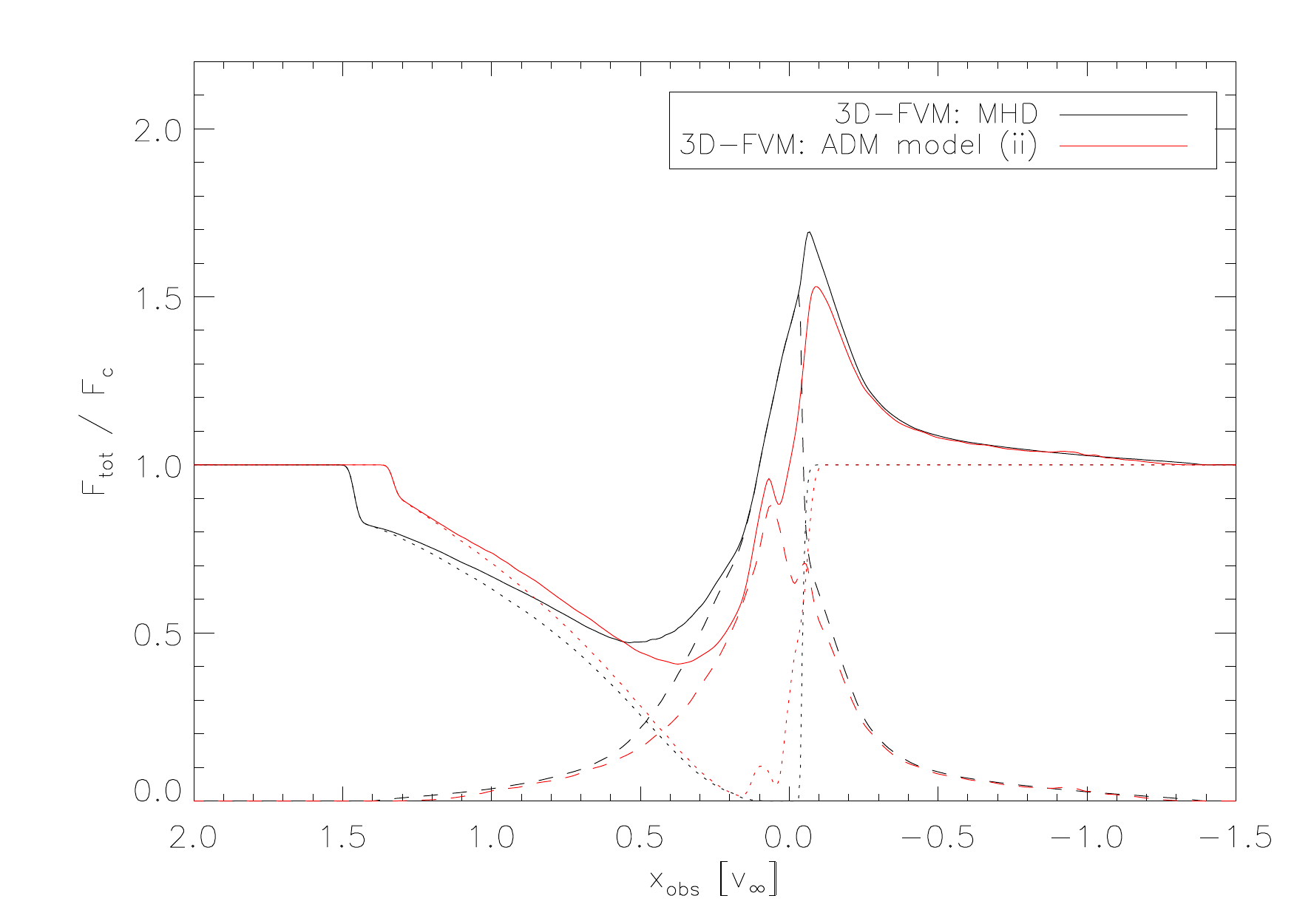}}
   \end{minipage}
   \begin{minipage}{0.32\hsize}
      \resizebox{\hsize}{!}{\includegraphics{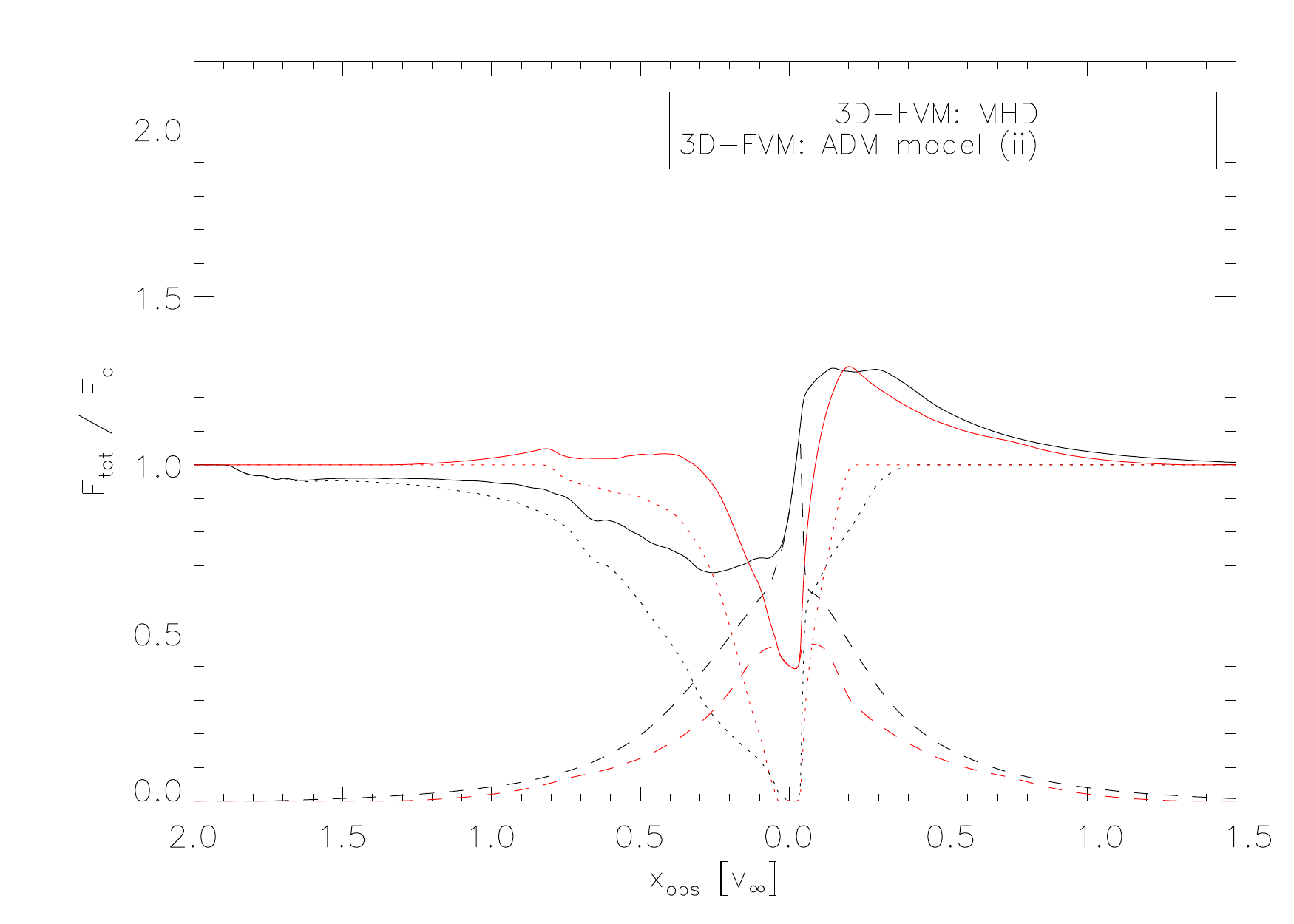}}
   \end{minipage}
}
\\
\resizebox{\hsize}{!}{
   \begin{minipage}{0.36\hsize}
      \resizebox{\hsize}{!}{\includegraphics{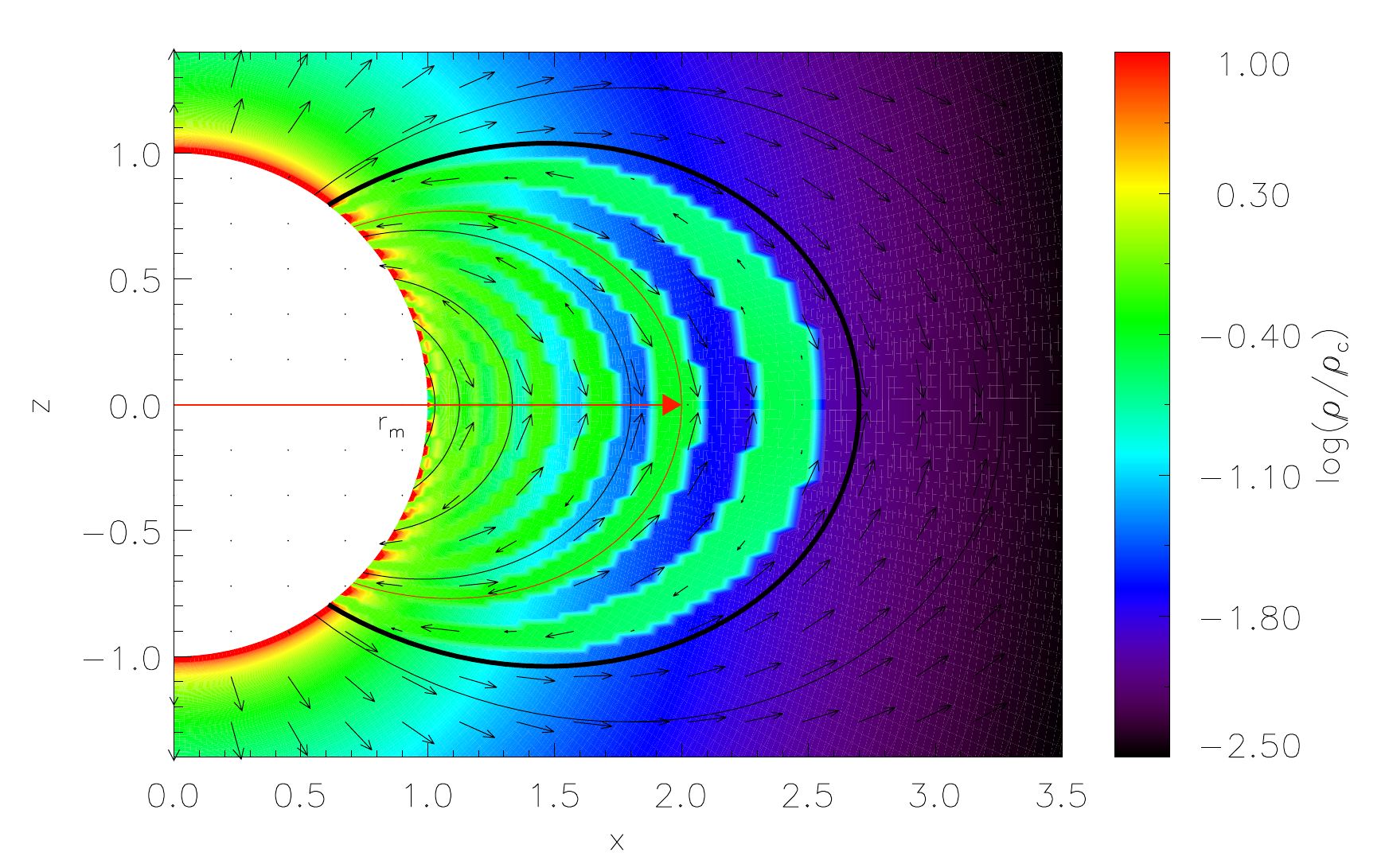}}
   \end{minipage}
   \begin{minipage}{0.32\hsize}
      \resizebox{\hsize}{!}{\includegraphics{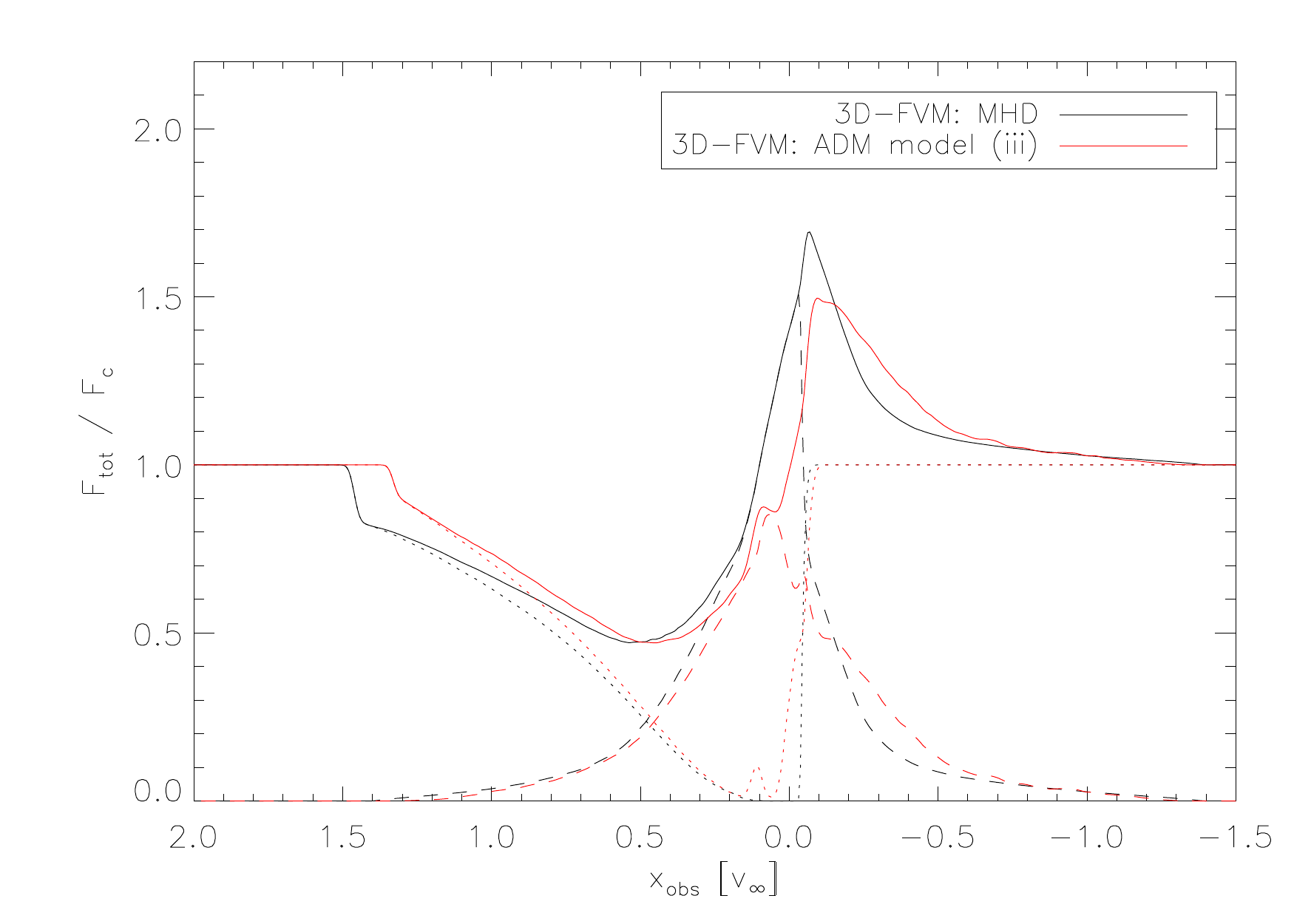}}
   \end{minipage}
   \begin{minipage}{0.32\hsize}
      \resizebox{\hsize}{!}{\includegraphics{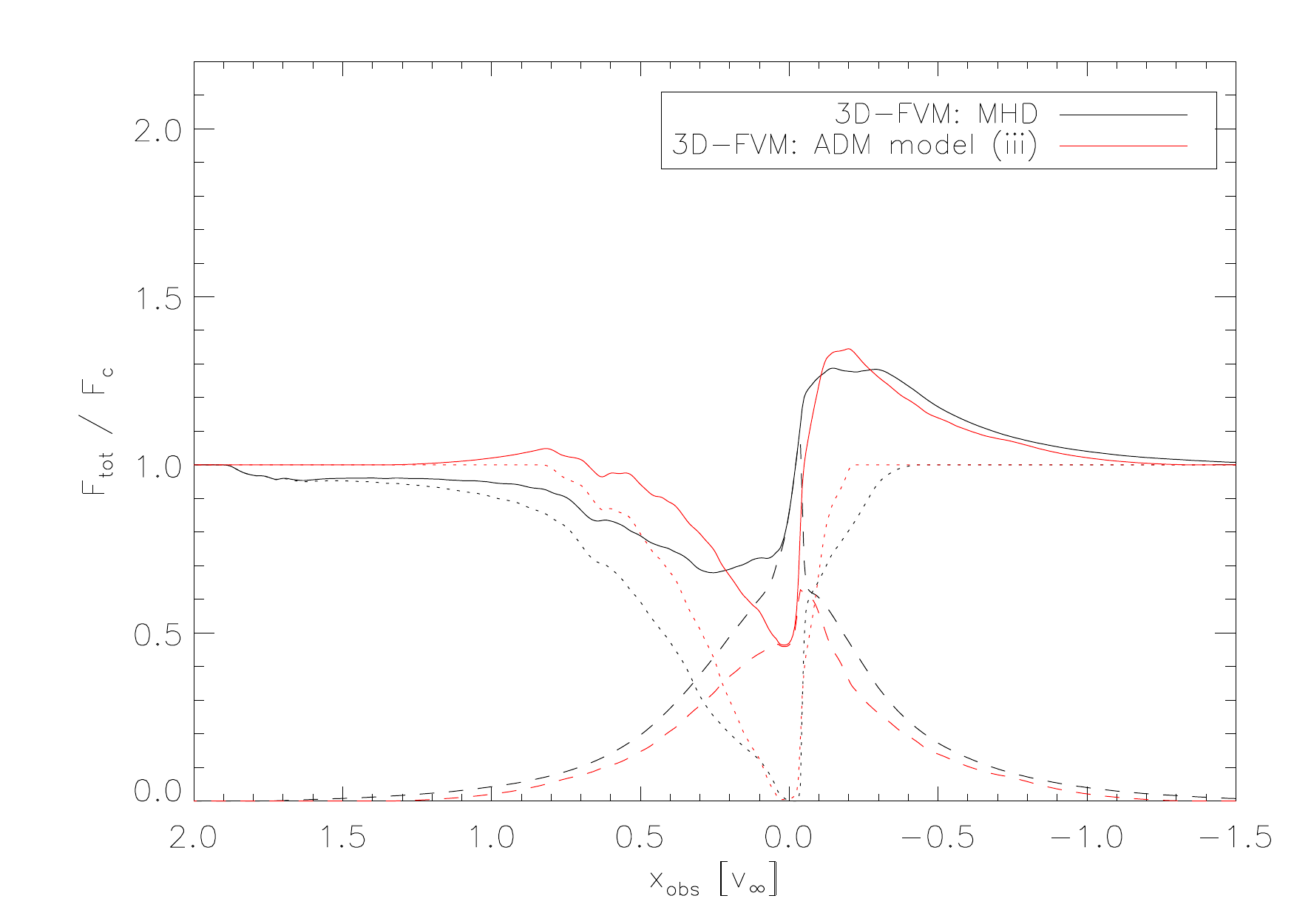}}
   \end{minipage}
}
\\
\resizebox{\hsize}{!}{
   \begin{minipage}{0.36\hsize}
      \resizebox{\hsize}{!}{\includegraphics{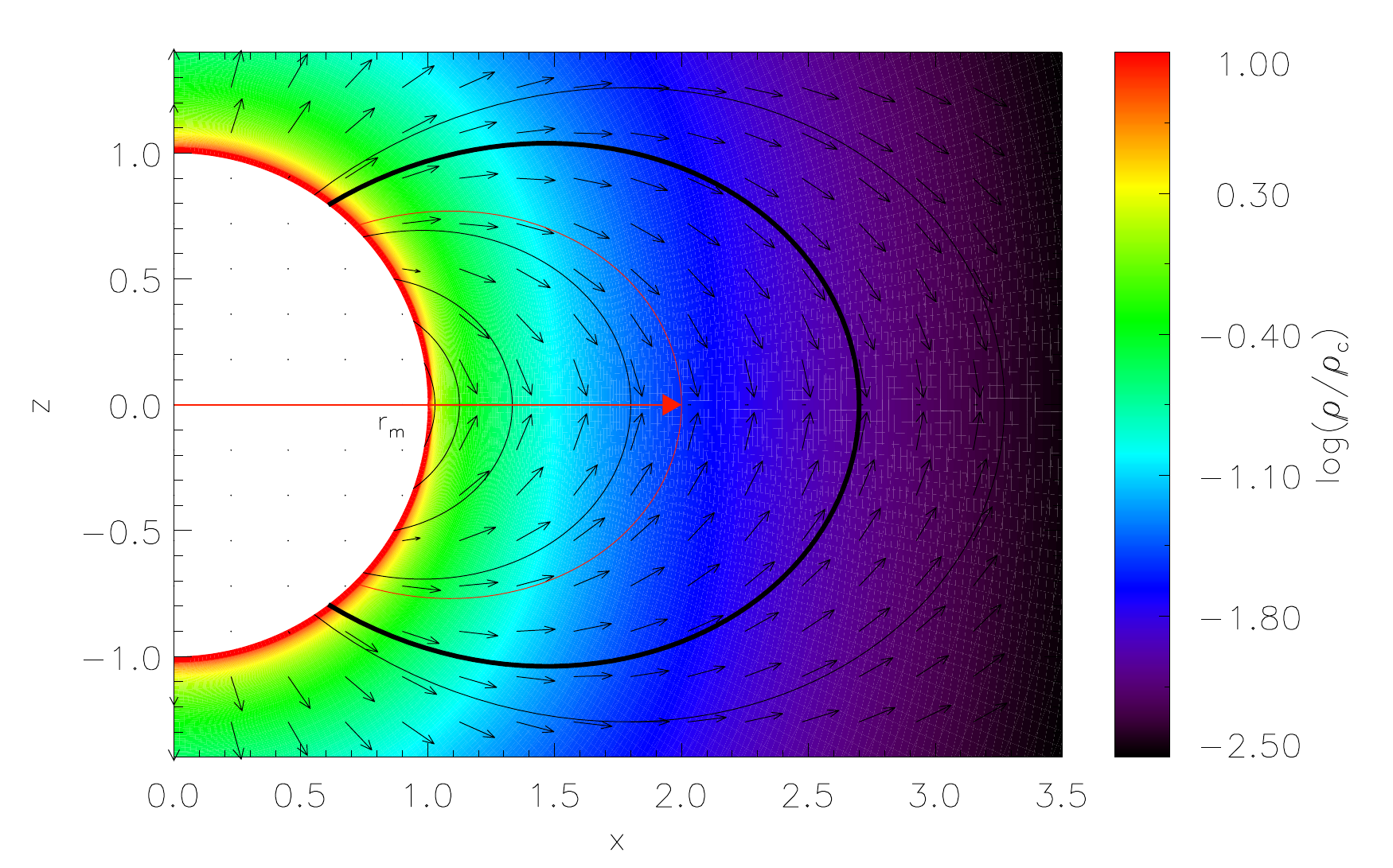}}
   \end{minipage}
   \begin{minipage}{0.32\hsize}
      \resizebox{\hsize}{!}{\includegraphics{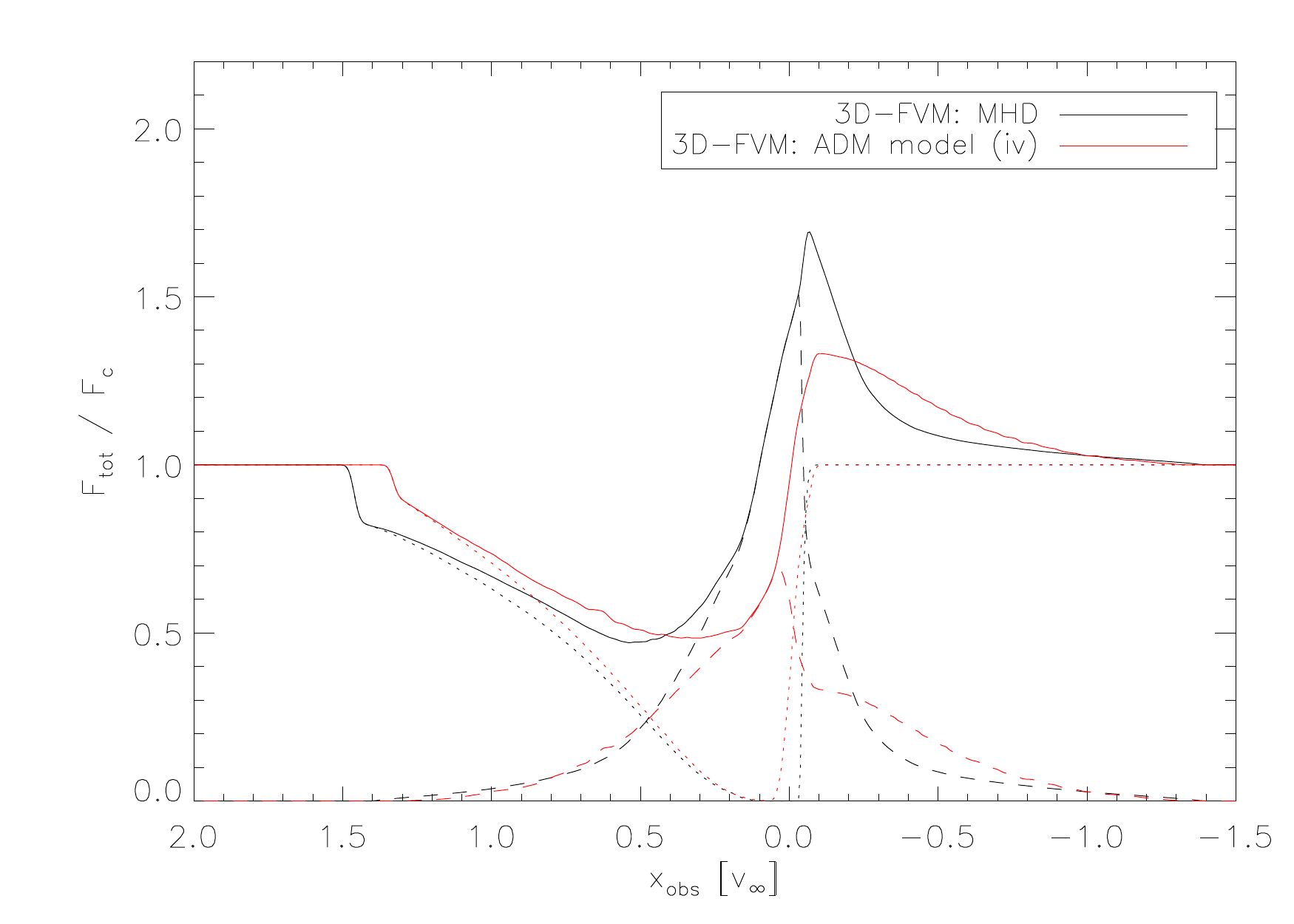}}
   \end{minipage}
   \begin{minipage}{0.32\hsize}
      \resizebox{\hsize}{!}{\includegraphics{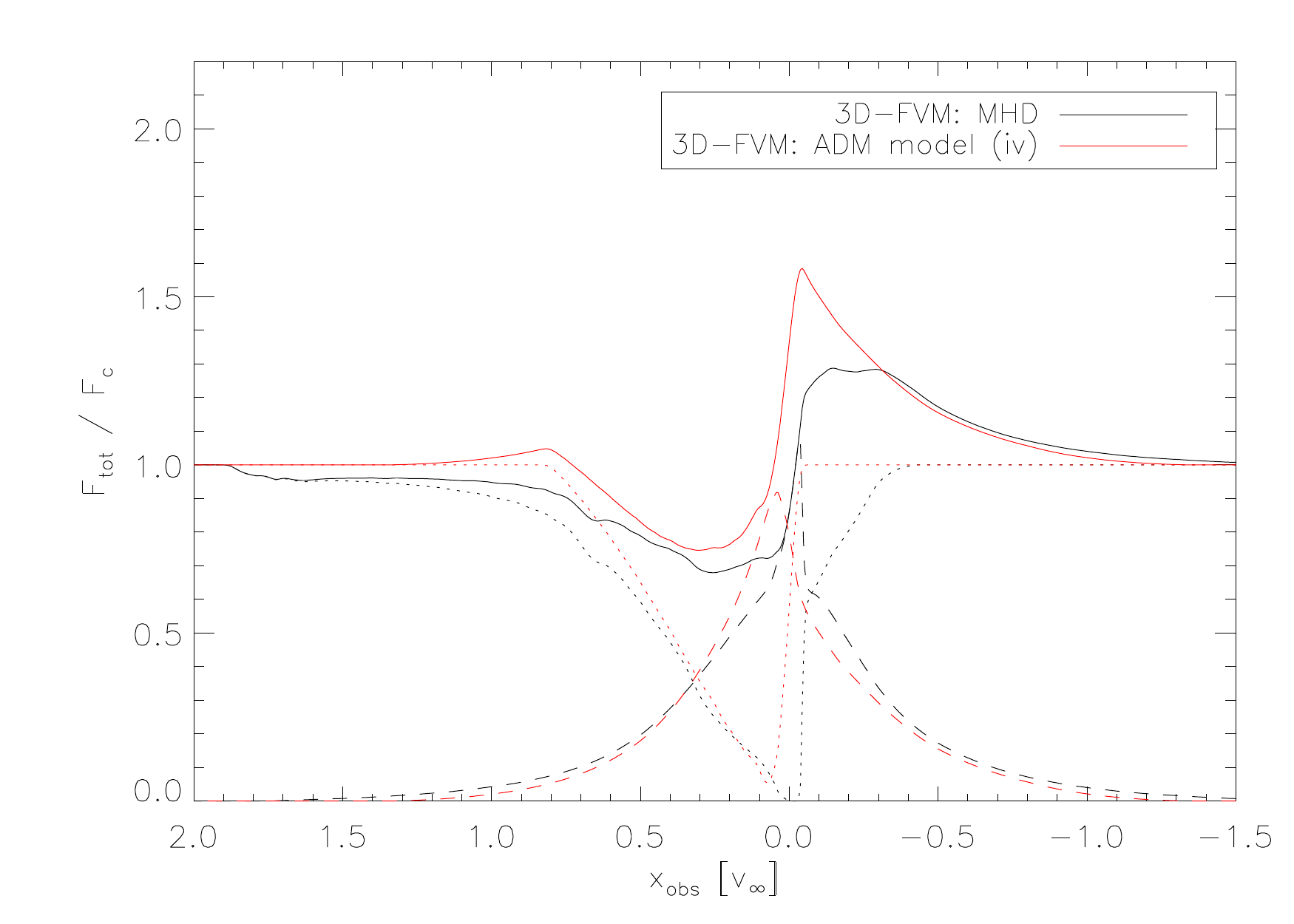}}
   \end{minipage}
}
\caption{\textit{Left panel:} As Fig.~\ref{fig:density_mhd}, but for the
  corresponding ADM model structures. To clarify the definition of the apex
  radius, \rapex, we have displayed a specific value, \rapex=2~\Rstar, as a
  red arrow, where this value corresponds to all points located on the red
  magnetic-field line. In the closed-field region (inside \rapex=\ralf,
  displayed by a thick line), the models contain, from top to bottom: (i) The
  cooled-downflow component alone. (ii) A statistical approach for the
  downflow and upflow component. (iii) Alternating flux tubes with
  cooled-downflow and wind-upflow component. (iv) The wind-upflow component
  alone. \textit{Middle panel:} As Fig. \ref{fig:profile_adm}, for pole-on
  observers, and for the different ADM models (i) to (iv). The dashed and
  dotted lines display the emission part and the absorption part of the line
  profiles, respectively.  \textit{Right panel:} As middle panel, but for
  equator-on observers.}
\label{fig:profiles_density_adm}
\end{figure*}

\end{document}